\documentclass[11pt]{article}
\usepackage{latexsym,array,theorem,mathrsfs,bm,float}
\usepackage{psfrag}
\usepackage{amsfonts,amsmath,amssymb,latexsym,array,afterpage,
theorem,mathrsfs,bm,float,epsfig,color,graphicx,tabularx,here,multirow}
\usepackage[dvipdfmx]{hyperref}
\usepackage{fancybox}

  \makeatletter
    
    \@addtoreset{equation}{section}
  \makeatother

\usepackage[top=30mm,bottom=35mm,left=25mm,right=25mm]{geometry}

\newcommand{\bea}{\begin{eqnarray}}
\newcommand{\ena}{\end{eqnarray}}
\newcommand{\beann}{\begin{eqnarray*}}
\newcommand{\enann}{\end{eqnarray*}}

\newcommand{\ma}[1]{\mbox{$\mathcal{#1}$}}

\newcommand{\calhR}[1]{\raisebox{2ex}{\tiny ({\em h})}\hspace{-0.8em}{\ma R}}
\newcommand{\pd}{\partial}
\newcommand{\BS}{\boldsymbol}
\newcommand{\MC}{\mathcal}

\newcommand{\MB}{\mathbb}

\newcommand{\p}{\prime}

\newif\iffigure
\figurefalse
\figuretrue

\begin{document}

\title{\LARGE{\bf{
Gravitational and Gravitoscalar Thermodynamics }}
}

\author{
\\
Shoichiro Miyashita\thanks{e-mail address : s-miyashita''at''aoni.waseda.jp} 
\\ \\ 
{\it Department of Physics, Waseda University, 
Okubo 3-4-1, Shinjuku, Tokyo 169-8555, Japan} 
\\ \\
}

\date{~}

\thisfancyput(13.cm,1cm){
{\fbox{WUCG-21-08}}}
\maketitle
\begin{abstract}
Gravitational thermodynamics and gravitoscalar thermodynamics with $S^2 \times \MB{R}$ boundary geometry are investigated through the partition function, assuming that all Euclidean saddle point geometries contribute to the path integral and dominant ones are in the $B^3 \times S^1$ or $S^2 \times Disc$ topology sector. In the first part, I concentrate on the purely gravitational case with or without a cosmological constant and show there exists a new type of saddle point geometry, which I call the ``bag of gold(BG) instanton,'' only for the $\Lambda>0$ case. Because of this existence, thermodynamical stability of the system and the entropy bound are absent for $\Lambda>0$, these being universal properties for $\Lambda \leq 0$. In the second part, I investigate the thermodynamical properties of a gravity-scalar system with a $\varphi^2$ potential. I show that when $\Lambda \leq 0$ and the boundary value of scalar field $J_{\varphi}$ is below some value, then the entropy bound and thermodynamical stability do exist. When either condition on the parameters does not hold, however, thermodynamical stability is (partially) broken. The properties of the system and the relation between BG instantons and the breakdown  are discussed in detail.
\end{abstract}

\clearpage

\tableofcontents

\clearpage

\section{Introduction}
At the moment, we do not know how to track the fine-grained dynamics of quantum gravity. However, one promising thing is that gravity would reach (thermal) equilibrium as do usual systems with many degrees of freedom. The outset is the discovery of black hole (BH) thermodynamics \cite{BardeenCarterHawking, Hawking}, which states that BHs can be treated as if they have temperature and entropy and obey the thermodynamical laws. To uncover this mysterious phenomena, Gibbons and Hawking tried to derive these laws by a statistical treatment; that is, they defined the partition function by the path integral of gravity, and actually re-derived the thermodynamical laws \cite{GibbonsHawking}. Indeed, the system they considered is thermodynamically unstable in the sense that it has a negative heat capacity. However, what they considered is the partition function of gravity (not of BHs) with the asymptotically flat boundary condition. Later, Hawking and Page introduced a negative cosmological constant and changed the boundary condition to the asymptotically Anti-de Sitter (AdS) one \cite{HawkingPage}. They showed that the system is thermodynamically stable and exhibits a phase transition, which is now called the Hawking--Page phase transition. Therefore, gravity actually thermalizes and we could elucidate the coarse-grained dynamics of quantum gravity through the partition function.  

For pure gravity, the thermodynamical properties of the following four cases are known in the literature: \\
\hspace{1cm} $\bullet$ $\Lambda=0$, asymptotically flat boundary condition \cite{GibbonsHawking} \\
\hspace{1cm} $\bullet$ $\Lambda<0$, asymptotically AdS boundary condition \cite{HawkingPage} \\
\hspace{1cm} $\bullet$ $\Lambda=0$, finite radius $S^{2} \times \MB{R}$ boundary condition \cite{York2} \\
\hspace{1cm} $\bullet$ $\Lambda<0$, finite radius $S^{2} \times \MB{R}$ boundary condition \cite{BrownCreightonMann} \\
The first one, as mentioned above, is thermodynamically unstable. The second, third, and fourth ones share qualitatively the same properties: they are thermodynamically stable and exhibit a Hawking--Page phase transition. In this paper, after reviewing the third and fourth cases, I investigate the case of $\Lambda>0$ and finite radius $S^{2} \times \MB{R}$ boundary condition to know completely the thermodynamical properties of pure gravity. After that, as a simple extension, I consider a gravity-scalar system with a $\varphi^2$ potential to obtain intuitions about the coarse-grained dynamics/properties of quantum gravity when it couples to matter fields. In addition, in light of holography \cite{tHooft, Susskind}, these will be some clues to specify what class of quantum field theory corresponds to gravity when it is not with the asymptotically AdS boundary condition.

The paper is organized as follows. In section 2, the path integral representation of the partition function of a gravity(-scalar) system is briefly reviewed and I also mention the minisuperspace used in this paper. In section 3, gravitational thermodynamics with or without a cosmological constant is discussed. Subsections 3.1 and 3.2 are devoted to reviewing the thermodynamics of gravity with $\Lambda \leq 0$ and finite radius $S^2 \times \MB{R}$ boundary condition.  In subsection  3.3, thermodynamics with $\Lambda > 0$ and finite radius $S^2 \times \MB{R}$ boundary condition is investigated. I show that there exists a new type of Euclidean saddle point geometry that makes the thermodynamical properties drastically different from those in the $\Lambda \leq 0$ cases. In section 4, I introduce a scalar field with a $\varphi^2$  potential into the system and investigate its thermodynamical properties. Before discussing thermodynamical properties, in subsection 4.1 I show the conditions for the existence of saddle points. In subsection 4.2, thermodynamical properties are discussed. In the system, in order to obtain saddle point geometries, we have to employ numerical calculation. The detail of the calculation and some examples of saddle points are shown in Appendix A. Section 5 is devoted to a summary and discussions.

\section{Partition functions for gravity-scalar system}
The partition function of canonical ensembles for gravity may be obtained by Euclidean gravitational path integral, which includes the summation of manifold topology $\MC{M}$ \cite{GibbonsHawking, York2}: 
\bea
Z_{c}= \sum_{\MC{M}} \int \MC{D} \BS{g} ~ e^{-I_{c}^{E}[\BS{g}]}
\label{EQ2path}
\ena
$I^{E}_{c}[\BS{g}]$ is the Euclidean gravitational action functional for canonical ensembles, that is, for Dirichlet type boundary conditions:
\bea
I_{c}^{E}[\BS{g}]= \frac{-1}{16\pi G}\int_{\MC{M}} d^4 x \sqrt{g} (\MC{R}-2\Lambda)+ \frac{-1}{8\pi G} \int_{\pd \MC{M}} d^3 y \sqrt{\gamma} (\Theta- \Theta_{sub})
\ena
$\Theta$ is the extrinsic curvature of $\pd \MC{M}$ and $\Theta_{sub}$ is the subtraction term needed for regularization in case on-shell actions diverge. Although it is not necessary for the analysis in this paper, for convenience I use it for setting the (free) energy of the ground states to zero. I adopt the Mann counterterm \cite{Mann} for the subtraction term, which has been shown to work well for the purpose for spacetimes with $S^{2}\times \MB{R}$ boundary geometry \cite{Miyashita2, KrausLarsenSiebelink}:
\footnote{
Precisely, in \cite{Miyashita2}, the Mann counterterm and its straightforward generalizations were shown to set the ground state energy to zero for spacetimes with $S^{d-2} \times \MB{R}$ boundary geometry.
}
\bea
\Theta_{sub}=\Theta_{Mann}(\BS{\gamma})=\sqrt{2\MC{R}^{(3)}-\frac{4}{3}\Lambda } \label{Mann}
\ena
I concentrate only on the case that the boundary manifold has $S^2 \times S^1$ topology and whose geometry is the product of a geometric 2-sphere and a circle. This corresponds to the partition function of thermal equilibrium of a quantum spacetime whose boundary geometry is $S^2 \times \MB{R}$. Suppose the dominant paths come from the saddle points of relatively simple topology sectors, say $S^2 \times D(isc)$ and $B^3 \times S^1$; then, Eq. (\ref{EQ2path})
 can be approximated 
\bea
Z_{c}(\beta, A; \Lambda)\simeq \sum_{{\rm saddles ~ of ~} S^2 \times D} e^{-I^{E,os}_{c}} + \sum_{{\rm saddles ~ of ~} B^3 \times S^1} e^{-I^{E,os}_{c}}
\ena
``os'' represents on-shell. Additionally, I assume the dominant paths are in a simple minisuperspace,
\footnote{Originally, superspace is the configuration space of general relativity  and minisuperspace means some subset of it. I will abuse the word ``minisuperspace'' in the context of gravitational path integral.   }
a class of 4-dim.\ metrics whose member takes the following form:
\bea
\BS{g}(x)= f(r)d\tau^2 + \frac{N^2}{f(r)}dr^2 + R(r)^2 (d\theta^2 + \sin^2 \theta d\phi^2 )
\ena
In this setup, the canonical partition function and microcanonical partition function of finite boundary volume were investigated previously in \cite{York2, BCMMWY, BradenWhitingYork, HalliwellLouko, LoukoWhiting, MelmedWhiting, Miyashita1, TucciHellerLehners}. I will review $\Lambda \leq 0$ cases and give some new results on the $\Lambda > 0$ case in section 3.

As long as we concentrate only on saddle points, we can fix the gauge as
\bea
\BS{g}(x)= f(r)d\tau^2 + \frac{1}{f(r)}dr^2 + r^2 (d\theta^2 + \sin^2 \theta d\phi^2 ) ~ .
\ena
I will use $r_{b}$ for the boundary value of the $r$ coordinate that satisfies $A=4\pi r_{b}^2$. Energy can be defined by the Brown--York tensor \cite{BrownYork1} and, according to my choice of counterterm, it is given by
\bea
E= \frac{r_{b}}{G}\left[ \sqrt{1-\frac{\Lambda }{3}r_{b}} -\sqrt{f(r_{b})} \right]
\ena
when the $r$ coordinate decreases from the boundary $r=r_{b}$ and
\bea
E= \frac{r_{b}}{G}\left[ \sqrt{1-\frac{\Lambda }{3}r_{b}} +\sqrt{f(r_{b})} \right]
\ena
when it increases from the boundary. 

The objective of this paper is to investigate the thermodynamical properties of a gravity-scalar system. Especially, I will consider a scalar field with a $\varphi^2$ potential. The action functional I will use in section 4 is
\bea
I_{c}^{E}[\BS{g},\varphi]=I^{E}_{c}[\BS{g}]+\int_{\MC{M}} d^4 x \sqrt{g} \left[ \frac{1}{2}g^{\mu\nu} \pd_{\mu}\varphi \pd_{\nu} \varphi  + \frac{1}{2} M^2 \varphi^2 \right] \label{EQ2grasca}
\ena
where $M^2$ is the mass of the scalar field and $\frac{\Lambda}{8\pi G}$ can be considered as the minimum of the potential $V(\varphi)\equiv \frac{\Lambda}{8\pi G} + \frac{1}{2}M^2 \varphi^2 $. The corresponding canonical partition function of gravitoscalar thermal equilibrium is given by
\footnote{
Although I employ Dirichlet type boundary conditions to the Euclidean path integral of the gravity-scalar system, it still represents the canonical partition function. We can confirm $I^{E}_{c} \simeq \beta E -S $ at the saddle points (\ref{EQ4ansatz}) by using the equation of motion (\ref{EQAbasic}). This fact is in contrast to the case of the Einstein-Maxwell system, where $I^{E}_{c} \simeq \beta E -S -\beta \mu Q $ at the saddle points when we employ Dirichlet type boundary conditions \cite{HawkingRoss} (here, $\mu$ is the electric potential and $Q$ is the electric charge).
}
\bea
Z_{c}(\beta, A; \Lambda, M, J_{\varphi})=\sum_{\MC{M}} \int \MC{D} \BS{g} \MC{D} \varphi ~ e^{-I^{E}_{c}[\BS{g},\varphi]}
\ena
where $J_{\varphi}$ is the boundary value of $\varphi$.
I will approximate this by saddle points in a minisuperspace
\bea
\BS{g}(x)= f(r)e^{-2\delta(r)}d\tau^2 + \frac{1}{f(r)}dr^2 + r^2 (d\theta^2 + \sin^2 \theta d\phi^2 ) ~ , \label{EQ2minispGRA} \\
\varphi(x)=\varphi(r) ~ . \label{EQ2minispSCA} \hspace{7.2cm}
\ena

Throughout this paper, I assume all Euclidean saddle points in the minisuperspace contribute to the path integral. 
\footnote{
This assumption is not so trivial because there exists the contour problem in the Euclidean path integral of gravity \cite{GibbonsHawkingPerry, HalliwellHartle}. I will mention it in the final section.
}

\section{Gravitational thermodynamics}
In this section, I firstly review the thermodynamical properties of gravity alone without or with a negative cosmological constant in subsections 3.1 and 3.2. These are quantitatively different of course but qualitatively the same. We will see this is not the case for a positive cosmological constant in subsection 3.3.
\subsection{$\Lambda=0$}
As is well known, Gibbons and Hawking showed that gravitational thermodynamics in asymptotically flat spacetime is {\it ill-defined} in the sense that quantum spacetimes with the asymptotics never realize thermal equilibrium states \cite{GibbonsHawking}. After the work by Hawking and Page \cite{HawkingPage} that claimed that those with AdS asymptotics do reach thermal equilibrium, York claimed that quantum spacetimes without $\Lambda$ having a time-like boundary with finite spatial volume also reach thermal equilibrium \cite{York2}. He considered a gravitational partition function of a quantum spacetime with an $S^2\times \MB{R}$ Lorentzian boundary geometry, which may be given by summing over Euclidean geometries with an $S^2 \times S^1$ Euclidean boundary geometry. It was approximated by contributions from the following saddle points
\bea
\BS{g}(x)= f(r)d \tau^2 + \frac{1}{f(r)}dr^2 + r^2 (d\theta^2 + \sin^2 \theta d\phi^2) \hspace{3cm} \\
 \hspace{4cm} f(r)=1-\frac{2Gm}{r} \hspace{1.1cm} {\rm for } ~ S^2\times D ~ {\rm topology} \hspace{0.2cm} \label{EQ3no1BH} \\
 \hspace{4.1cm} f(r)=1 \hspace{2.5cm} {\rm for } ~ B^3\times S^1 ~ {\rm topology}
\ena
where the coordinate ranges of $r$ and $\tau$, $[ 2Gm, r_{b}  ]$ and $[0, \beta_{0}]$, and the mass parameter $m$ should be chosen appropriately for given $\beta$ and $A$. Namely, they must satisfy
\bea
\frac{4\pi}{f^{\p}(2Gm)}= \beta_{0}, ~~~ \sqrt{f(r_{b})}\beta_{0}=\beta, ~~~ 4\pi r_{b}^2 =A \hspace{1cm} {\rm for } ~ S^2\times D ~ {\rm topology}  \hspace{0.2cm} \label{EQ3cond1} \\
M=0, ~~~ \beta_{0}=\beta, ~~~ 4\pi r_{b}^2 =A \hspace{1cm} {\rm for } ~ B^3\times S^1 ~ {\rm topology} \hspace{0.cm}
\ena
where the first condition for $S^2 \times D$ topology is for a conical singularity not to appear at the center of $D$, and the second and third for both topologies are for matching to boundary conditions. These geometries are shown in Fig. \ref{FIG3no1no1}.
                                                 %
\iffigure
\begin{figure}
\begin{center}
	\includegraphics[width=4.cm]{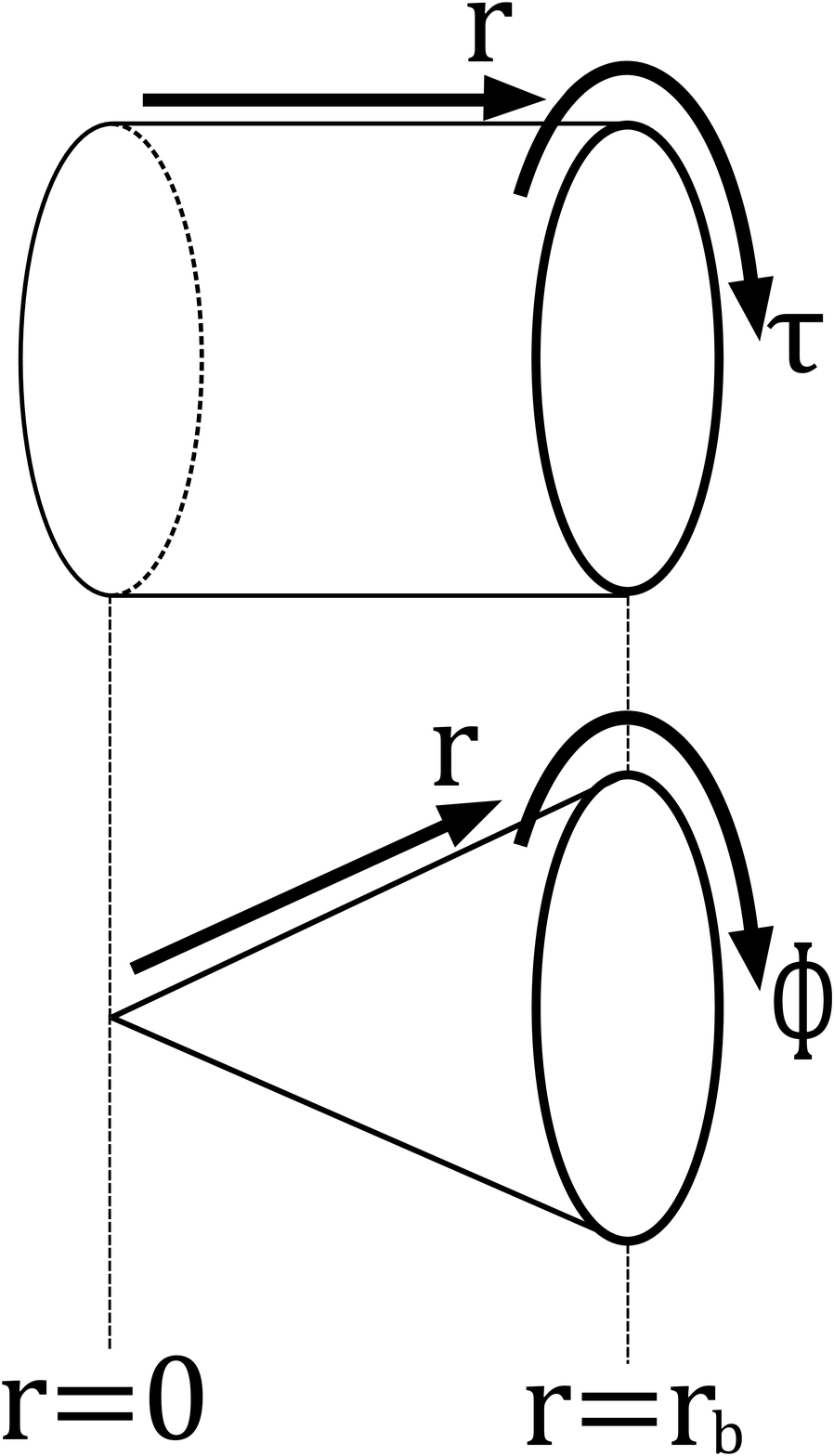} \hspace{3cm} 	\includegraphics[width=4.cm]{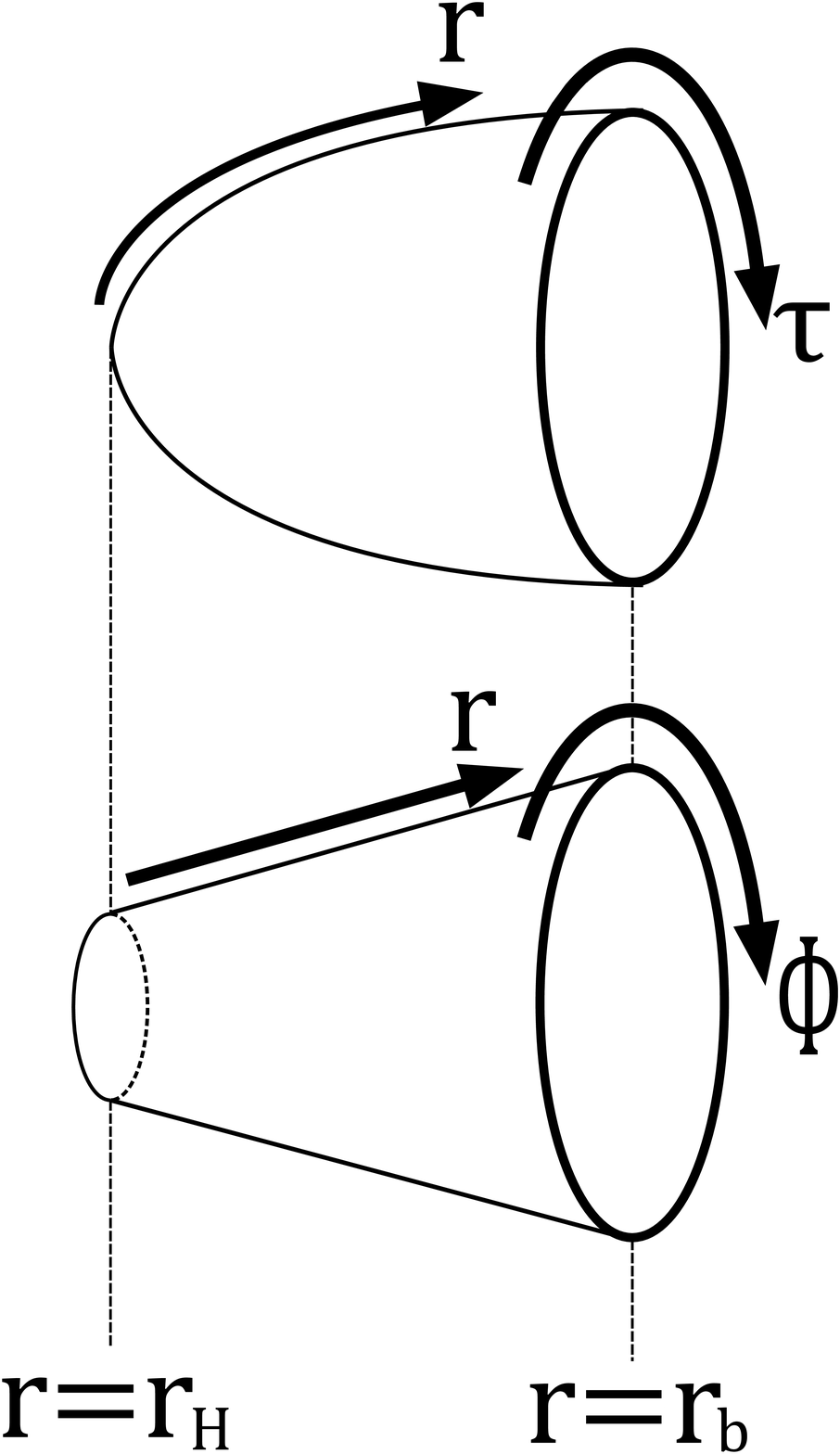}
	\caption{($\BS{{\rm Left}}$) Saddle point geometry of $B^3 \times S^1$ topology. The upper is the $r$-$\tau$ section and the lower is the $r$-$\phi$ section. Note that the $r$-$\phi$(-$\theta$) section is just a disc $D$, that is, 2(3)-ball $B^{2(3)}$, and is not conic as the figure shows, and especially there is no singularity at the center $r=0$. ($\BS{{\rm Right}}$) Saddle point geometry of $S^2 \times D$ topology (Euclidean BH). The $r$-$\tau$ section is smoothly capped at $r=r_{H}$.}
\label{FIG3no1no1}
\end{center}
\end{figure}
\fi
                                                 %
The resulting saddle point contributions are 
\bea
{\rm Branch ~1A:} \hspace{4.15cm} 1 \hspace{5.1cm} {\rm for ~ any ~ value ~ of ~ } \beta \\
{\rm Branch ~1B:} ~~{\rm exp} \left[ -\frac{1}{G}\left( \frac{3}{16\pi} \beta_{0}^{-}(\beta,r_{b})^2 - r_{b} \beta_{0}^{-}(\beta,r_{b}) + r_{b} \beta \right) \right] \hspace{1cm} {\rm for ~ } \beta \leq \frac{8\pi r_{b}}{3\sqrt{3}} \hspace{0.9cm} \\
{\rm Branch ~1C:} ~~ {\rm exp} \left[ -\frac{1}{G}\left( \frac{3}{16\pi} \beta_{0}^{+}(\beta,r_{b})^2 - r_{b} \beta_{0}^{+}(\beta,r_{b}) + r_{b} \beta \right) \right] \hspace{1cm} {\rm for ~ } \beta \leq \frac{8\pi r_{b}}{3\sqrt{3}} \hspace{0.9cm}  
\ena
where $\beta^{+}_{0}$ and $\beta^{-}_{0}$ are the positive roots of 
\bea
\beta_{0}^3 -4\pi r_{b} \beta_{0}^2 + 4\pi r_{b} \beta^2 =0
\label{EQ3beta}
\ena
                                                 %
\iffigure
\begin{figure}
\begin{center}
	\includegraphics[width=8.cm]{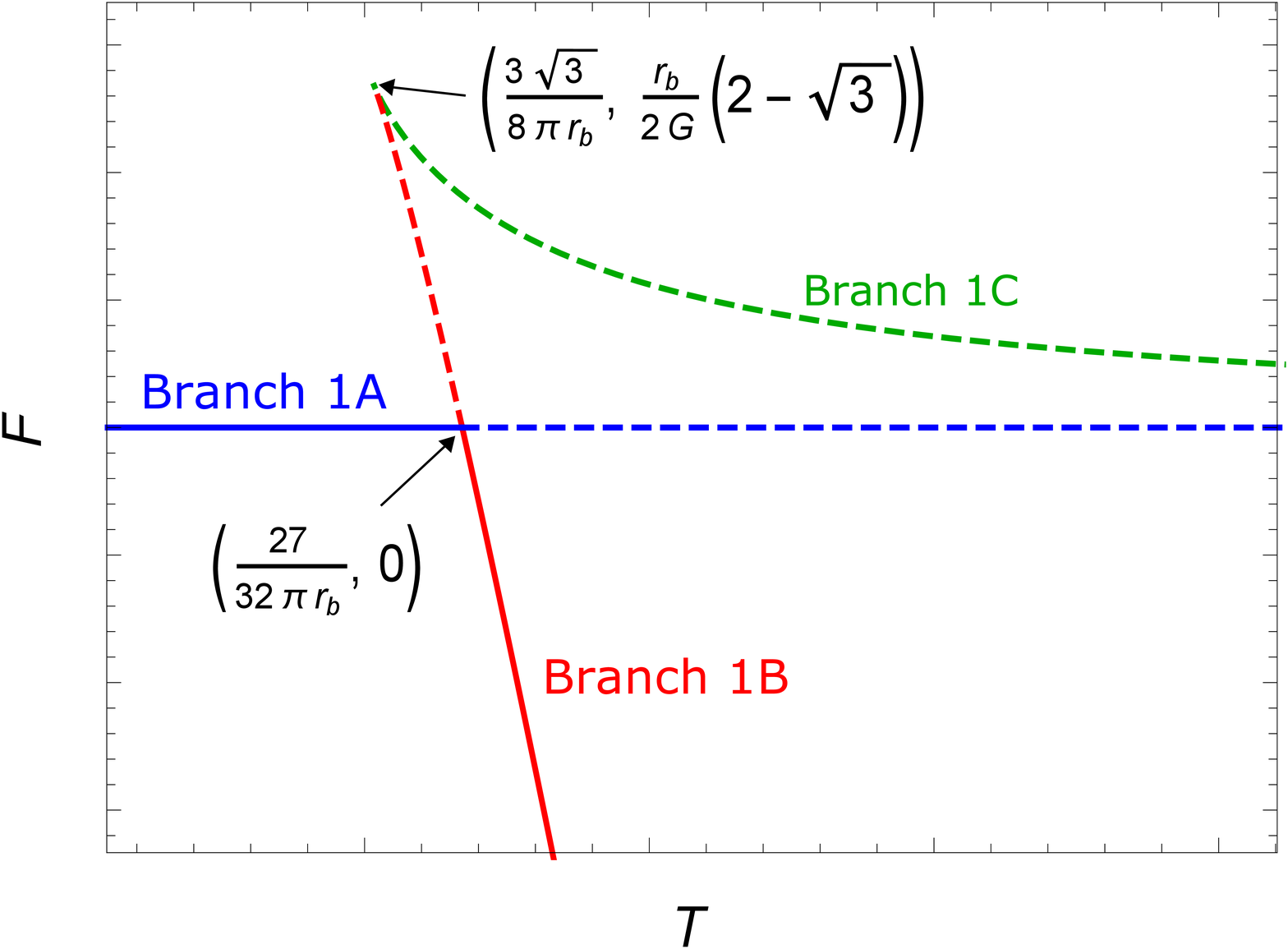} ~~ 	\includegraphics[width=8.cm]{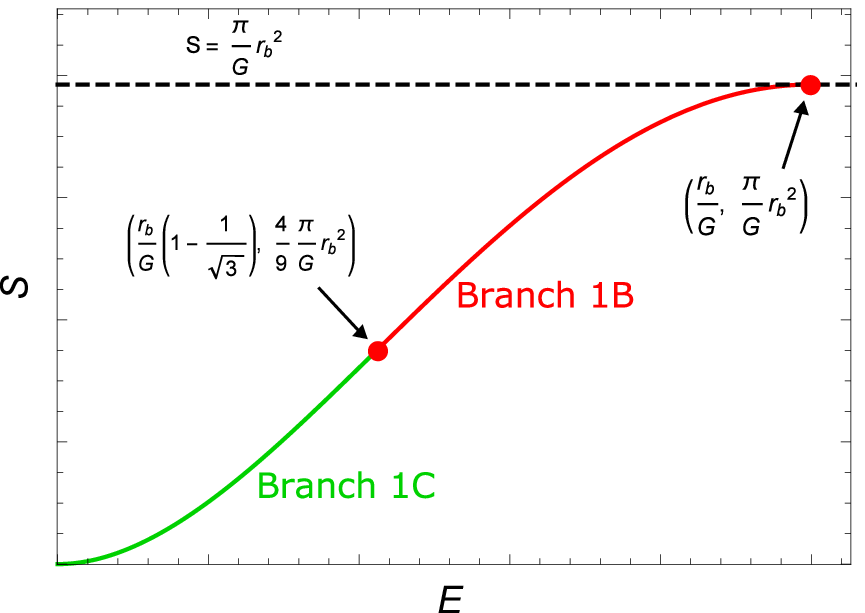}
	\caption{($\BS{{\rm Left}}$) Qualitative behavior of free energies of saddles 1A, 1B, and 1C. The portion that does not give dominant contribution is dashed.  ($\BS{{\rm Right}}$) Qualitative behavior of entropy as the function of energy. The maximum energy is $\displaystyle \frac{r_{b}}{G}$ and the maximum entropy is $\displaystyle \frac{\pi}{G}r_{b}^2$. }
\label{FIG3no1no2}
\end{center}
\end{figure}
\fi
                                                 %
for $\displaystyle \beta \leq \frac{8\pi r_{b}}{3\sqrt{3}}$, and $\beta^{+}_{0}$ is defined to be the larger one and $\beta^{-}_{0}$ is the smaller one. 
\footnote{
Equation (\ref{EQ3beta}) has one negative root, and two complex roots for  $\displaystyle \beta > \frac{8\pi r_{b}}{3\sqrt{3}}$ and two positive roots for  $\displaystyle \beta < \frac{8\pi r_{b}}{3\sqrt{3}}$. Although I ignore the former case, it will be important when we consider the integration contour of the path integral, and it is essentially the source of the difficulty for the well-definedness of the canonical partition function \cite{HalliwellLouko}.
}
Equation (\ref{EQ3beta}) comes from the first and second conditions for $S^2 \times D$ topology (\ref{EQ3cond1}). Branch 1A is from the $B^3 \times S^1$ topology sector and branches 1B and 1C are from the $S^2 \times D$ topology sector. The behaviors of the corresponding free energies  against temperature are shown in Fig. \ref{FIG3no1no2}. Branch 1A is dominant when $T\in \left[0, \frac{27}{32\pi r_{b}} \right]$ and branch 1B is when $ T\in \left[ \frac{27}{32\pi r_{b}}, \infty \right)$. The former may be interpreted as flat spacetime with thermal radiations and the latter as BH spacetime (with thermal radiations). Branch 1C is never dominant, its heat capacity computed from its free energy is always negative, and it will be the only BH branch when we take the limit $r_{b} \to \infty$, which implies non-existence of gravitational thermal equilibrium in asymptotically flat spacetime.

We can also plot the $E$-$S$ diagram. The rightmost endpoint of 1B is located at $\left( \frac{r_{b}}{G}, \frac{\pi}{G} r_{b}^2 \right)$, which corresponds to the infinite temperature state in the canonical ensembles. We can interpret this diagram in two ways. One way is to interpret it as simply representing the energy and entropy of thermal equilibrium states that may be described by the canonical ensembles. In that case, only the information of some restricted region of 1B is important since the other regions of 1B and 1C are fictitious in the sense that they are not realized in thermal equilibrium. The other way is to think of this microcanonical ensemble as describing equilibrium states that are realized when the system is isolated.  In any case, there exists maximum energy and entropy. Especially, entropy is always bounded by the entropy bound and saturated at the maximum energy.
\footnote{
There have been several attempts to derive the $E$-$S$ relationship beyond the maximum energy $\displaystyle \frac{r_{b}}{G}$ \cite{BradenWhitingYork, LoukoWhiting, MelmedWhiting, Miyashita1}. Although the detailed behaviors of their entropy functions differ, all their maximum entropies are $\displaystyle \frac{\pi}{G}r_{b}^2$, which is located at  $\displaystyle E=\frac{r_{b}}{G}$.
}

\subsection{$\Lambda<0$}
Asymptotically AdS spacetime is the first example where there exists gravitational thermal equilibrium, shown by Hawking and Page \cite{HawkingPage}. After their work and York's work, which I reviewed in the previous subsection, Brown et al. generalized York's work to the case with negative $\Lambda$ and with a time-like boundary of finite boundary volume \cite{BrownCreightonMann}.
The saddle points they considered are
\bea
\BS{g}(x)= f(r)d \tau^2 + \frac{1}{f(r)}dr^2 + r^2 (d\theta^2 + \sin^2 \theta d\phi^2) \hspace{3cm} \\
 \hspace{4.1cm} f(r)=1-\frac{\Lambda}{3}r^2 \hspace{2.5cm} {\rm for } ~ B^3\times S^1 ~ {\rm topology} \\
 \hspace{4.1cm} f(r)=1-\frac{2Gm}{r}- \frac{\Lambda}{3}r^2 \hspace{1.1cm} {\rm for } ~ S^2\times D ~ {\rm topology} \hspace{0.15cm}  \label{EQ3no2AdSBH}
\ena
As before, the coordinate ranges of $r$ and $\tau$, $[ r_{H}, r_{b}  ]$ and $[0, \beta_{0}]$, and the mass parameter $m$ should be chosen appropriately for given $\beta$ and $A$. They must satisfy
\bea
f(r_{H})=0, ~~~ \frac{4\pi}{f^{\p}(r_{H})}= \beta_{0}, ~~~ \sqrt{f(r_{b})}\beta_{0}=\beta, ~~~ 4\pi r_{b}^2 =A \hspace{1cm} {\rm for } ~ S^2\times D ~ {\rm topology}  \hspace{0.1cm}  \\
r_{H}=0, ~~~ \sqrt{f(r_{b})}\beta_{0}=\beta, ~~~ 4\pi r_{b}^2 =A \hspace{4cm} {\rm for } ~ B^3\times S^1 ~ {\rm topology} \hspace{0.cm}
\ena
Qualitatively, these geometries are the same as those of the $\Lambda=0$ case (Fig. \ref{FIG3no1no1}). The saddle point contributions are
\bea
{\rm Branch ~2A:} \hspace{2.4cm} 1 \hspace{2.25cm} {\rm for ~ any ~ value ~ of ~ } \beta \hspace{0.4cm} \\
{\rm Branch ~2B:} ~~{\rm exp} \left[ -I^{os, -}(\beta, r_{b},\Lambda) \right] \hspace{1cm} {\rm for ~ } \beta\leq\beta_{cr}(r_{b},\Lambda) \hspace{0.5cm} \\
{\rm Branch ~2C:} ~~ {\rm exp}  \left[ -I^{os, +}(\beta, r_{b},\Lambda) \right]  \hspace{1cm} {\rm for ~ } \beta\leq\beta_{cr}(r_{b},\Lambda) \hspace{0.5cm}  
\ena
where the on-shell action functions $I^{os, -}$ and $I^{os, +}$ are given by
\bea
I^{os, \pm}(\beta, r_{b},\Lambda) \equiv \frac{-1}{G}\left[ \frac{4\pi}{1-\Lambda r_{H}^{\pm}(\beta, r_{b},\Lambda)^2} \left\{ r_{H}^{\pm}(\beta, r_{b},\Lambda) \left( r_{b}-\frac{\Lambda}{3}r_{b}^3 \right) \right. \right. \hspace{4.1cm} \notag \\
\left. \left. -\frac{3}{4}r_{H}^{\pm}(\beta, r_{b},\Lambda)^2 + \frac{\Lambda}{12}r_{H}^{\pm}(\beta, r_{b},\Lambda)^4\right\}  -r_{b}\beta \sqrt{1-\frac{\Lambda}{3}r_{b}^2}  \right] \label{EQ3no2IOS}
\ena
and $r_{H}^{-}$ and $r_{H}^{+}$ are the positive roots of 
\bea
\frac{16\pi^2 \Lambda}{3 r_{b}} r_{H}^5 -\Lambda^2 \beta^2 r_{H}^4 - \frac{16\pi^2}{r_{b}} r_{H}^3 + \left( 16\pi^2 - \frac{16\pi^2 \Lambda r_{b}^2}{3} + 2\Lambda \beta^2 \right) r_{H}^2 -\beta^2=0 \label{EQ3no2rHpm}
\ena
for $\beta<\beta_{cr}(r_{b},\Lambda)$. $r_{H}^{-}$ is the smaller one and  $r_{H}^{+}$ is the larger one. 
$\beta_{cr}$ is defined by
\bea
\beta_{cr}(r_b,\Lambda)= \frac{4\pi r_{H,cr}(r_b,\Lambda)}{1-\Lambda r_{H,cr}(r_b,\Lambda)^2} \sqrt{\left(1-\frac{\Lambda}{3}r_{b}^2 \right) - \frac{r_{H,cr}(r_b,\Lambda)}{r_{b}} + \frac{\Lambda}{3r_{b}}r_{H,cr}(r_b,\Lambda)^3 }
\ena
and $r_{H,cr}$ is the positive root of
\bea
-\frac{\Lambda^2}{6r_{b}}r_{H}^5 + \frac{\Lambda}{3r_{b}} r_{H}^3 + \Lambda\left( 1-\frac{\Lambda}{3}r_{b}^2 \right)r_{H}^2 -\frac{3}{2 r_{b}}r_{H} + \left(1-\frac{\Lambda}{3}r_{b}^2  \right)=0 \label{EQ3no2RHcr}
\ena
The transition inverse temperature is given by
\bea
\beta_{tr}(r_b,\Lambda)= \frac{4\pi r_{H,tr}(r_b,\Lambda)}{1-\Lambda r_{H,tr}(r_b,\Lambda)^2} \sqrt{\left(1-\frac{\Lambda}{3}r_{b}^2 \right) - \frac{r_{H,tr}(r_b,\Lambda)}{r_{b}} + \frac{\Lambda}{3r_{b}}r_{H,tr}(r_b,\Lambda)^3 }
\ena
and $r_{H,tr}$ is the positive root of
\bea
\frac{\Lambda^2}{144} r_{H}^6 - \frac{\Lambda}{8}r_{H}^4 - \frac{\Lambda}{6} \left( 1-\frac{\Lambda}{3}r_{b}^2 \right)r_{b} r_{H}^3 + \frac{9}{16}r_{H}^2 - \frac{1}{2}\left( 1-\frac{\Lambda}{3}r_{b}^2 \right)r_{b} r_{H}  =0
\ena
For general values of $\Lambda<0$, we cannot obtain exact expressions for $\beta_{cr}, \beta_{tr}$, and so on, contrary to the case of $\Lambda=0$. We could know them, however, for large $|\Lambda|$. For example,
\bea
r_{H,cr}(r_{b},\Lambda) \simeq \frac{1}{\sqrt{-\Lambda}} \hspace{1cm} {\rm for ~ } |\Lambda|>>1   \hspace{0.4cm}   \\
r_{H,tr}(r_{b},\Lambda) \simeq \sqrt{\frac{3}{-\Lambda}} \hspace{1cm} {\rm for ~ } |\Lambda|>>1   \hspace{0.4cm} 
\ena
and correspondingly,
\bea
\beta_{cr} \to \frac{2\pi}{\sqrt{3}} r_{b} ~~ (\Lambda \to -\infty) \\
\beta_{tr} \to \pi r_{b} ~~ (\Lambda \to -\infty) 
\ena
The qualitative behavior of free energies and the behaviors of the critical and transition points are shown in Fig.  \ref{FIG3no2no1}.
                                                 %
\iffigure
\begin{figure}
\begin{center}
	\includegraphics[width=8.cm]{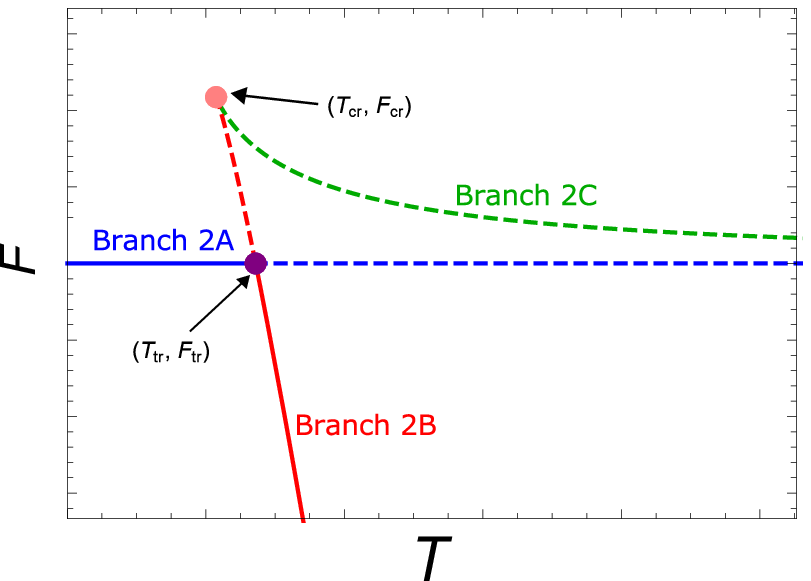} ~~ 	\includegraphics[width=8.cm]{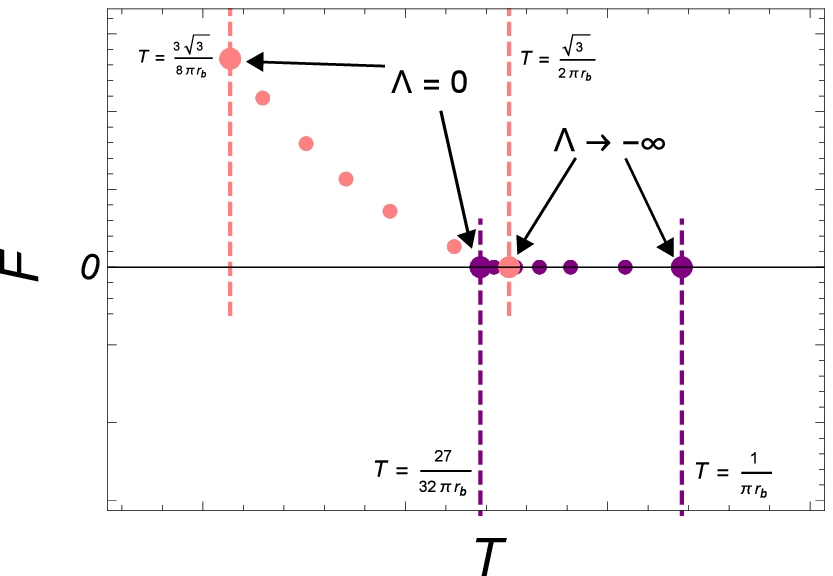}
	\caption{($\BS{{\rm Left}}$) Qualitative behavior of free energies of saddles 2A, 2B, and 2C. The portion that does not give a dominant contribution is dashed. The critical point and the transition point are marked. The temperature and free energy at those points are $T_{cr (tr)}=1/\beta_{cr (tr)}(r_{b}, \Lambda)$ and $F_{cr (tr)}=I^{os,+}(\beta_{cr (tr)}, r_{b},\Lambda)/\beta_{cr (tr)}(r_{b}, \Lambda)=I^{os,-}(\beta_{cr (tr)}, r_{b},\Lambda)/\beta_{cr (tr)}(r_{b}, \Lambda). ~~   $($\BS{{\rm Right}}$) Behaviors of critical point and transition point against $\Lambda$ on $T$-$F$ plane.}
\label{FIG3no2no1}
\end{center}
\end{figure}
\fi
                                                 %
The behavior is qualitatively the same as that in the $\Lambda=0$ case; branch 2A is dominant when $T\in \left[ 0, 1/\beta_{tr}(r_{b},\Lambda) \right]$, branch 2B is when $T\in \left[1/\beta_{tr}(r_{b},\Lambda), \infty  \right]$, and branch 2C is never dominant. The behaviors of the critical point and the transition point on the $T$-$F$ plane are also shown in Fig. \ref{FIG3no2no1}. The $\Lambda$ dependence of $\beta_{cr}$ and $\beta_{tr}$ is weak and they are inversely proportional to $r_{b}$,  i.e. , $1/\beta_{cr}(r_{b},\Lambda)\in \left[ \frac{3\sqrt{3}}{8\pi r_{b}}, \frac{\sqrt{3}}{2\pi r_{b}} \right]$ and $1/\beta_{tr}(r_{b},\Lambda)\in \left[ \frac{27}{32\pi r_{b}}, \frac{1}{\pi r_{b}} \right]$.

The behavior of entropy against energy is shown in Fig. \ref{FIG3no2no2}. This is also qualitatively the same as in the previous case. 
                                                 %
\iffigure
\begin{figure}
\begin{center}
	\includegraphics[width=8.cm]{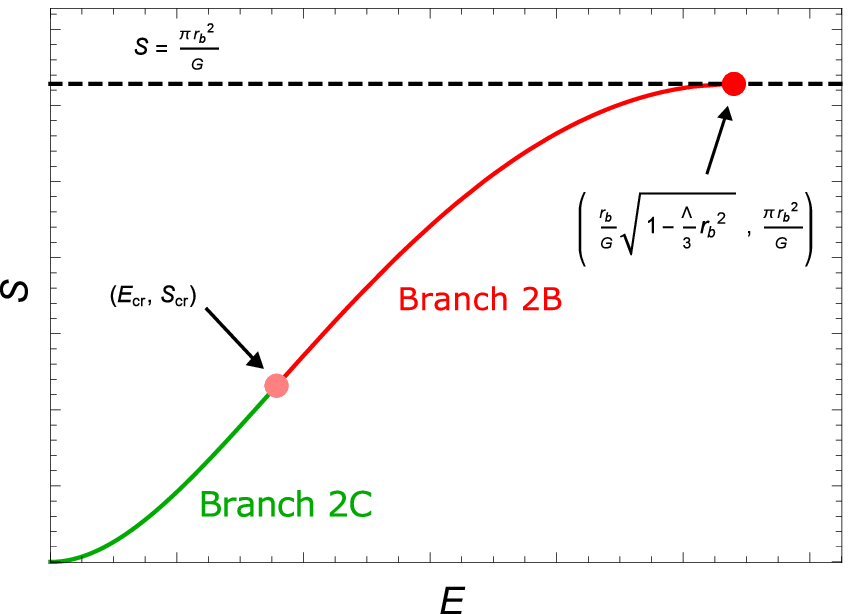} ~~ 	\includegraphics[width=8.cm]{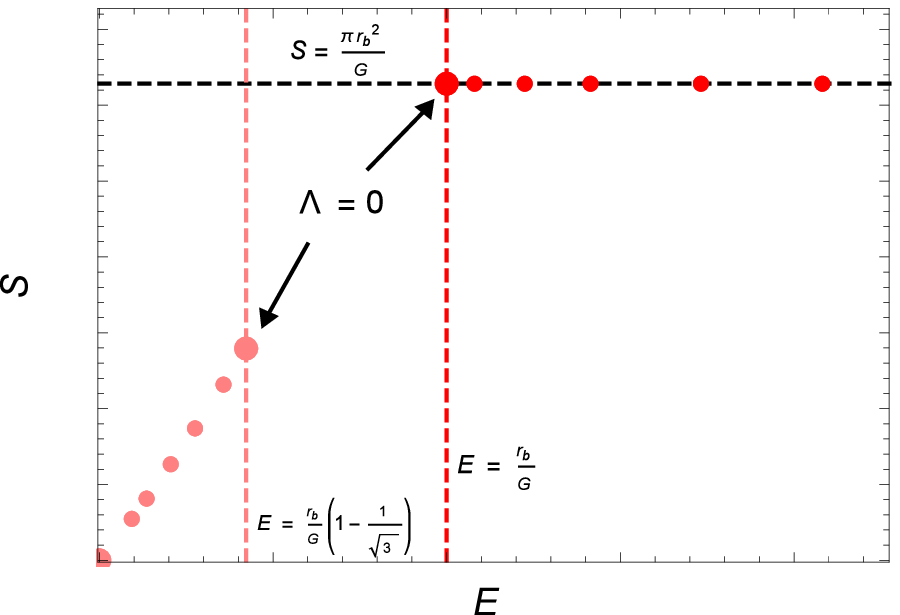}
	\caption{($\BS{{\rm Left}}$) Qualitative behavior of entropy as a function of energy. The critical point and the maximum entropy point are marked. The energy and entropy at the critical point are given by $E_{cr}=E_{cr}(r_{b},\Lambda)$ and $S_{cr}=\frac{\pi r_{H, cr}(r_{b}, \Lambda)^2}{G}$ (see the main text). The maximal energy and entropy are $\frac{r_{b}}{G}\sqrt{1-\frac{\Lambda}{3}r_{b}^2}$ and $\frac{\pi r_{b}^2}{G}$. ($\BS{{\rm Right}}$) Behaviors of critical point and maximum entropy point against $\Lambda$ on $E$-$S$ plane.}
\label{FIG3no2no2}
\end{center}
\end{figure}
\fi
                                                 %
The maximum entropy does not depend on $\Lambda$. Therefore, the entropy bound may be given by the area of the boundary over $4G$, independent of $\Lambda \leq 0$. On the other hand, the maximum energy is highly dependent on $\Lambda$ and goes to infinity as we increase $|\Lambda|$. The energy and entropy at the critical point are given by
\bea
E_{cr}(r_{b},\Lambda)= \frac{r_{b}}{G}\left[ \sqrt{1-\frac{\Lambda}{3}r_{b}^2} - \sqrt{\left(1-\frac{\Lambda}{3}r_{b}^2 \right) - \frac{r_{H,cr}(r_b,\Lambda)}{r_{b}} + \frac{\Lambda}{3r_{b}}r_{H,cr}(r_b,\Lambda)^3 } \right] \\
S_{cr}(r_{b},\Lambda) = \frac{\pi}{G}r_{H, cr}(r_{b}, \Lambda)^2 \hspace{8.6cm}
\ena
The behaviors of the critical point and the maximum entropy point are also shown in Fig. \ref{FIG3no2no2}.  As I mentioned above, the maximum energy goes to infinity, with fixing the maximum entropy, as we increase $|\Lambda|$. On the other hand, the critical point goes to $(0,0)$. Therefore, the ratio of the thermodynamically stable energy range $\left[ E_{cr}(r_{b},\Lambda), \frac{r_{b}}{G}\sqrt{1-\frac{\Lambda}{3}r_{b}^2} \right]$ to the whole energy range $\left[ 0, \frac{r_{b}}{G}\sqrt{1-\frac{\Lambda}{3}r_{b}^2} \right]$ increases.

We have seen the thermodynamical properties for the $\Lambda \leq 0$ case. Some are universal, irrelevant of the precise value of $\Lambda$, and others are highly dependent on it.  Let me summarize them here: \\
~ \\
$\bullet$ Universal \\
~~~ $\cdot$ Thermodynamically stable. \\
~~~ $\cdot$ There exist two phases: radiation and BH. \\
~~~ $\cdot$ The transition temperature and the critical temperature are always given by $\frac{a(\Lambda, r_{b})}{r_{b}}$, \\
~~~~~  where $ 0.206<a(\Lambda, r_{b})<0.319$.  \\
~~~ $\cdot$ The maximum entropy is given by $\frac{\pi r_{b}^2}{G}$. \\
 ~ \\
$\bullet$ $\Lambda$ dependent \\
~~~ $\cdot$ The maximum energy goes to infinity as $|\Lambda|$ is increased. \\
~~~ $\cdot$ The critical energy (and the critical entropy) decreases as $|\Lambda|$ is increased. \\
~~~~~ Consequently, the ratio of the thermodynamically stable region increases and goes to $1$\\
~~~~~ when $\Lambda \to - \infty$. \\

\subsection{$\Lambda>0$}
Contrary to the $\Lambda<0$ case, there are few studies of the $\Lambda>0$ case. (From different perspectives, ``thermodynamics of de Sitter space'' or ``thermodynamics of de Sitter BH'' is investigated in the literature, for instance, \cite{HuangLiuWang, WangHuang, Dehghani, Sekiwa, Saida, KuviznakSimovic}.) I restrict the analysis to $0<\Lambda<\frac{3}{r_{b}^2}$. At the end of this subsection, I will give very brief comments on the $\frac{3}{r_{b}^2}\leq \Lambda$ case.
\subsubsection{$\displaystyle 0<\Lambda<\frac{3}{r_{b}^2}$}
Firstly, there exist similar saddle points to those in the $\Lambda \leq 0$ case:
\bea
\BS{g}(x)= f(r)d \tau^2 + \frac{1}{f(r)}dr^2 + r^2 (d\theta^2 + \sin^2 \theta d\phi^2) \hspace{3cm} \\
 \hspace{4.1cm} f(r)=1-\frac{\Lambda}{3}r^2 \hspace{2.5cm} {\rm for } ~ B^3\times S^1 ~ {\rm topology} \\
 \hspace{4cm} f(r)=1-\frac{2Gm}{r}- \frac{\Lambda}{3}r^2 \hspace{1.1cm} {\rm for } ~ S^2\times D ~ {\rm topology} \hspace{0.15cm} \label{EQ3no3dSBH}
\ena
where the coordinate ranges of $r$ and $\tau$, $[ r_{H}, r_{b}  ]$ and $[0, \beta_{0}]$, and the mass parameter $m$ satisfy the following conditions (see Fig. \ref{FIG3no1no1}):
\bea
f(r_{H})=0, ~~~ \frac{4\pi}{f^{\p}(r_{H})}= \beta_{0}, ~~~ \sqrt{f(r_{b})}\beta_{0}=\beta, ~~~ 4\pi r_{b}^2 =A \hspace{1cm} {\rm for } ~ S^2\times D ~ {\rm topology}  \hspace{0.1cm} \\
r_{H}=0, ~~~ \sqrt{f(r_{b})}\beta_{0}=\beta, ~~~ 4\pi r_{b}^2 =A \hspace{4.1cm} {\rm for } ~ B^3\times S^1 ~ {\rm topology} \hspace{0.cm}
\ena
Additionally, there also exists the following saddle point in the $S^2 \times D$ topology sector:
\bea
\BS{g}(x)= f(r)d \tau^2 + \frac{1}{f(r)}dr^2 + r^2 (d\theta^2 + \sin^2 \theta d\phi^2) \hspace{3cm} \label{EQ3no3BG} \\
 \hspace{4cm} f(r)=1-\frac{2Gm}{r}- \frac{\Lambda}{3}r^2 \hspace{4.5cm} \\
{\rm coordinate}: ~~~ \tau \in [0,\beta_{0}], ~~ r\in [r_{b}, r_{c}] \hspace{5.6cm} \\
{\rm condition}: ~~~ f(r_{c})=0, ~~~ \frac{4\pi}{f^{\p}(r_{c})}= -\beta_{0}, ~~~ \sqrt{f(r_{b})}\beta_{0}=\beta, ~~~ 4\pi r_{b}^2 =A
\ena
The difference from Eq. (\ref{EQ3no3dSBH}) is that the boundary is now located at the smaller end of $r$ and the bolt \cite{GibbonsHawking2} is at the larger end $r=r_{c}$ . 
\footnote{
The reason why I use ``c'' is because it is the Euclidean counterpart of the cosmological horizon. 
}
Since the position of the bolt is at the larger end of the $r$ coordinate, we have a minus sign in the smoothness condition there (Fig. \ref{FIG3no3no1}).
                                                 %
\iffigure
\begin{figure}
\begin{center}
	\includegraphics[width=4.cm]{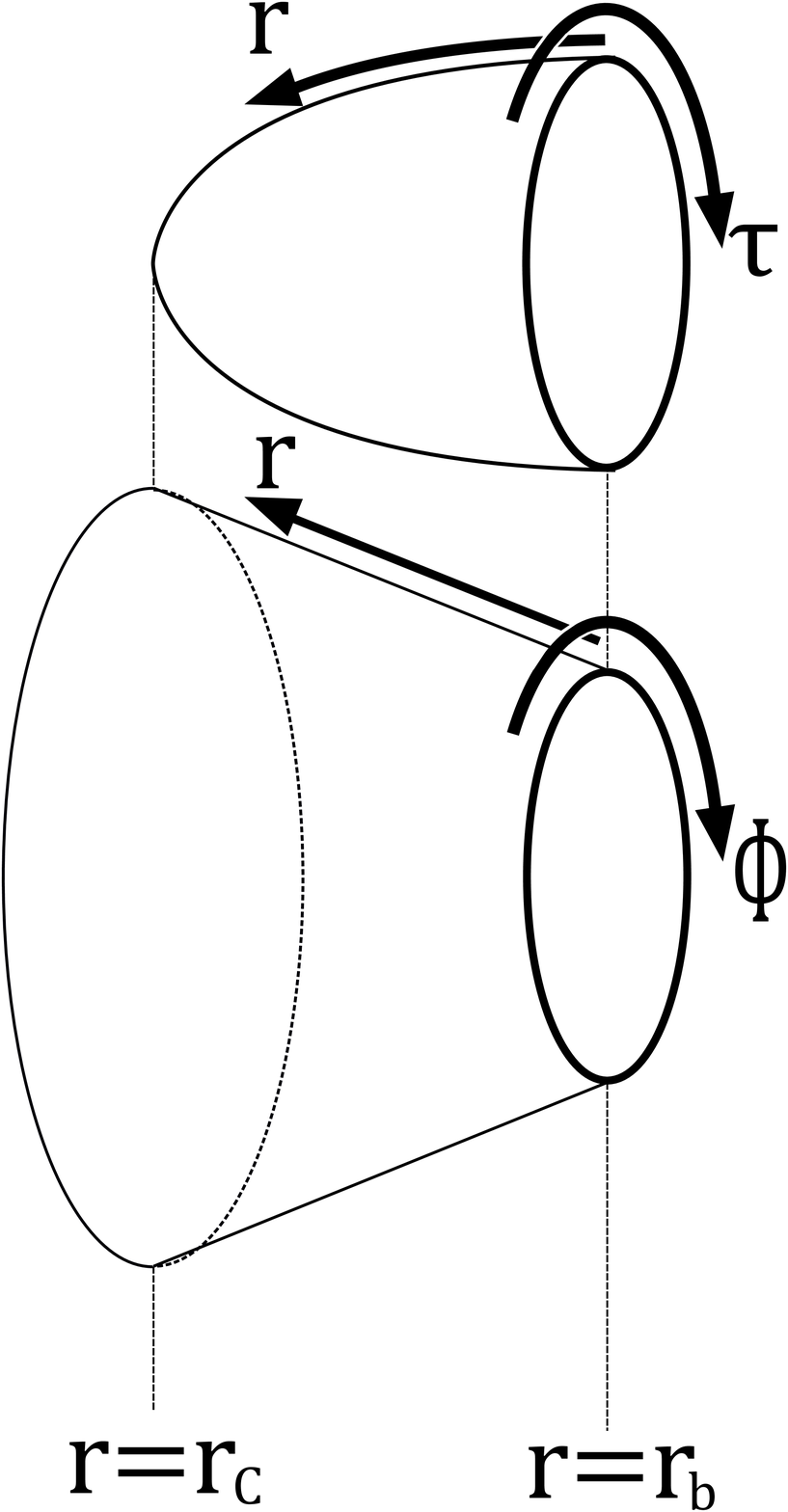}
	\caption{Extra saddle point geometry of $S^2 \times D$ topology (BG instanton). The area of the bolt $4\pi r_{c}^2$ is larger than that of the boundary $4\pi r_{b}^2$. For fixed $r_{b}$, $r_{c}$ can be arbitrarily large depending on $\beta$. Branch 3D consists of this type of geometry. }
\label{FIG3no3no1}
\end{center}
\end{figure}
\fi
                                                 %
For fixed $r_{b}$, $r_{c}$ can be arbitrarily large depending on $\beta$. This is in contrast to the saddle point (\ref{EQ3no3dSBH}) (and (\ref{EQ3no2AdSBH}) or (\ref{EQ3no1BH}) in the case of $\Lambda \leq 0$), whose entropy, that is, the area of the bolt, is always bounded by $4\pi r_{b}^2$. Since this saddle point enables the system to contain arbitrarily large entropy, as we will see shortly, I call such a Euclidean geometry, of which the area of the bolt is greater than the area of the boundary, the ``bag of gold (BG) instanton,'' after Wheeler's bag of gold spacetime. 
\footnote{
Wheeler's bag of gold spacetime can contain an arbitrary amount of entropy in space. The BG instanton can contain an arbitrary amount of ``entropy'' in the ``Euclidean holographic plate.'' (However, attention must be paid to the former statement because it may not be true when the matter couples to the dynamical gravity \cite{AlmheiriMahajanMaldacenaZhao}.)
}
\footnote{
Note that the solution itself is not new or surprising. I do not know who first introduced this kind of solution into physics. For example, in \cite{CaiMyungZhang}, they presented the Lorentzian (and topologically generalized) version of (\ref{EQ3no3BG}) (when $m<0$) without the boundary $r=r_{b}$ but with a singularity at the center $r=0$. They called it ``topological de Sitter.''
}
Eventually, except for the special case that I will explain shortly, the saddle point contributions are
\bea
{\rm Branch ~3A:} \hspace{2.4cm} 1 \hspace{2.25cm} {\rm for ~ any ~ value ~ of ~ } \beta \hspace{0.4cm} \\
{\rm Branch ~3B:} ~~~{\rm exp} \left[ -I^{os, -}(\beta, r_{b},\Lambda) \right] \hspace{0.9cm} {\rm for ~ } \beta\leq\beta_{cr}(r_{b},\Lambda) \hspace{0.5cm} \\
{\rm Branch ~3C:} ~~~ {\rm exp}  \left[ -I^{os, +}(\beta, r_{b},\Lambda) \right]  \hspace{0.9cm} {\rm for ~ } \beta\leq\beta_{cr}(r_{b},\Lambda) \hspace{0.5cm}  \\
{\rm Branch ~3D:} ~~ {\rm exp}  \left[ -I^{os, BG}(\beta, r_{b},\Lambda) \right]  \hspace{0.75cm} {\rm for ~ any ~ value ~ of ~ } \beta \hspace{0.4cm}  
\ena
The definition of $I^{os,+}, I^{os,-}$ is the same as (\ref{EQ3no2IOS}), and that of $I^{os,BG}$ is (\ref{EQ3no2IOS}) with $r_{H}^{\pm}$ replaced by $r_{c}$:
\bea
I^{os, BG}(\beta, r_{b},\Lambda) \equiv \frac{-1}{G}\left[ \frac{4\pi}{1-\Lambda r_{c}(\beta, r_{b},\Lambda)^2} \left\{ r_{c}(\beta, r_{b},\Lambda) \left( r_{b}-\frac{\Lambda}{3}r_{b}^3 \right) \right. \right. \hspace{4.1cm} \notag \\
\left. \left. -\frac{3}{4}r_{c}(\beta, r_{b},\Lambda)^2 + \frac{\Lambda}{12}r_{c}(\beta, r_{b},\Lambda)^4\right\}  -r_{b}\beta \sqrt{1-\frac{\Lambda}{3}r_{b}^2}  \right] \hspace{0.0cm} \label{EQ3no3IOSBG}
\ena
$r_{c}, r^{+}_{H}, $ and $r^{-}_{H}$ in (\ref{EQ3no3IOSBG}) and (\ref{EQ3no2IOS}) are defined as the largest, the middle, and the smallest of the positive roots of Eq. (\ref{EQ3no2rHpm}) for $\beta < \beta_{cr}(r_{b}, \beta)$, and  for $\beta > \beta_{cr}(r_{b}, \beta)$, $r_{c}$ is the positive root of the equation. The definition of $\beta_{cr}$ is the same as before. When $\Lambda=\frac{1}{r_{b}^2}$, branch 3B ceases to exist and the behaviors of branches 3C and 3D change:
\footnote{
When $\Lambda$ takes this value (or, when we put the boundary $r=r_{b}$ appropriately for a given $\Lambda$), both the $r_{H} \to r_{b}$ limit of Euclidean BH and the $r_{c} \to r_{b}$ limit of BG instanton corresponds to the Nariai limit, that is, the resulting geometry is half of Euclidean Nariai geometry in the static patch. Therefore, two branches 3C and 3D are smoothly connected and their thermodynamical quantities do not diverge unlike the $\Lambda \neq \frac{1}{r_{b}^2}$ case.
}
\bea
{\rm Branch ~3A:} \hspace{2.4cm} 1 \hspace{2.4cm} {\rm for ~ any ~ value ~ of ~ } \beta \hspace{0cm} \\
{\rm Branch ~3C:} \hspace{0.4cm} {\rm exp} \left[ -I^{os, -}(\beta, r_{b},\Lambda) \right] \hspace{1.0cm} {\rm for ~ } \beta \leq \frac{2\pi}{\sqrt{\Lambda}}   \hspace{1.0cm} \\
{\rm Branch ~3D:} ~~ {\rm exp}  \left[ -I^{os, BG}(\beta, r_{b},\Lambda) \right]   \hspace{0.9cm} {\rm for ~ } \beta > \frac{2\pi}{\sqrt{\Lambda}}   \hspace{1.0cm}
\ena
The qualitative behaviors of free energies for each case are shown in Fig. \ref{FIG3no3no2}. 
                                                 %
\iffigure
\begin{figure}
\begin{center}
	\includegraphics[width=8.cm]{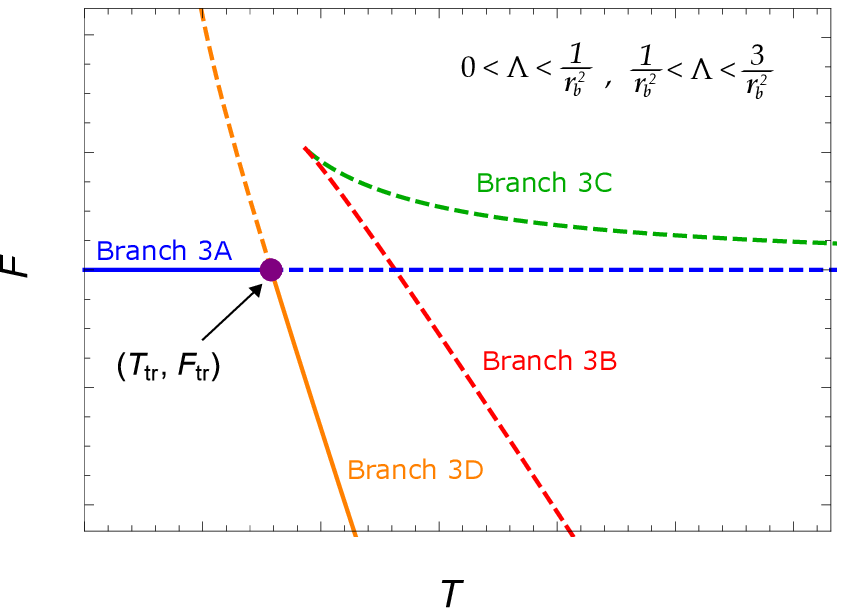} ~~~
	\includegraphics[width=8.cm]{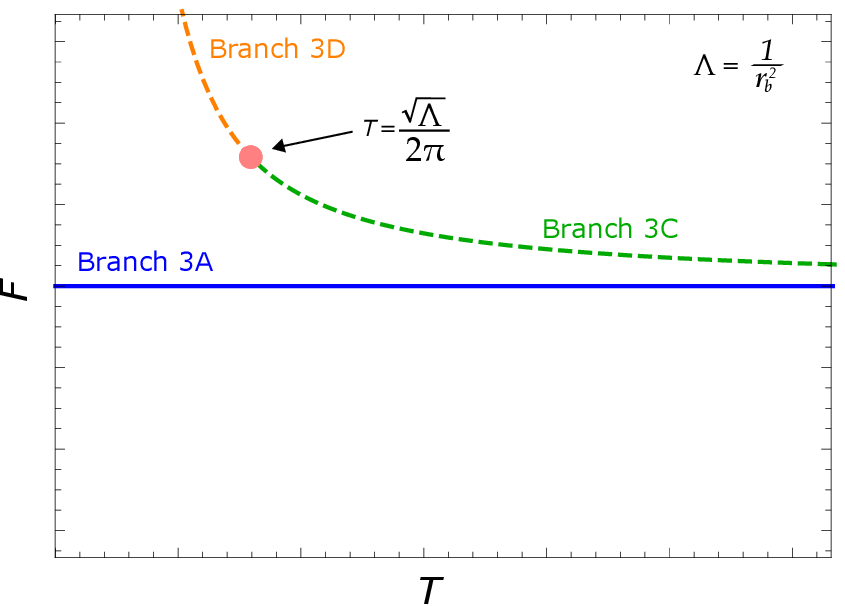}
	\caption{ Qualitative behaviors of free energies for $ 0<\Lambda<\frac{3}{r_{b}^2}$ case. ($\BS{{\rm Left}}$) $ \Lambda \neq \frac{1}{r_{b}^2}$ case. Below $T_{tr}$, branch 3A dominates; above $T_{tr}$, branch 3D dominates. ($\BS{{\rm Right}}$) $ \Lambda = \frac{1}{r_{b}^2}$ case. Branch 3A always dominates and there exist (thermodynamically unstable) subdominant branches for all $T$. This structure is similar to that of asymptotically flat spacetime. }
\label{FIG3no3no2}
\end{center}
\end{figure}
\fi
                                                 %
For the $ \Lambda \neq \frac{1}{r_{b}^2}$ case, although the behaviors of 3A, 3B, and 3C are qualitatively the same as before, the dominant saddle at high temperature becomes 3D, that is, the BG instanton. For the $ \Lambda = \frac{1}{r_{b}^2}$ case, branches 3C and 3D become subdominant for all $T$ and 3A is always dominant.

I also show the corresponding entropy behaviors against energy in Fig. \ref{FIG3no3no3}. For $ 0<\Lambda<\frac{1}{r_{b}^2}$, the entropy bound $S=\frac{\pi r_{b}^2}{G}$ is attained by the rightmost of 3B. This is exactly the same as the $\Lambda \leq 0$ case, where the energy and entropy at the rightmost of branch 2B are given by $ \left( \frac{r_{b}}{G}\sqrt{1-\frac{\Lambda}{3}r_{b}^2}, \frac{\pi r_{b}^2}{G} \right)$. The difference is the existence of branch 3D and it starts from the point $ \left( \frac{r_{b}}{G}\sqrt{1-\frac{\Lambda}{3}r_{b}^2}, \frac{\pi r_{b}^2}{4G}\left( \sqrt{\frac{12}{\Lambda r_{b}^2}-3}-1 \right)^2 \right)$. The entropy difference between the rightmost of 3B and the leftmost of 3D becomes small as we increase $\Lambda$, and finally becomes zero at $ \Lambda = \frac{1}{r_{b}^2}$. Above that value, the leftmost of 3D is fixed to the maximum entropy line and the rightmost of 3B detaches and starts to decrease. They go to $ \left( 0, \frac{\pi r_{b}^2}{G} \right)$ and $(0, 0)$ respectively as we increase $\Lambda$ to $ \frac{3}{r_{b}^2}$.
                                                 %
\iffigure
\begin{figure}
\begin{center}
	\includegraphics[width=8.cm]{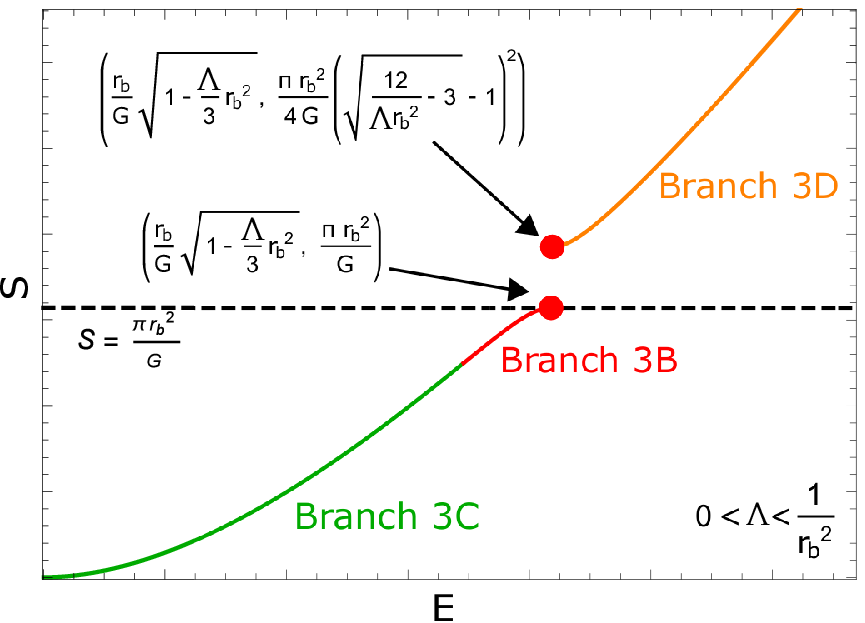} ~~~
	\includegraphics[width=8.cm]{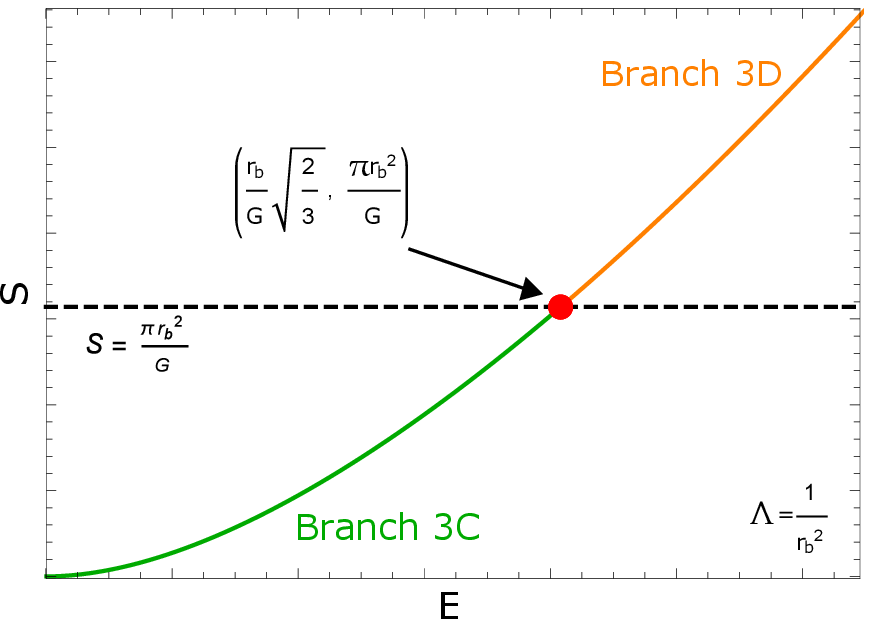} \\
~~\\
	\includegraphics[width=8.cm]{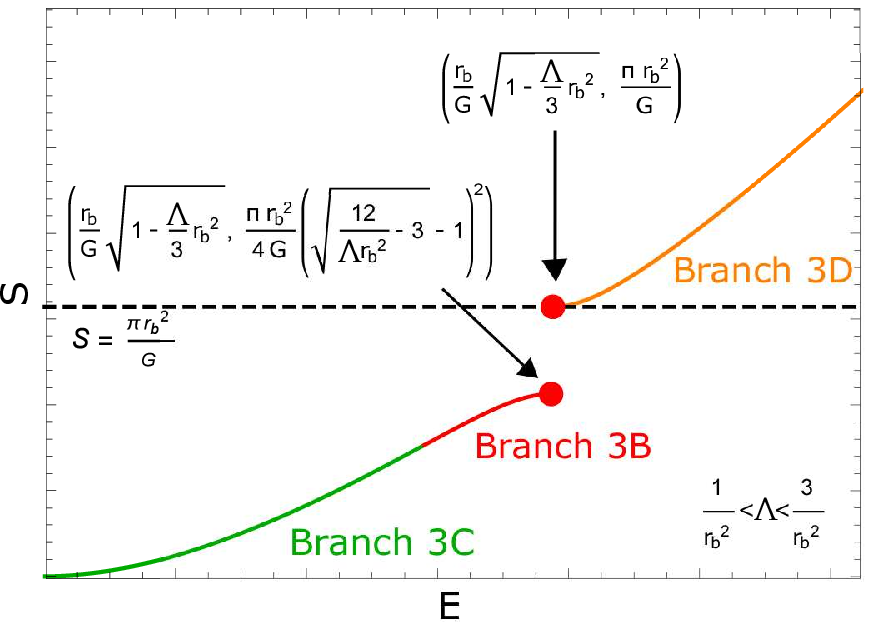}
	\caption{ Qualitative behaviors of entropy for $ 0<\Lambda<\frac{3}{r_{b}^2}$ case. The dashed lines represent the entropy bound $S=\frac{\pi r_{b}^2}{G}$.  ($\BS{{\rm Left}}$) $ 0<\Lambda<\frac{1}{r_{b}^2}$ case. The rightmost of 3B is at $ \left( \frac{r_{b}}{G}\sqrt{1-\frac{\Lambda}{3}r_{b}^2}, \frac{\pi r_{b}^2}{G} \right)$, the leftmost of 3D is at $ \left( \frac{r_{b}}{G}\sqrt{1-\frac{\Lambda}{3}r_{b}^2}, \frac{\pi r_{b}^2}{4G}\left( \sqrt{\frac{12}{\Lambda r_{b}^2}-3}-1 \right)^2 \right)$. ($\BS{{\rm Right}}$) $ \Lambda=\frac{1}{r_{b}^2}$ case. For only this value of $\Lambda$, the endpoints of the two branches meet on the entropy bound line. ($\BS{{\rm Bottom}}$) $ \frac{1}{r_{b}^2}<\Lambda<\frac{3}{r_{b}^2}$ case. The leftmost of 3D attaches to the line at $ \left( \frac{r_{b}}{G}\sqrt{1-\frac{\Lambda}{3}r_{b}^2}, \frac{\pi r_{b}^2}{G} \right)$ and the rightmost of 3B detaches from the line and is located at $ \left( \frac{r_{b}}{G}\sqrt{1-\frac{\Lambda}{3}r_{b}^2}, \frac{\pi r_{b}^2}{4G}\left( \sqrt{\frac{12}{\Lambda r_{b}^2}-3}-1 \right)^2 \right)$.   }
\label{FIG3no3no3}
\end{center}
\end{figure}
\fi
                                                 %

In previous subsections, we have seen that $\it{thermodynamical ~ stability}$ and $\it{the ~ entropy ~ bound}$ are universal properties for $\Lambda\leq 0$. I will argue that both are absent for $\Lambda>0$.

\subsubsection*{Absence of thermodynamical stability}
Firstly, let us see properties of branch 3D for the $ \Lambda \neq \frac{1}{r_{b}^2}$ case. Since $ \frac{\pd E}{\pd T}<0$ always holds on branch 3D,
\footnote{
Branch 3D is parameterized by the radius of the cosmological horizon $r_{c} \in \left[ \max\left( r_{b}, \frac{r_{b}}{2}\left( \sqrt{\frac{12}{\Lambda r_{b}^2}-3 }-1 \right) \right), \infty \right)$ for given $r_{b}$ and $\Lambda$. We can easily show that $ E_{3D}(r_{c})=\frac{r_{b}}{G}\left[ \sqrt{1-\frac{\Lambda}{3}r_{b}^2} - \sqrt{\left(1-\frac{\Lambda}{3}r_{b}^2 \right) - \frac{r_{c}}{r_{b}} + \frac{\Lambda}{3r_{b}}r_{c}^3 } \right] $ and $ \beta_{3D}(r_{c})=-\frac{4\pi r_{c}}{1-\Lambda r_{c}^2} \sqrt{\left(1-\frac{\Lambda}{3}r_{b}^2 \right) - \frac{r_{c}}{r_{b}} + \frac{\Lambda}{3r_{b}}r_{c}^3 }$ are monotonically increasing functions in that range.
}
it is thermodynamically unstable. This means that, above $T_{tr}$, the system does not settle down to thermal equilibrium. One might think, then, that thermal equilibrium would be realized below $T_{tr}$ as thermal dS spacetime. But that is not the case. Even if the thermal dS spacetime seems to be stable, it may transition to the BG spacetime
\footnote{
This is {\it not} Wheeler's bag of gold spacetime but the Lorentzian continuation of the BG instanton. 
}
with a small probability $\sim e^{-I^{os, BG}}$. Because, as stated above, it is thermodynamically unstable, it can induce eternal growth of the cosmological horizon and will not maintain equilibrium any longer. Therefore, even if the thermodynamically stable branch 3A dominates below $T_{tr}$, the system cannot be thermodynamically stable owing to the instanton effect. This is essentially the same reasoning by which the authors concluded that {\it hot flat space is unstable} in \cite{GrossPerryYaffe}. Applying the latter discussion to the $ \Lambda = \frac{1}{r_{b}^2}$ case, we can also conclude that the system is thermodynamically unstable for that case.

Generally, if there exists a thermodynamically unstable saddle, the endpoints of its branch must be bounded by thermodynamically stable branches in order for the system to be thermodynamically stable, for example, as is branch 2C in Fig. \ref{FIG3no2no1}.

\subsubsection*{Absence of the entropy bound}
Since I argued that the system is thermodynamically unstable, we may not be able to regard Fig. \ref{FIG3no3no3} as an $E-S$ relation in thermal equilibrium. Instead, we should view it as the relation of equilibrium states of isolated setting or the density of states of the system. From that point of view, the relation Fig. \ref{FIG3no3no3} tells us the system can contain arbitrary entropy, or the density of states indefinitely grows as we increase the energy. In this sense, the system does not have the entropy bound.

\subsubsection{$\displaystyle \frac{3}{r_{b}^2}\leq \Lambda $}
Above $\Lambda = \frac{3}{r_{b}^2}$, branches 3A, 3B, and 3C cease to exist. The only remaining Euclidean saddle is 3D. Owing to the choice of the subtraction term (\ref{Mann}), the energy of the branch becomes complex. I do not know how to interpret or remedy this. Additionally, I do not know whether we should include the contribution from the saddle 3D because Euclidean path integrals of general relativity may ultimately be defined by some purely complex integration contour \cite{HalliwellHartle}, as I will comment in the final section. I hope these points will be clarified in the future work \cite{Miyashita3}.

\section{Gravitoscalar thermodynamics with a simple potential}
Assuming all the Euclidean saddle points contribute to the partition function, I have shown that the properties of the system drastically change depending on $\Lambda$. The next thing I want to do is to add a   new entity into the system and investigate its properties, especially focusing on thermodynamical stability and the existence of the entropy bound. In this paper, as a simple extension, I consider a gravity-scalar system with $\varphi^2$ potential (\ref{EQ2grasca}). In addition to the parameter $M$, the mass of the scalar field, we have the  external field $J_{\varphi}(y)\equiv \varphi(y) ~(y\in \pd \MC{M})$ for parameterizing the theory. 
\footnote{
$J_{\varphi}(y)$ is, of course, just a boundary condition in terms of the gravitational theory. However, it is an external field in terms of 
the holographic field theory.
}
\footnote{
Although, together with the operator
\bea
\MC{O}_{\varphi} \equiv \int_{S^2} d^2 z \sqrt{\sigma} n_{\nu} \nabla^{\nu} \varphi ~ ,
\ena
$(J_{\varphi}, \MC{O}_{\varphi})$ form conjugate variables on the boundary, they cannot be regarded as thermodynamical variables, unlike the boundary conjugate pair of the $U(1)$ gauge field. This is because $\MC{O}_\varphi$ is not a conserved charge. In essentially the same sense, I do not regard the cosmological constant and thermodynamical volume \cite{KastorRayTraschen, CveticGibbonsKubiznakPope, Sekiwa} as thermodynamical variables in this paper. (See also footnote 3.) 
}
In order to realize thermal equilibrium in the Lorentzian theory, I take $J_{\varphi}$ to be constant over (boundary) space and time. Correspondingly, it is also for the Euclidean boundary. I mainly focus on the $J_{\varphi}$ dependence of thermodynamical properties rather than $M$ because non-zero $J_{\varphi}$ will generally induce non-trivial scalar field configurations on the saddle point geometries. I will discuss qualitatively the conditions for the existence of saddle point geometries in subsection 4.1 and the behaviors of thermodynamic potentials in subsection 4.2. 

\subsection{Conditions for the existence of saddle points}
As we saw in subsection 3.3, the horizon radius $r_{H}$ of a Euclidean BH is in the range $(0, r_{b})$ when $0<\Lambda \leq \frac{1}{r_{b}^2}$, in $(0, \frac{r_{b}}{2} \sqrt{\frac{12}{\Lambda r_{b}^2}-3  }-1 )$ when $\frac{1}{r_{b}^2}<\Lambda \leq \frac{3}{r_{b}^2}$, and Euclidean BHs cease to exist when $\Lambda > \frac{3}{r_{b}^2} $ in the pure gravity system. On the other hand, the range of the horizon radius of the BG instanton is $( \frac{r_{b}}{2} \sqrt{\frac{12}{\Lambda r_{b}^2}-3  }-1 ,\infty)$ when $0<\Lambda \leq \frac{1}{r_{b}^2}$ and $(r_{b}, \infty)$ when $\frac{1}{r_{b}^2}<\Lambda$.   I would like to discuss similar conditions on $J_{\varphi}$ in this system. 

The types of saddle point geometry are essentially the same as before, but generally with a scalar cloud. The forms of metric and scalar are as follows:
\bea
\begin{array}{l}
\displaystyle \BS{g}(x)= f(r) e^{-2\delta(r)}d \tau^2 + \frac{1}{f(r)}dr^2 + r^2 (d\theta^2 + \sin^2 \theta d\phi^2) \\\varphi(x)=\varphi(r) \hspace{7.1cm}
\end{array}
\label{EQ4ansatz}
\ena
where the coordinate ranges of $r$ and $\tau$, $[ r_{H}, r_{b}  ]$ and $[0, \beta_{0}]$, and the functions $f$, $\delta$, and $\varphi$ must satisfy the following conditions:
\bea
\begin{array}{l}
\displaystyle f(r_{H})=0, ~~~ \frac{4\pi}{f^{\p}(r_{H})}e^{\delta(r_{H})}= \beta_{0}, ~~~ \varphi^{\p}(r_{H}) = \frac{M^2 \varphi(r_{H})}{f^{\prime}(r_{H}) },    \\
~~ \\
\displaystyle \sqrt{f(r_{b})}e^{-\delta(r_{b})}\beta_{0}=\beta, ~~~ 4\pi r_{b}^2 =A, ~~~ \varphi(r_{b})=J_{\varphi}
\end{array}
 \hspace{1cm} {\rm for } ~ S^2\times D ~ {\rm topology}  \hspace{0.1cm} \\
~ \notag \\
~ \notag \\
\begin{array}{l}
\displaystyle r_{H}=0, ~~~ f(0)=1, ~~~ \varphi^{\p}(0)=0,   \\
~~ \\
\displaystyle \sqrt{f(r_{b})}e^{-\delta(r_{b})}\beta_{0}=\beta, ~~~ 4\pi r_{b}^2 =A, ~~~ \varphi(r_{b})=J_{\varphi}
\end{array}
\hspace{2.2cm} {\rm for } ~ B^3\times S^1 ~ {\rm topology} \hspace{0.cm} 
\ena
Since the latter is the Euclidean version of solitonic solutions, I call it a Euclidean soliton.
For some case, the BG instanton may exist. In that case, the coordinate range of $r$ is $[r_{b}, r_{c}]$ and the variables must satisfy the following:
\bea
\begin{array}{l}
\displaystyle f(r_{c})=0, ~~~ \frac{4\pi}{f^{\p}(r_{c})}e^{\delta(r_{c})}= -\beta_{0}, ~~~ \varphi^{\p}(r_{c}) = \frac{M^2 \varphi(r_{c})}{f^{\prime}(r_{c}) },    \\
~~ \\
\displaystyle \sqrt{f(r_{b})}e^{-\delta(r_{b})}\beta_{0}=\beta, ~~~ 4\pi r_{b}^2 =A, ~~~ \varphi(r_{b})=J_{\varphi}
\end{array}
 \hspace{1cm} 
\begin{array}{l} 
{\rm for } ~ S^2\times D ~ {\rm topology}  \\
\hspace{0.5cm}  {\rm (BG ~ instanton)} 
\end{array}
\ena
These saddles are shown schematically in Fig. \ref{FIG4no1no1}.
                                                 %
\iffigure
\begin{figure}
\begin{center}
	\includegraphics[width=4.cm]{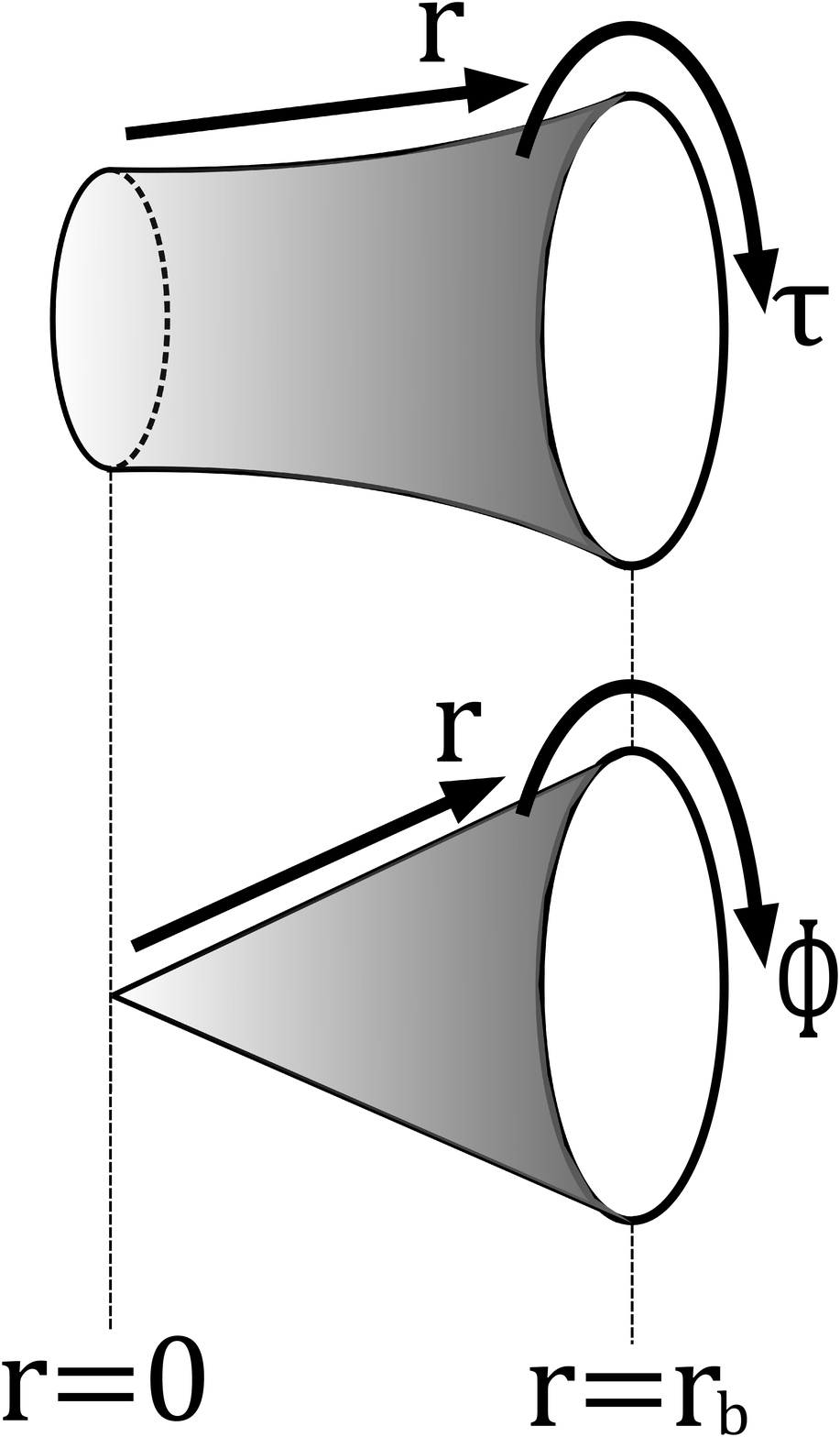} \hspace{1.5cm} 	\includegraphics[width=4.cm]{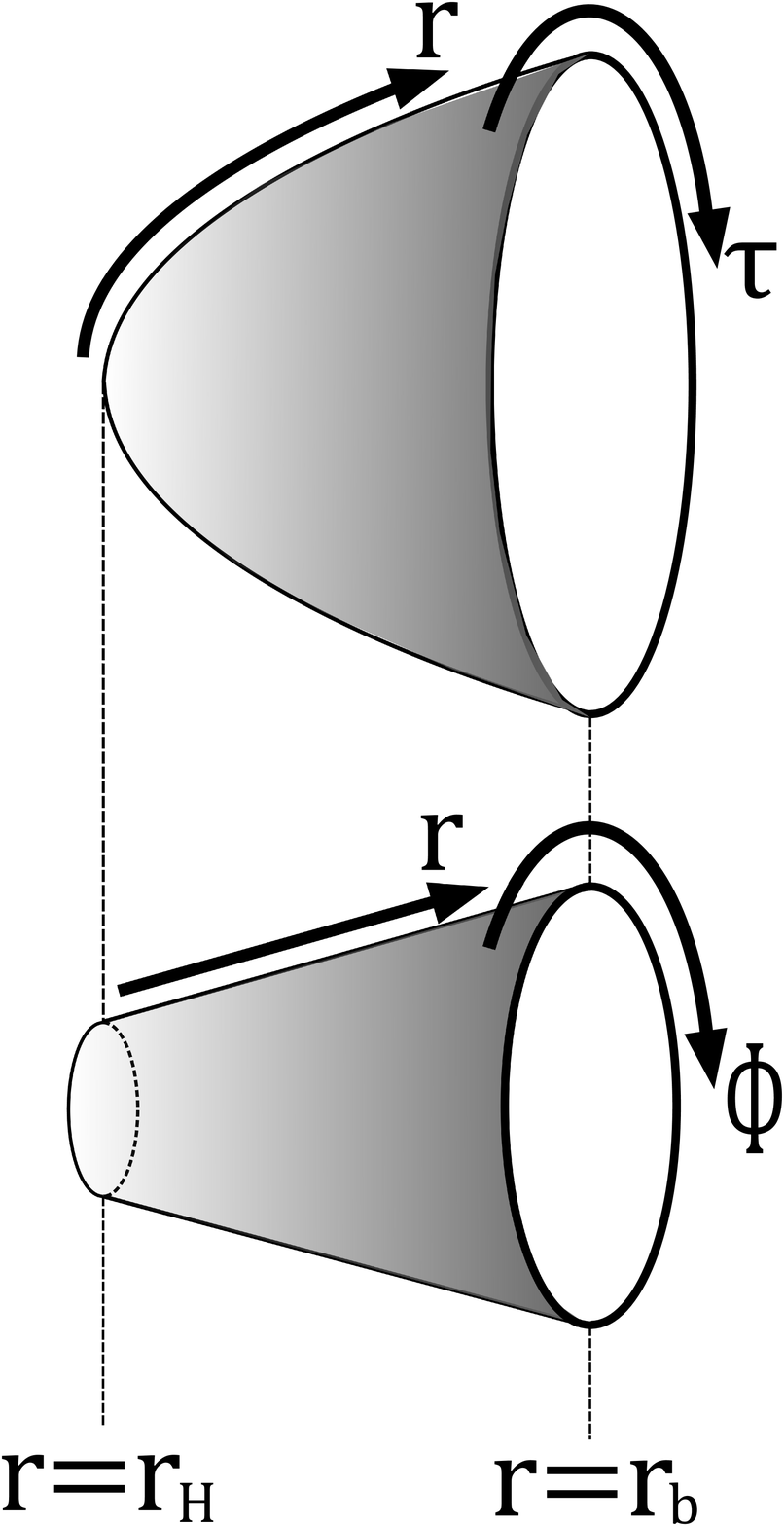} \hspace{1.5cm} 	\includegraphics[width=4.cm]{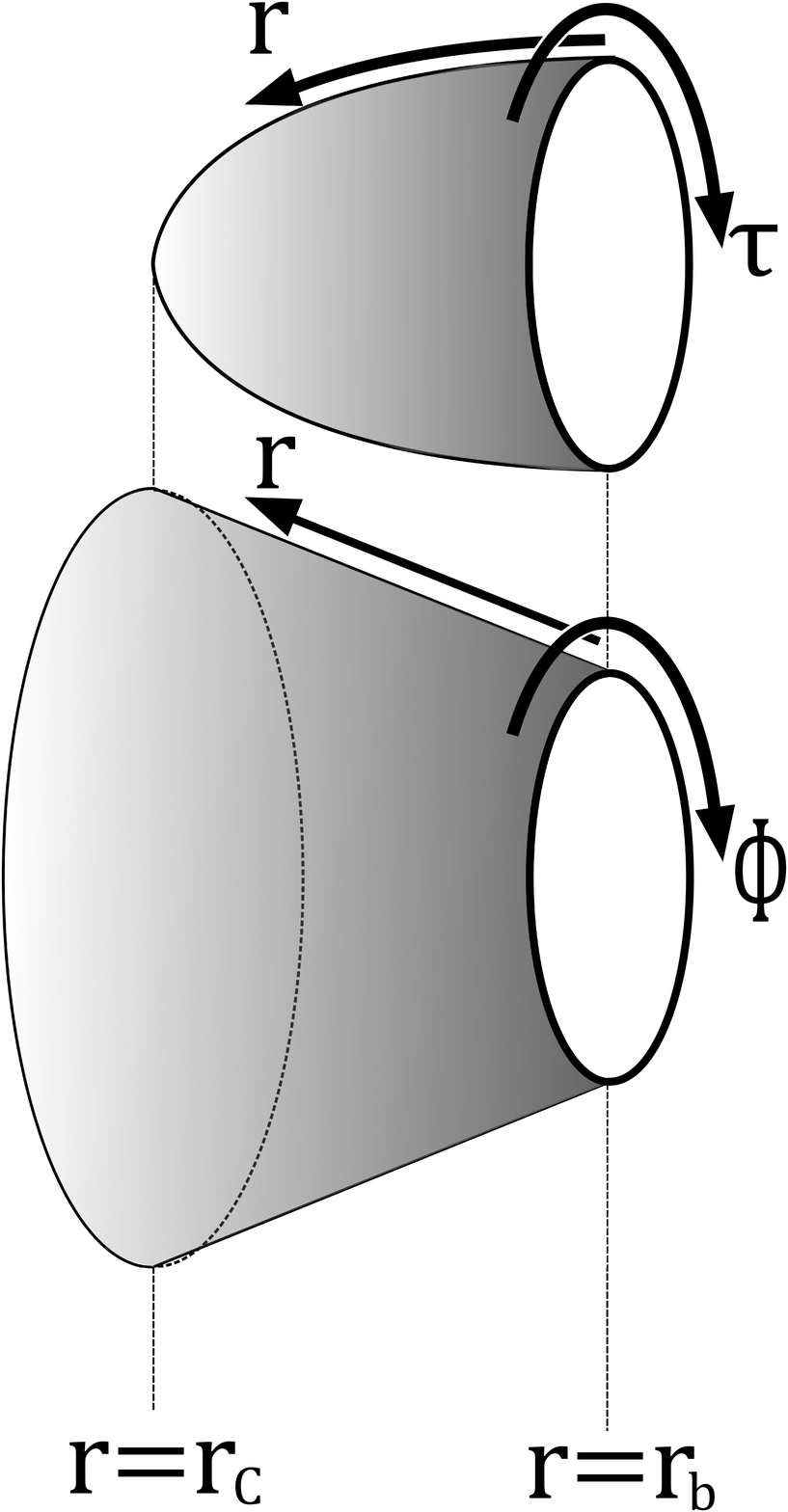}
	\caption{Types of saddle point. Their geometries are qualitatively similar to those of the pure gravity case. The gradation schematically represents the value of the scalar field.  ($\BS{{\rm Left}}$) Euclidean soliton ($B^3 \times S^1$ topology). ($\BS{{\rm Middle}}$) Euclidean BH ($S^2 \times D$ topology).   ($\BS{{\rm Right}}$) BG instanton ($S^2 \times D$ topology). }
\label{FIG4no1no1}
\end{center}
\end{figure}
\fi
                                                 %

The radius of $S^2$ of all solutions described above is monotonically increasing or decreasing. This behavior is the same as that for pure gravity with $\Lambda>0$. However, in this gravity-scalar system, there exist other types of solutions, of which the radius of $S^2$ firstly increases and then decreases (Fig. \ref{FIG4no1no1PL}). I call these types of solutions ``Euclidean PL-soliton,'' ``Euclidean PL-BH,'' and ``PL-BG instanton,'' respectively.
\footnote{
PL denotes ``Python's Lunch'' \cite{BrownGharibyanPeningtonSusskind}, {\it only} due to its shape.  
}
                                                 %
\iffigure
\begin{figure}
\begin{center}
	\includegraphics[width=3.cm]{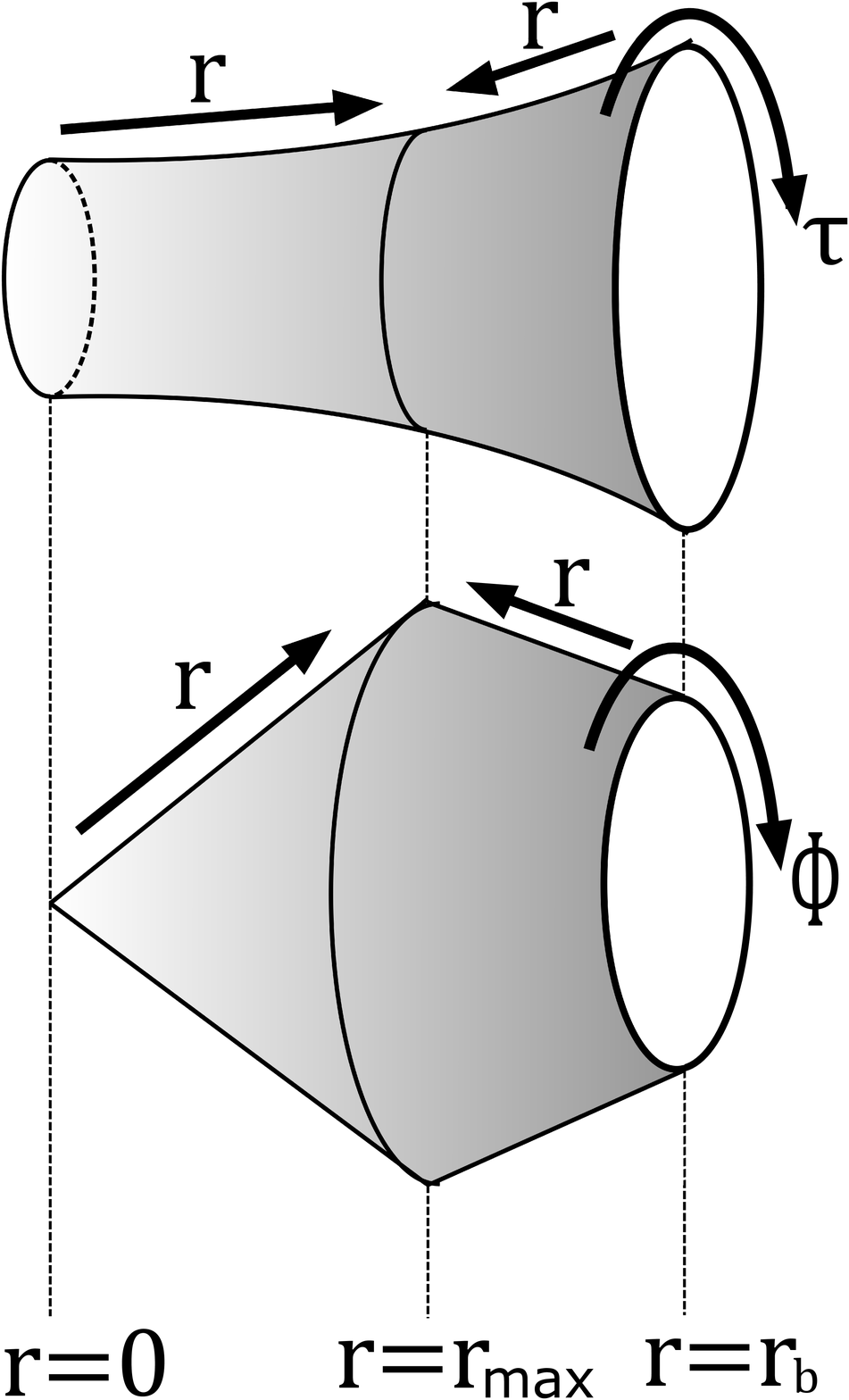} \hspace{1.5cm} 	\includegraphics[width=3.cm]{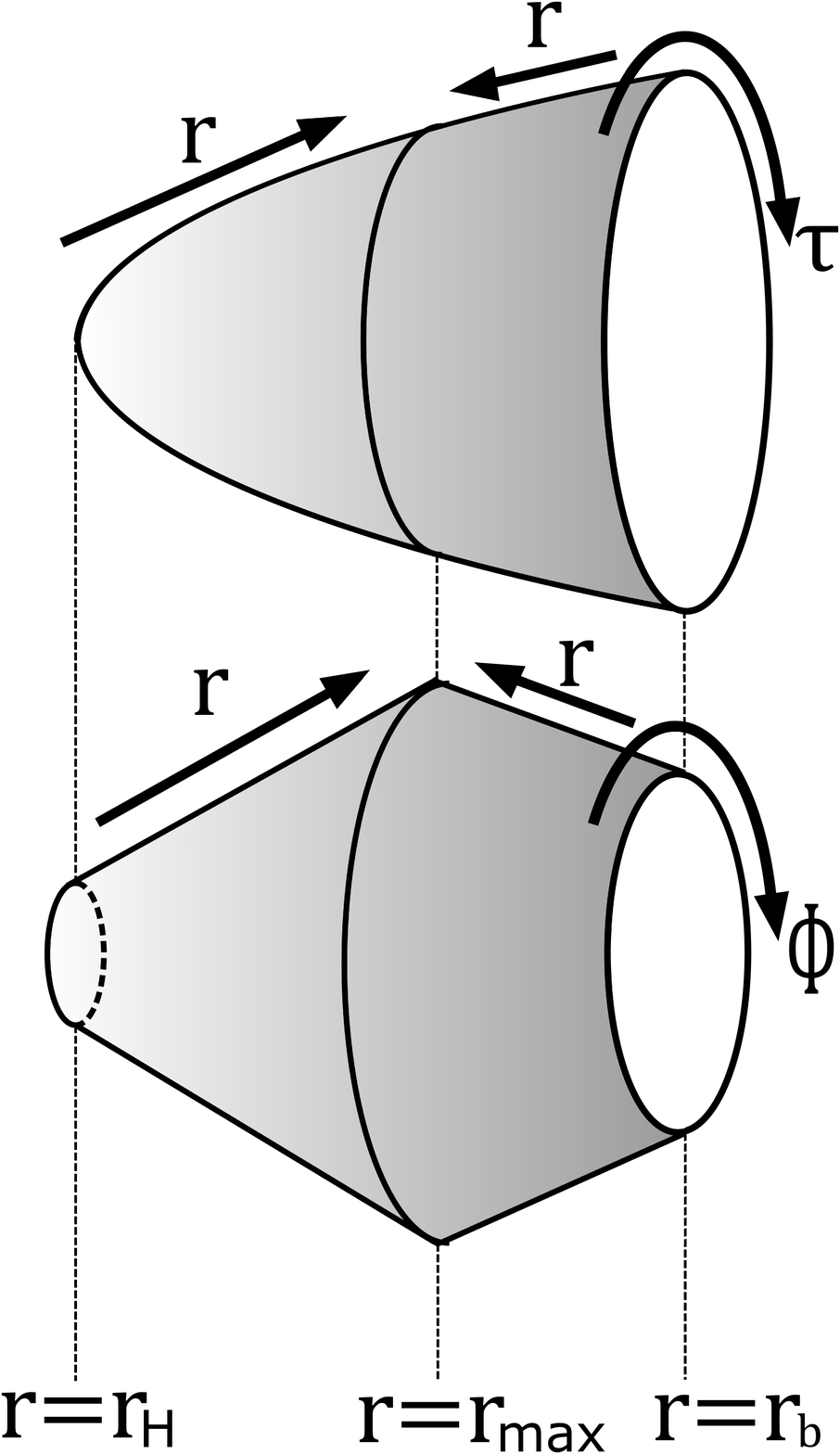} \hspace{1.5cm} 	\includegraphics[width=3.cm]{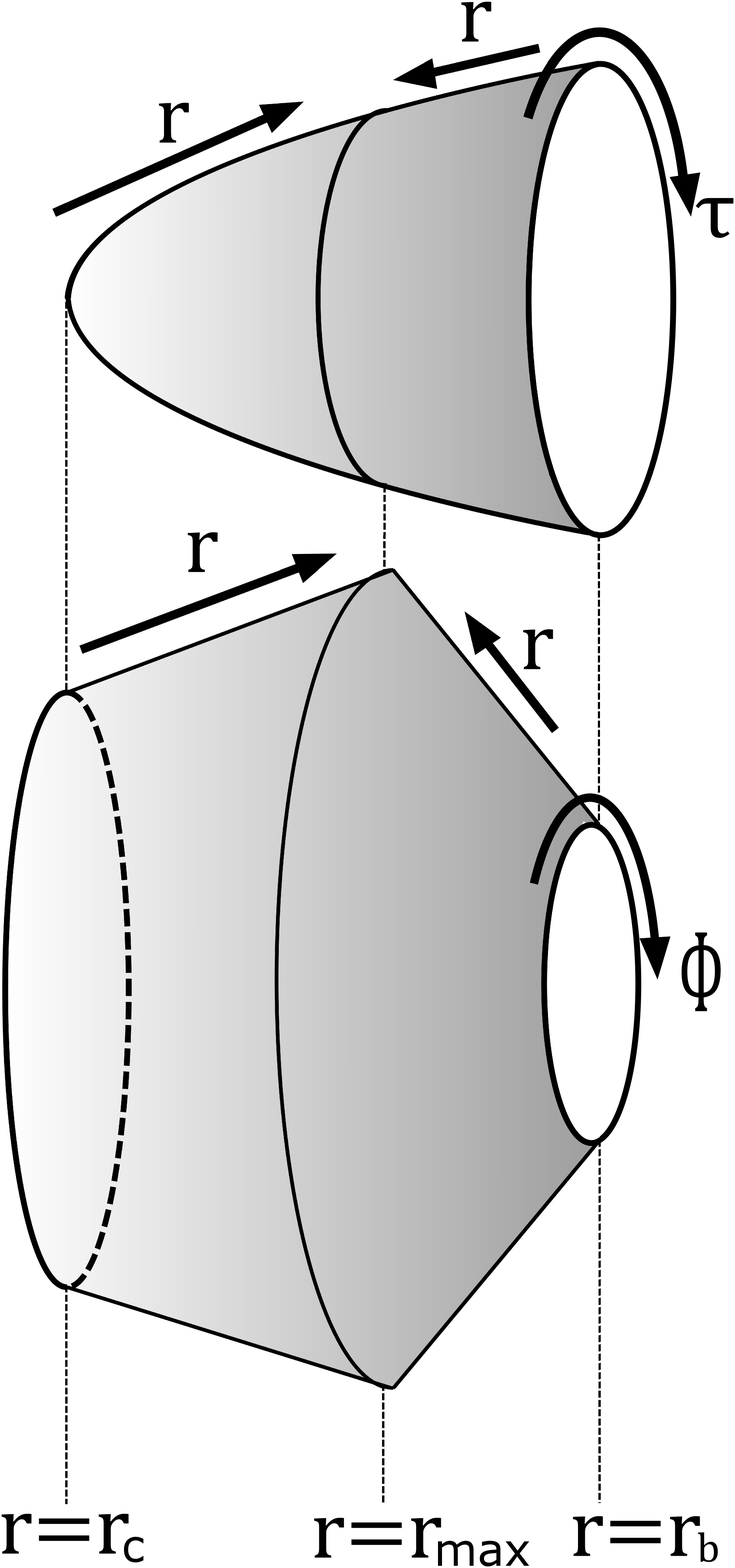}
	\caption{Three types of PL geometry whose $S^2$ radius grows from the boundary, reaches some maximum radius $r_{max}$, and shrinks toward the center or the bolt. The gradation schematically represents the value of the scalar field.  ($\BS{{\rm Left}}$) Euclidean PL-soliton ($B^3 \times S^1$ topology). ($\BS{{\rm  Middle}}$) Euclidean PL-BH ($S^2 \times D$ topology).  ($\BS{{\rm Right}}$) PL-BG instanton ($S^2 \times D$ topology).  }
\label{FIG4no1no1PL}
\end{center}
\end{figure}
\fi
                                                 %

 All the solutions described above will be obtained by integrating the equation of motion numerically. The details and some examples of numerical solutions are given in Appendix A. 
In addition, rigorously speaking, PL solutions cannot be described by the form of the metric (\ref{EQ4ansatz}); that is, we have to use other coordinates and choices of metric functions. These details will also be explained in Appendix A.  

Below, I will classify the structure of the existence of solutions on $J_{\varphi}-r_{H(c)}$ plane by the value of $\Lambda$. For some cases, however, it highly depends on $r_{b}$ and $M$ so that it is difficult to show its qualitative properties exhaustively. For these cases, I just give some examples of the structure and I mark these cases with $(\ast)$.

\subsubsection*{Euclidean BH and Euclidean soliton}
Similar to the $\Lambda > 0$ case, $J_{\varphi}$ does not affect the existence of Euclidean BHs if it is sufficiently small but does if it is not. I define $r_{H,max}(r_{b}, \Lambda, J_{\varphi}, M)$ as the maximum horizon radius for given parameters. The behavior of $r_{H,max}(r_{b}, \Lambda, J_{\varphi}, M)$ against $J_{\varphi}$ for fixed $r_{b}, \Lambda, M$ may be classified into three cases depending on $\Lambda$:\\
~\\
$\bullet$ $-\infty<\Lambda<\frac{1}{r_{b}^2}$ \\

$r_{H,max}=r_{b} ~ {\rm for} ~ 0 \leq J_{\varphi} < \sqrt{\frac{1-\Lambda r_{b}^2}{4\pi G}}\frac{1}{r_{b} M } $ and it monotonically decreases until reaching some critical value $J_{\varphi, BH cri}(r_{b}, \Lambda, M)$ where $r_{H,max}(r_{b}, \Lambda, J_{\varphi, BH cri}(r_{b}, \Lambda, M), M)=0$. $r_{H,max}=0$ for $J_{\varphi, BH cri}(r_{b}, \Lambda, M) \leq J_{\varphi}$. 
\footnote{
For $-\infty < \Lambda < 0$ and $0 < J_{\varphi}< \sqrt{ \frac{-\Lambda}{4\pi G M^2} } $, there always exist solutions for arbitrary $r_{b}>0$. One might guess there also exist hairy solutions for asymptotically AdS ($r_{b} \to \infty$) case. However, since the energy of these solutions diverges as we increase $r_{b}$, the limiting solutions are not physical, or do not contribute to the partition function. This is also true for other types of solution. 
}
~\\
~\\
$\bullet$ $\frac{1}{r_{b}^2}<\Lambda<\frac{3}{r_{b}^2}$ \\

For $0 \leq J_{\varphi} <J_{\varphi, BH cri}(r_{b}, \Lambda, M) $, $r_{H,max} $ monotonically decreases from $r_{H.max}(r_{b},\Lambda, 0, M)= \frac{r_{b}}{2} \left( \sqrt{\frac{12}{ \Lambda r_{b}^2}-3 }-1 \right)  $. $r_{H,max}=0$ for $J_{\varphi, BH cri}(r_{b}, \Lambda, M) \leq J_{\varphi}$.
~\\
~\\
$\bullet$ $\frac{3}{r_{b}^2}<\Lambda<\infty$ \\

$r_{H,max}=0$, that is, Euclidean BHs and Euclidean solitons do not exist for any $J_{\varphi}$.

~\\
The qualitative behaviors of $r_{H,max}(r_{b}, \Lambda, J_{\varphi}, M)$ against $J_{\varphi}$ for the first and second cases are shown in Fig. \ref{FIG4no1no2}. 
                                                 %
\iffigure
\begin{figure}
\begin{center}
	\includegraphics[width=7.cm]{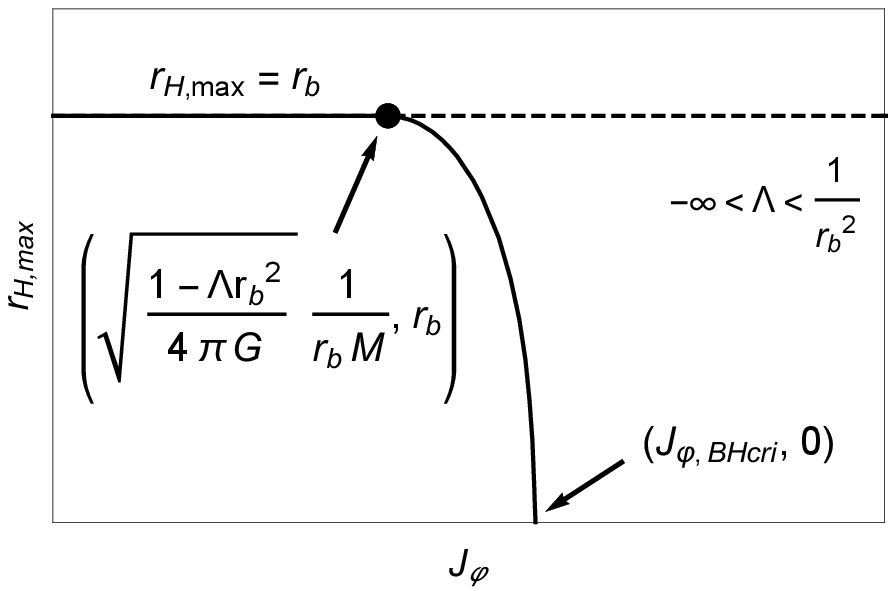} \hspace{1.5cm} 	\includegraphics[width=7.cm]{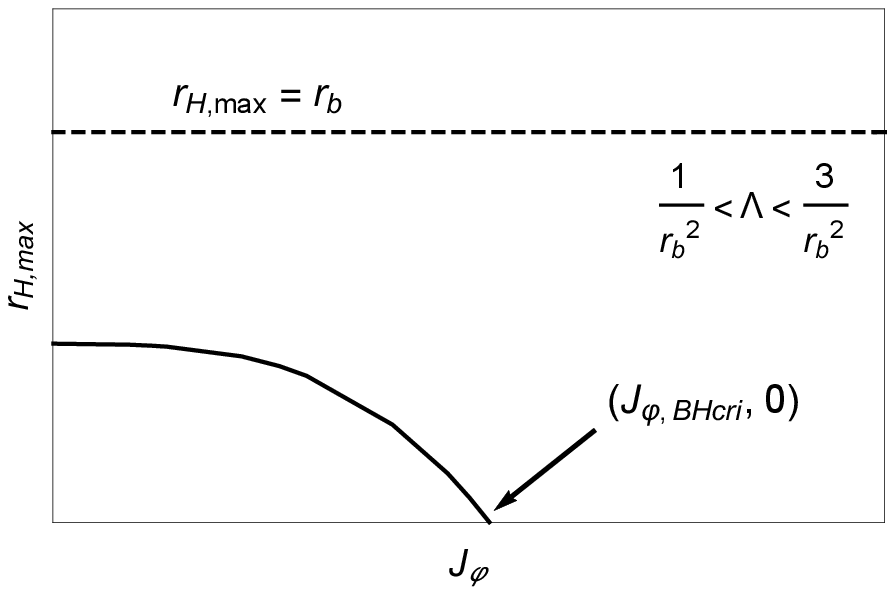} 
	\caption{Qualitative behaviors of $r_{H,max}(r_{b}, \Lambda, J_{\varphi}, M)$ against $J_{\varphi}$. ($\BS{{\rm Left}}$)  $-\infty<\Lambda<\frac{1}{r_{b}^2}$ case.  ($\BS{{\rm Right}}$) $\frac{1}{r_{b}^2}<\Lambda<\frac{3}{r_{b}^2}$ case. As we increase $\Lambda$, the curve shrinks and collapses to the point $(0,0)$ when $\Lambda= \frac{3}{r_{b}^2}$.}
\label{FIG4no1no2}
\end{center}
\end{figure}
\fi
                                                 %

\subsubsection*{BG instanton}
The range of possible size of the cosmological horizon $r_{c}$ of the BG instanton depends on not only $J_{\varphi}$ but also $r_{b}, \Lambda, M$. The behavior of the range against $J_{\varphi}$ for fixed $r_{b}, \Lambda, M$ may be roughly classified into three cases depending on $\Lambda$: \\
~\\
$\bullet$ $-\infty<\Lambda\leq0$ \\

There are no BG instantons for $0\leq J_{\varphi}\leq \sqrt{\frac{1-\Lambda r_{b}^2}{4\pi G}}\frac{1}{r_{b}M}$. For $\sqrt{\frac{1-\Lambda r_{b}^2}{4\pi G}}\frac{1}{r_{b}M}< J_{\varphi}$, the range is $(r_{b}, r_{c,max}(r_{b}, \Lambda,J_{\varphi}, M))$, where $r_{c,max}(r_{b}, \Lambda, J_{\varphi}, M)$ is a monotonically increasing function for $J_{\varphi}$. There also exists some critical radius $r_{b}<r_{c,cr}(r_{b},  \Lambda,J_{\varphi}, M)<r_{c,max}(r_{b}, \Lambda,J_{\varphi}, M)$, which is also a monotonically increasing function for $J_{\varphi}$. Between $r_{c,max}$ and $r_{c,cr}$, there exist two different solutions with the same $r_{c}$. The qualitative behavior of the range is shown in Fig. \ref{FIG4no1no3}.

~\\
~\\
                                                 %
\iffigure
\begin{figure}
\begin{center}
	\includegraphics[width=7.cm]{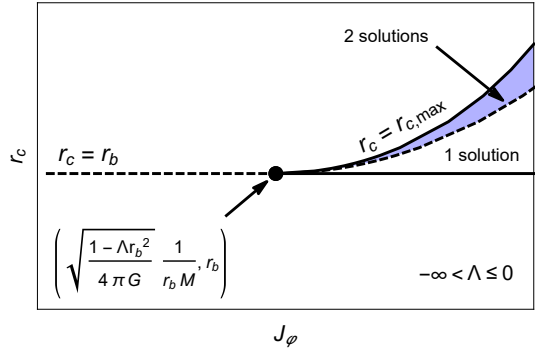} 
	\caption{Qualitative behavior of range of possible size of cosmological horizon $r_{c}$ of BG instanton against $J_{\varphi}$ when $-\infty < \Lambda \leq 0$. For $0\leq J_{\varphi}\leq \sqrt{\frac{1-\Lambda r_{b}^2}{4\pi G}}\frac{1}{r_{b}M}$, there are no BG instantons. Above $\sqrt{\frac{1-\Lambda r_{b}^2}{4\pi G}}\frac{1}{r_{b}M}$, there exist BG instantons between $r_{c,max}(r_{b}, \Lambda,J_{\varphi}, M)$ and $r_{b}$. There exist two solutions with the same $r_{c}$ in the shaded region.}
\label{FIG4no1no3}
\end{center}
\end{figure}
\fi
                                                 %
                                                 %
\iffigure
\begin{figure}
\begin{center}
	\includegraphics[width=7.cm]{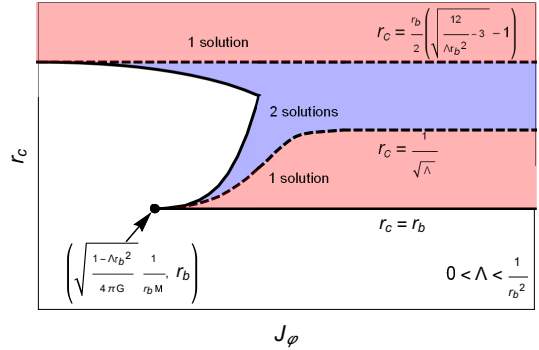} 
	\caption{Example of qualitative behavior of range of possible size of cosmological horizon $r_{c}$ of BG instanton against $J_{\varphi}$ when $0 < \Lambda \leq \frac{1}{r_{b}^2}$. This structure may appear when both $\Lambda/M^2$ and $r_{b}M$ are large enough. This is the simplest structure. If we choose parameters that do not satisfy the above conditions, a more complicated structure will appear, such as Fig. \ref{FIG4no1no5}. (An example of parameters that exhibit this structure is $(r_{b}, \Lambda,  M )=(5\sqrt{G}, 0.012/G, 0.1/\sqrt{G})$).}
\label{FIG4no1no4}
\end{center}
\end{figure}
\fi
                                                 %
                                                 %
\iffigure
\begin{figure}
\begin{center}
	\includegraphics[width=7.cm]{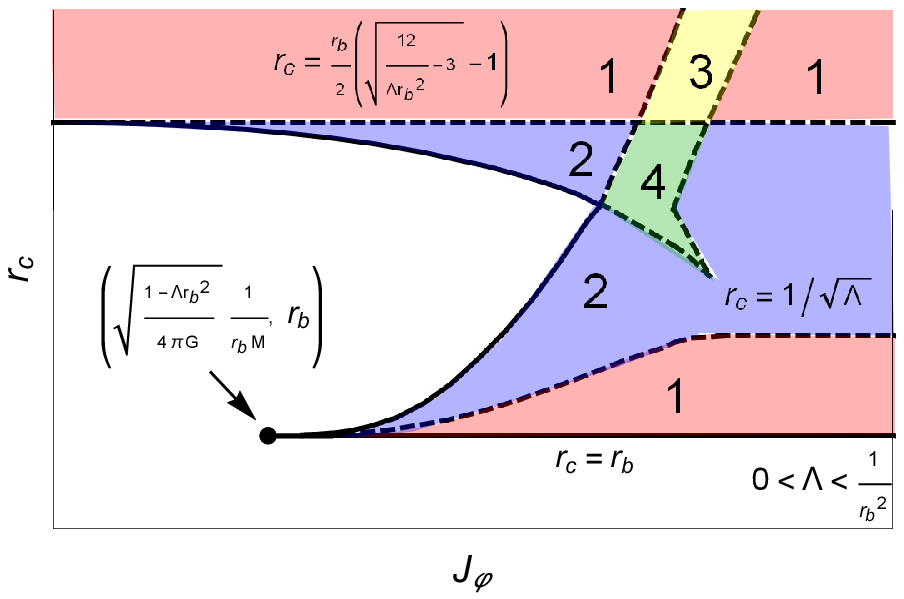} ~~~~~~
		\includegraphics[width=7.cm]{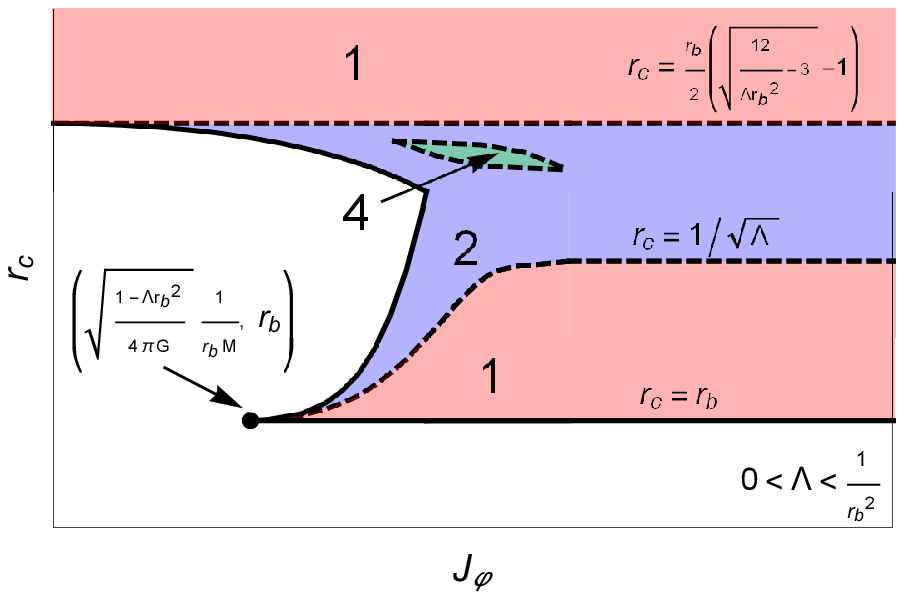} 
	\caption{Other examples of qualitative behavior of range of possible size of cosmological horizon $r_{c}$ of BG instanton against $J_{\varphi}$ when $0 < \Lambda \leq \frac{1}{r_{b}^2}$.  The numbers in the figures represent the number of solutions with the same $r_{c}$ and $J_{\varphi}$. (Examples of parameters that exhibit these structures are $(r_{b}, \Lambda, M )=(7\sqrt{G}, 0.002/G, 0.1/\sqrt{G})$ ($\BS{{\rm Left}}$) and $(r_{b}, \Lambda, M )=(3\sqrt{G}, 0.0056/G, 0.1/\sqrt{G})$ ($\BS{{\rm Right}}$).)}
\label{FIG4no1no5}
\end{center}
\end{figure}
\fi
                                                 %
~\\
$\bullet$ $0<\Lambda \leq \frac{1}{r_{b}^2}$ $(\ast)$ \\

Contrary to the above case, there exist BG instantons for all $J_{\varphi}$, and the structure of the existence domain and the number of instanton solutions are complicated and depend on the other parameters $r_{b}$ and $M$. The qualitative behaviors that do not depend on $r_{b}$ and $M$ are as follows: \\
$\cdot$ for sufficiently small $J_{\varphi}$, the existence range is $[r_{c,min}(r_{b}, \Lambda, J_{\varphi}, M), \infty)$, where $r_{c,min}$ is a  monotonically decreasing function for $J_{\varphi}$, $r_{c,min}(r_{b}, \Lambda, 0,  M)= \frac{r_{b}}{2} \left(\sqrt{\frac{12}{\Lambda r_{b}^2}-3} -1\right)$ and greater than $r_{b}$,\\
$\cdot$ for sufficiently small $J_{\varphi}$, between $\frac{r_{b}}{2} \left(\sqrt{\frac{12}{\Lambda r_{b}^2}-3} -1\right)$ and $r_{c,min}(r_{b}, \Lambda, J_{\varphi}, M)$, there exist two different solutions with the same $r_{c}$, \\
$\cdot$ for $\sqrt{\frac{1-\Lambda r_{b}^2}{4\pi G}}\frac{1}{r_{b}M}\leq J_{\varphi}$, the smallest cosmological horizon is $r_{b}$, \\
$\cdot$ when $J_{\varphi}$ is greater than some critical value, the range of cosmological horizon is $[r_{b}, \infty)$. \\
Some examples of the qualitative behaviors of the range are shown in Fig. \ref{FIG4no1no4}, \ref{FIG4no1no5}. The simplest structure would be Fig. \ref{FIG4no1no4}. This may be realized when the ratio $\Lambda/M^2$ is sufficiently large and the ratio of the boundary radius $r_{b}$ to the Compton wavelength $\frac{1}{M}$ is also sufficiently large, that is, when the effect of the scalar field is minimum (but still non-trivial). When one of the conditions ceases to hold, the structure becomes complicated, such as Fig. \ref{FIG4no1no5}. 
\footnote{This is not the complete list. Depending on the parameters $r_{b}, \Lambda$, and $M$, there are many structures. I will not try to make an exhaustive list of structures in this paper.}

~\\
~\\
$\bullet$ $\frac{1}{r_{b}^2} < \Lambda<\infty$ \\

There exist BG instantons for all $J_{\varphi}$. The range is $[r_{b}, \infty)$ and the number of solutions is one.

~\\
~\\

\subsubsection*{Euclidean PL-soliton, Euclidean PL-BH, and PL-BG instanton}
                                                 %
\iffigure
\begin{figure}
\begin{center}
	\includegraphics[width=7.cm]{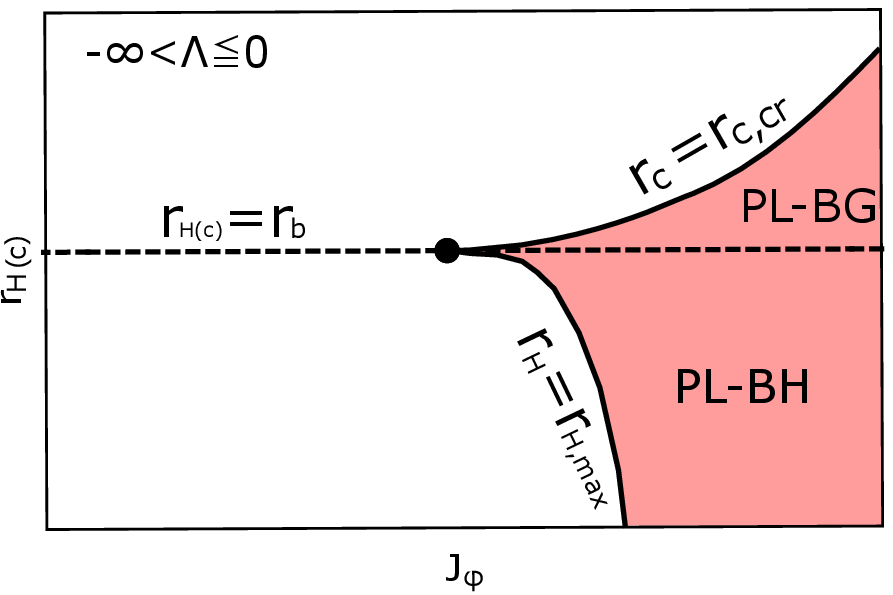} ~~~~~~ 	\includegraphics[width=7.cm]{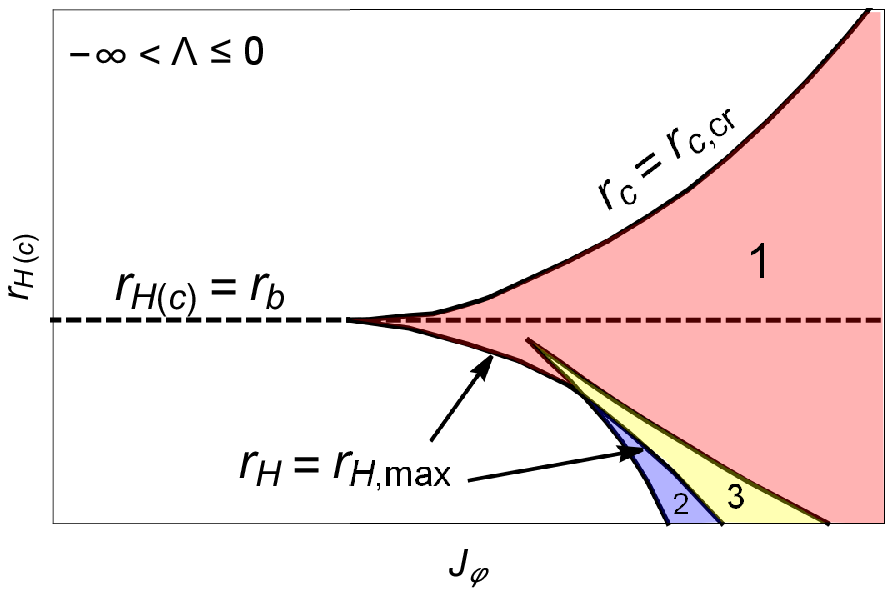}
	\caption{Existence region on $J_{\varphi}-r_{H(c)}$ plane of Euclidean PL-soliton, PL-BH, and PL-BG instanton when $\Lambda\leq 0$. (\textbf{Left}) Simple structure which may appear when both $|\Lambda|/M^2$  and $r_{b}M$ are large enough. (\textbf{Right}) Complicated structure which appear when the condition does not hold. (Examples of parameters that exhibit these structures are  $(r_{b}, \Lambda, M )=(5\sqrt{G}, -0.04/G, 0.3/\sqrt{G})$ ($\BS{{\rm Left}}$) and $(r_{b}, \Lambda, M )=(5\sqrt{G}, -0.004/G, 0.1/\sqrt{G})$ ($\BS{{\rm Right}}$).  )       }
\label{FIG4no1noSHIN6}
\end{center}
\end{figure}
\fi
                                                 %
The existence of these geometries also depends on the parameters. 
Actually, PL-BH and PL-BG are continuously deformable by changing parameters. Therefore, I will treat them in an unified manner. \\
~\\
$\bullet$ $-\infty<\Lambda\leq0$ $(\ast)$ \\

For this case, as shown before, the existence region on the $J_{\varphi}-r_{H}$ plane of Euclidean solitons or BHs is below the line $r_{H}=r_{b}$ and on the left of the curve $r_{H}=r_{H, max}(r_{b}, \Lambda, J_{\varphi}, M)$. When both $|\Lambda|/M^2$  and $r_{b}M$ are large enough, that is, when the effect of scalar filed is minimum, the existence region of the Euclidean PL-soliton or PL-BH is on the right of the curve $r_{H}=r_{H, max}(r_{b}, \Lambda, J_{\varphi}, M)$ and below $r_{H}=r_{b}$, the existence region of the PL-BG instanton is between the line $r_{c}=r_{b}$ and the curve $r_{c}=r_{c,cr}(r_{b}, \Lambda, J_{\varphi}, M)$ The number of solutions is always one. See Fig. \ref{FIG4no1noSHIN6} (\textbf{Left}). When the condition does not hold, the structure becomes complicated as Fig. \ref{FIG4no1noSHIN6} (\textbf{Right}).  

~\\
~\\

~\\
$\bullet$ $0<\Lambda\leq\frac{1}{r_{b}}$ $(\ast)$ \\

In this case, there exists an upper bound on $J_\varphi$ for a given $r_{H(c)}<\frac{1}{\sqrt{\Lambda}}$. Additionally, there exist several solutions with the same $r_{H(c)}$ and $J_{\varphi}$, and there are many types of structure of the existence domain depending on $r_{b}$ and $M$.  Let $r_{c,cr}$ denote the lower boundary of solution regions 2 and 1 in Fig. \ref{FIG4no1no4}.  The simplest example of the existence region is shown schematically in Fig. \ref{FIG4no1noSHIN7} (\textbf{Left}).  This type of structure appears when $\Lambda$ is sufficiently small compared to $1/r_{b}^2$ and $M$. When the condition does not hold, the structure becomes more complicated, for example, as Fig \ref{FIG4no1noSHIN7} (\textbf{Right}). Another example that shows a more complicated structure  is shown in Fig. \ref{FIG4no1noSHIN8}.

~\\
~\\
                                                 %
\iffigure
\begin{figure}
\begin{center}
	\includegraphics[width=6.cm]{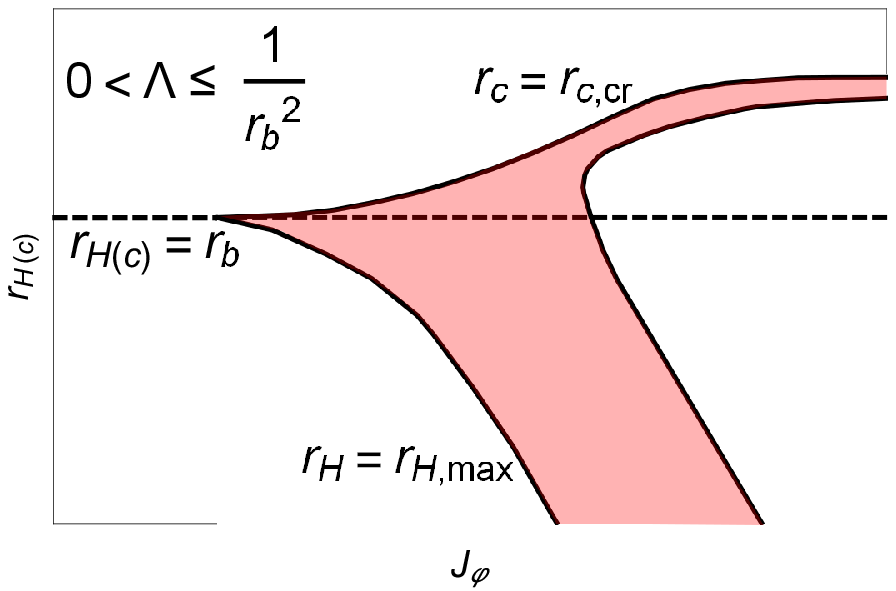} ~~~~~~
		\includegraphics[width=6.cm]{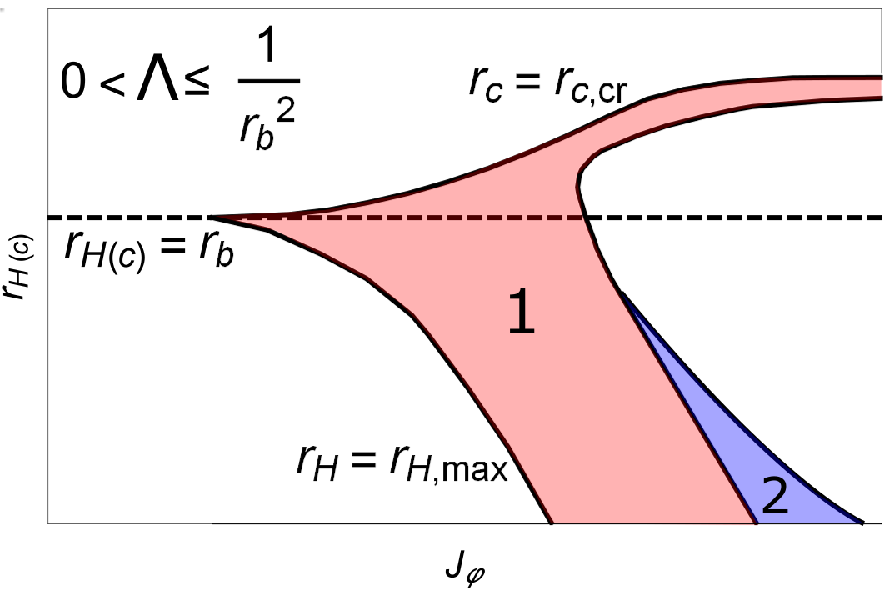} 
	\caption{Examples of the structure of existence region on $J_{\varphi}-r_{H(c)}$ plane of Euclidean PL-soliton, PL-BH, and PL-BG instanton when $0<\Lambda\leq \frac{1}{r_{b}^2}$. (\textbf{Left}) This type of structure appears when $\Lambda$ is sufficiently small compared to $1/r_{b}^2$ and $M$. This would be the simplest one.  (\textbf{Right}) When the condition no longer holds, the structure becomes complicated. The numbers in the figures represent the number of solutions with the same $r_{c}$ and $J_{\varphi}$. (Examples of parameters that exhibit these structures are  $(r_{b}, \Lambda, M )=(5\sqrt{G}, 0.012/G, 0.5/\sqrt{G})$ ($\BS{{\rm Left}}$) and $(r_{b}, \Lambda, M )=(5\sqrt{G}, 0.036/G, 0.5/\sqrt{G})$ ($\BS{{\rm Right}}$).  )  }
\label{FIG4no1noSHIN7}
\end{center}
\end{figure}
\fi
                                                 %
                                                 %
\iffigure
\begin{figure}
\begin{center}
	\includegraphics[width=7.cm]{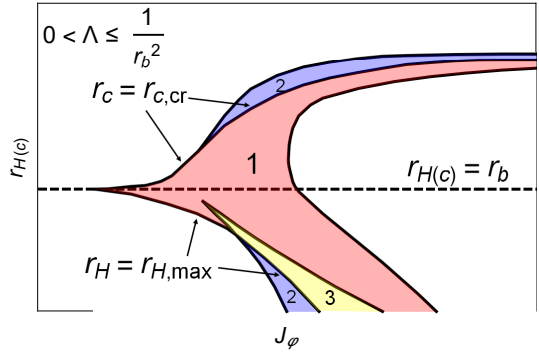} ~~
	\caption{Another example of the structure of existence region on $J_{\varphi}-r_{H(c)}$ plane of Euclidean PL-soliton, PL-BH, and PL-BG instanton when $0<\Lambda\leq \frac{1}{r_{b}^2}$.   (A example of parameters that exhibits this structure is $(r_{b}, \Lambda, M )=(5\sqrt{G}, 0.0012/G, 0.1/\sqrt{G})$.)  }
\label{FIG4no1noSHIN8}
\end{center}
\end{figure}
\fi
                                                 %

~\\
$\bullet$ $\frac{1}{r_{b}^2}<\Lambda\leq \frac{3}{r_{b}^2}$ $(\ast)$ \\

In this parameter range, there are no PL-BG instanton solutions and only PL-BHs and PL-solitons exist. There exist the upper bound and the lower bound of $J_{\varphi}$. The lower bound is always greater than $0$. It implies that PL type solutions does not exist in the pure gravity system, same as the previous cases. This case also has many types of structure. The simplest structure is shown in Fig. \ref{FIG4no1noSHIN9} (\textbf{Left}) and a example of complicated structure is Fig. \ref{FIG4no1noSHIN9}  (\textbf{Right}).

                                                 %
\iffigure
\begin{figure}
\begin{center}
	\includegraphics[width=7.cm]{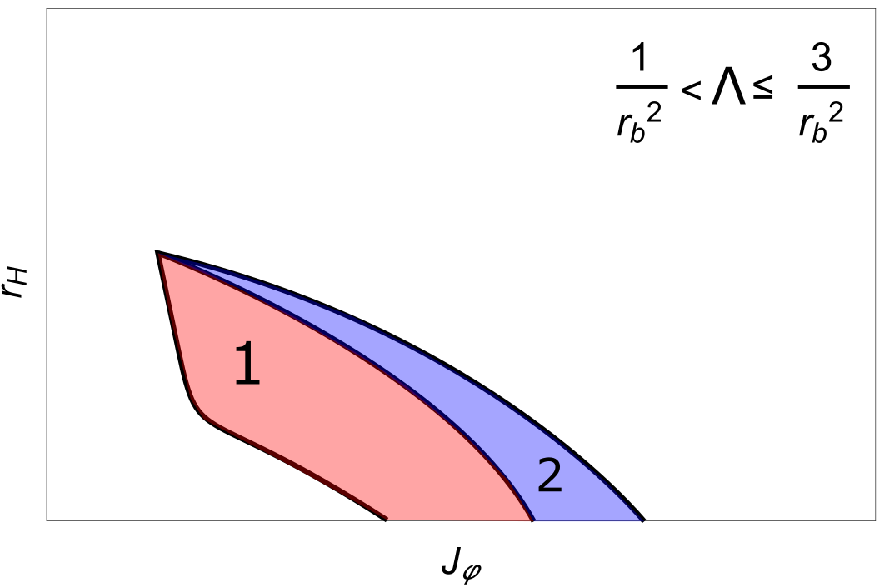} ~~~ 	\includegraphics[width=7.cm]{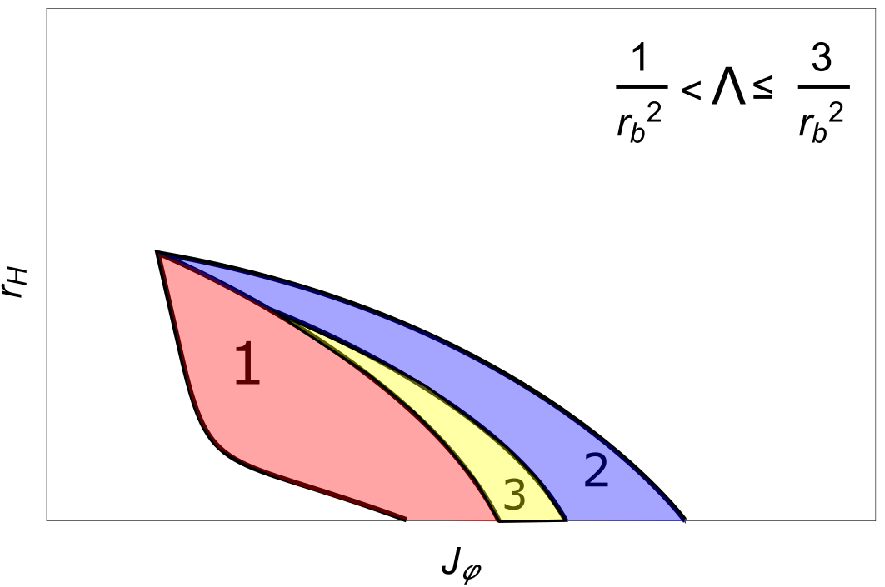} 
	\caption{Examples of the structure of existence region on $J_{\varphi}-r_{H(c)}$ plane of Euclidean PL-soliton and PL-BH when $\frac{1}{r_{b}^2} <\Lambda\leq \frac{3}{r_{b}^2}$. Note that this structure is below $r_{H}=r_{b}$ line, i.e., there are no PL-BG instanton solutions.  (\textbf{Left}) Simple structure which may appear when both $|\Lambda|/M^2$  and $r_{b}M$ are large enough. (\textbf{Right}) Complicated structure which appear when the condition does not hold. (Examples of parameters that exhibit these structures are  $(r_{b}, \Lambda, M )=(5\sqrt{G}, 0.044/G, 0.1/\sqrt{G})$ ($\BS{{\rm Left}}$) and $(r_{b}, \Lambda, M )=(5\sqrt{G}, 0.044/G, 0.5/\sqrt{G})$ ($\BS{{\rm Right}}$).  )   }
\label{FIG4no1noSHIN9}
\end{center}
\end{figure}
\fi
                                                 %

~\\
~\\
$\bullet$ $\frac{3}{r_{b}^2}<\Lambda < \infty$  \\

In this parameter range, there are no PL type solutions.

~\\
~\\

\subsection{Thermodynamic potentials and thermodynamic properties}
In the previous subsection, I showed the conditions on $J_{\varphi}$ for the existence of saddle points, and the dependence on $J_{\varphi}$ of the number of saddle points can be classified into some cases depending on $\Lambda$. Therefore, in this subsection, I will consider four cases separately, namely, $-\infty<\Lambda\leq 0$, $0<\Lambda \leq \frac{1}{r_{b}^2} $, $\frac{1}{r_{b}^2}< \Lambda <\leq \frac{3}{r_{b}^2}$, and $\frac{3}{r_{b}^2}  < \Lambda <\infty$, and I will discuss the $J_{\varphi}$ dependence of thermodynamic potentials and thermodynamic properties, particularly thermodynamical stability and entropy bound. 

\subsubsection{$-\infty<\Lambda\leq 0$}
$\bullet$ $0 \leq J_{\varphi} < \sqrt{\frac{1-\Lambda r_{b}^2}{4\pi G}} \frac{1}{r_{b}M} $\\
~\\
Only Euclidean BHs (and Euclidean solitons) exist, and their maximum horizon radius is always $r_{b}$.  The qualitative behaviors of free energy and entropy are almost the same as those for pure gravity with $\Lambda \leq 0$ (see Fig. \ref{FIG3no2no1} ($\BS{{\rm Left}}$) or Fig. \ref{FIG3no2no2} ($\BS{{\rm Left}}$)); being thermodynamically stable, the system exhibits a Hawking--Page phase transition, and the maximum entropy $\frac{\pi r_{b}^2}{G}$ is attained at the maximum energy $\frac{r_{b}}{G} \sqrt
{1-\frac{\Lambda}{3}r_b^2}$.  One difference would be that the free energy at low temperature or the ground state energy $E_{gs}(r_{b},\Lambda,J_{\varphi},M)$ is not zero but some finite value. $E_{gs}(r_{b},\Lambda,J_{\varphi},M)$ is a monotonically increasing function for $J_{\varphi}$ and satisfies $E_{gs}(r_{b},\Lambda, 0, M)=0$ and $E_{gs}(r_{b},\Lambda, J_{\varphi, BHcri}(r_{b}, \Lambda, M), M)=\frac{r_{b}}{G} \sqrt
{1-\frac{\Lambda}{3}r_b^2}$. 
For the pure gravity case, the transition temperature depends on $\Lambda \leq 0$ and it takes minimal value $T_{tr}=\frac{27}{32\pi r_{b}}$ when $\Lambda=0$. Non-zero $J_{\varphi}$ shifts the minimal temperature a little lower. Generally, the larger $J_{\varphi}$, the larger the amount of shift. Therefore, the lowest transition temperature can be obtained when $J_{\varphi}= \sqrt{\frac{1-\Lambda r_{b}^2}{4\pi G}} \frac{1}{r_{b}M}$ for fixed $r_{b}, \Lambda, M$. Of course, $M$ also affects the amount of the shift.
                                                 %
\iffigure
\begin{figure}
\begin{center}
	\includegraphics[width=9.cm]{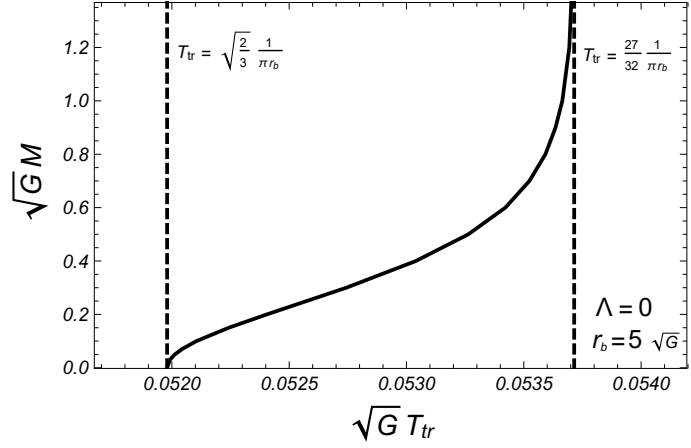} 
	\caption{Relation between transition temperature $T_{tr}$ and $M$ with fixing $J_{\varphi}=\sqrt{\frac{1-\Lambda r_{b}^2}{4\pi G}}\frac{1}{r_{b}M}$ when $\Lambda=0, r_{b}=5\sqrt{G}$. As we decrease $M$, $T_{tr}$ also decreases and $T_{tr}$ reaches the minimal value $\sqrt{\frac{2}{3}}\frac{1}{\pi r_{b}} $ in the limit $M\to 0$. For $J_{\varphi}$ less than $\sqrt{\frac{1-\Lambda r_{b}^2}{4\pi G}}\frac{1}{r_{b}M}$, the transition temperature is placed on the right of the curve and less than $\frac{27}{32\pi r_{b}}$. }
\label{FIG4no1no5ten5}
\end{center}
\end{figure}
\fi
                                                 %
As shown in Fig. \ref{FIG4no1no5ten5}, the minimal value of the transition temperature will decrease as we decrease $M$. Therefore, the limiting value of the minimal temperature can be obtained by taking the limit $M\to 0$ while keeping $J_{\varphi}= \sqrt{\frac{1-\Lambda r_{b}^2}{4\pi G}} \frac{1}{r_{b}M} $. This results in the saddle points becoming those of pure gravity with the positive cosmological constant $\frac{1}{r_{b}^2}$, and we can obtain the exact expression
\bea
T_{tr, mim}(r_{b}) = \sqrt{\frac{2}{3}} \frac{1}{\pi r_{b}} \label{EQ4no2mimT},
\ena
which is slightly smaller than $T_{tr}=\frac{27}{32\pi r_{b}}$.  This is the minimum transition temperature of the gravity-scalar system. Note that the transition temperature becomes the minimum one (\ref{EQ4no2mimT}) for any $\Lambda \leq 0$ when we take the limit $M\to 0$ while keeping $J_{\varphi}= \sqrt{\frac{1-\Lambda r_{b}^2}{4\pi G}} \frac{1}{r_{b}M} $.  \\
~ \\
$\bullet$ $\sqrt{\frac{1-\Lambda r_{b}^2}{4\pi G}} \frac{1}{r_{b}M} \leq J_{\varphi} $\\
~\\
In this range, both Euclidean BH, BG instanton and PL type solutions exist. The maximum entropy and energy still exist but are now due to the (PL-)BG instanton.  The qualitative behavior of entropies against energy is as shown in Fig. \ref{FIG4no1no6} (\textbf{Left}).
\footnote{
As I have shown in the previous section, for some parameter, the existence region on $J_\varphi - r_{H(c)}$ plane of PL type solution becomes complicated (Fig. \ref{FIG4no1noSHIN6} (\textbf{Right}) ). In that case, the behavior of branch is modified in the low entropy region. However, it does not affect the existence of entropy bound and thermodynamical (in)stability of the system.
}
                                                 %
\iffigure
\begin{figure}
\begin{center}
	\includegraphics[width=8.cm]{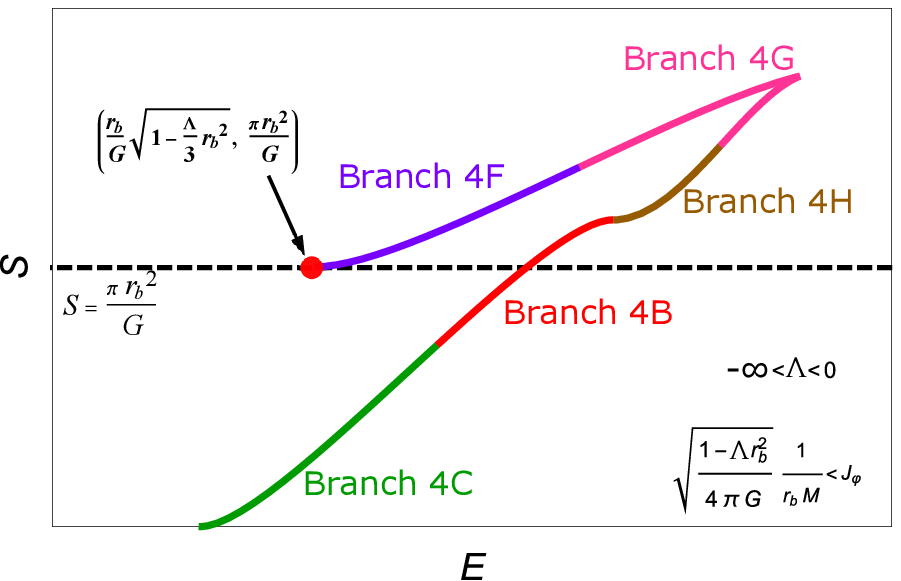} ~~ 	\includegraphics[width=8.cm]{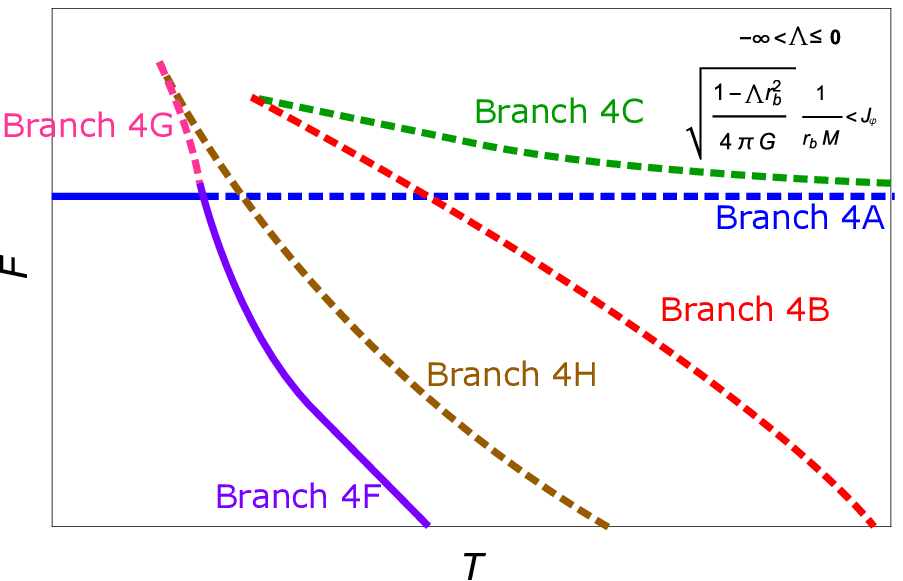} 
	\caption{Qualitative behavior of entropies against energy and that of free energy against temperature when $-\infty<\Lambda\leq 0$ and $\sqrt{\frac{1-\Lambda r_{b}^2}{4\pi G}} \frac{1}{r_{b}M} \leq J_{\varphi} $ (and when both $|\Lambda|/M^2$  and $r_{b}M$ are large enough, or $J_{\varphi}$ is enough large). (\textbf{Left}) The branches 4B and 4G are thermodynamically stable branches and 4C, 4F, and 4H are  thermodynamically unstable ones. 4F and 4G are consist of BG instantons. For 4C, 4B, and 4H, which part of the branches corresponds to which type of solution depends on parameters. (\textbf{Right}) At low temperature, branch 4A, which corresponds to Euclidean (PL-)soliton, dominates and, at high temperature, 4F dominates. Therefore, the system is thermodynamically stable at low temperature but unstable at high temperature.  }
\label{FIG4no1no6}
\end{center}
\end{figure}
\fi
                                                 %
There are 5 types of branch; the branches 4B and 4G are convex and 4C, 4F, and 4H are concave, i.e., the former correspond to thermodynamically stable branches and the latter to thermodynamically unstable ones. 4F and 4G are consist of BG instantons. For 4C, 4B, and 4H, which part of the branches corresponds to which type of solution depends on parameters. For example, when $J_{\varphi}$ is close to $\sqrt{\frac{1-\Lambda r_{b}^2}{4\pi G}} \frac{1}{r_{b}M}$, the most part of 4B and 4B consists of Euclidean BHs, the most part of 4H consists of BG instantons, and the small part near the junction of 4B and 4H consists of PL type solutions. On the other hand, when we choose sufficiently large $J_{\varphi}$, BH solutions cease to exist and 4C, 4B, and 4H consist of only PL type solutions and BG instantons. 
According to my choice of ground state energy, the energy at the rightmost of the Euclidean BH branch or at the leftmost of 4F is always $\frac{r_{b}}{G}\sqrt{1-\frac{\Lambda}{3}r_{b}^2}$. 

The qualitative behavior of free energies is shown in Fig. \ref{FIG4no1no6} (\textbf{Right}).
As we can see, the low temperature region is always dominated by thermodynamically stable branches and it does not suffer from the instability by subdominant unstable branches. In the high temperature region, there always exists the thermodynamically unstable branch 4F and it always gives the dominant contribution. Therefore, at high temperature, the system is thermodynamically unstable.

To summarize, when $\sqrt{\frac{1-\Lambda r_{b}^2}{4\pi G}} \frac{1}{r_{b}M} \leq J_{\varphi} $, the entropy bound still exists and the system is thermodynamically stable at low temperature, but thermodynamically unstable at high temperature. This strange behavior is due to the existence of (PL-)BG instanton. \\
~ \\

\subsubsection{$0<\Lambda \leq \frac{1}{r_{b}^2} $}
As I have shown in the previous subsection, the types of BG instanton and PL type solution are numerous and the behavior is complicated when $0<\Lambda \leq \frac{1}{r_{b}^2} $, such as in Figs. \ref{FIG4no1no4}, \ref{FIG4no1no5} and \ref{FIG4no1noSHIN7}, \ref{FIG4no1noSHIN8}. Here, I will consider only the case of Fig. \ref{FIG4no1no4} and Fig. \ref{FIG4no1noSHIN7} (\textbf{Left}), that is, when $\Lambda/M^2$ and $r_{b}M$ have sufficient magnitude. It is clear that, at least, the additional structures at small horizon radius region in Fig. \ref{FIG4no1noSHIN7}, \ref{FIG4no1noSHIN8} does not affect the existence of the entropy bound.
\footnote{
I confirmed, by numerical calculation, that it does not affect thermodynamical stability either.
}

~\\
$\bullet$ $0 < J_{\varphi} < \sqrt{\frac{1-\Lambda r_{b}^2}{4\pi G}} \frac{1}{r_{b}M} $\\
~\\
Both Euclidean BH and BG instanton exist. The qualitative behaviors of entropies and free energies are shown in Fig. \ref{FIG4no1no8}. 
                                                 %
\iffigure
\begin{figure}
\begin{center}
	\includegraphics[width=8.cm]{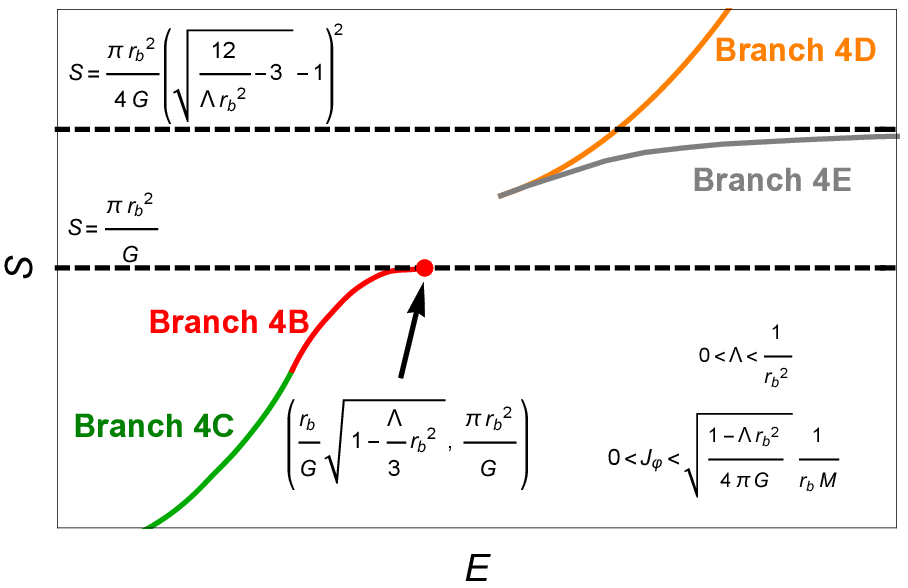} ~~~ \includegraphics[width=8.cm]{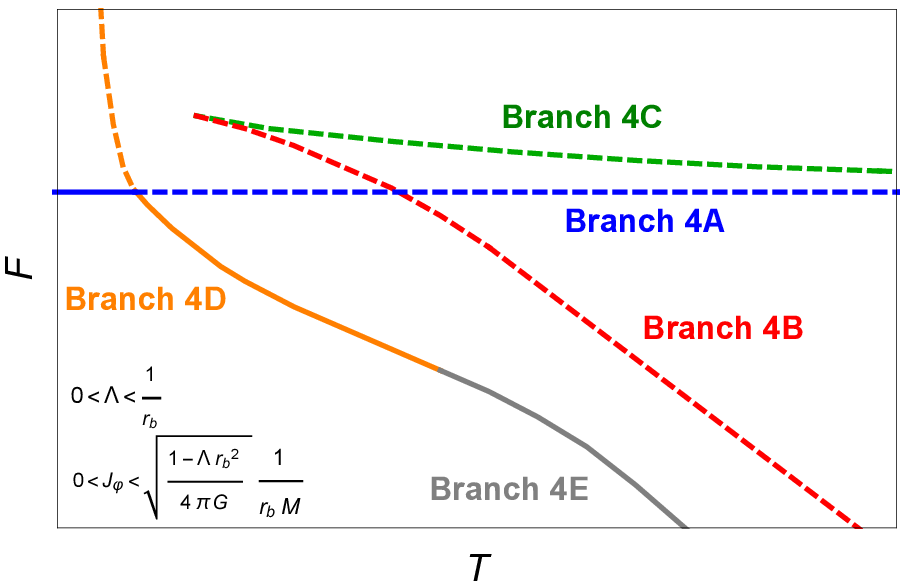} 
	\caption{Qualitative behaviors of entropies (left panel) and free energies (right panel) when $0<\Lambda< \frac{1}{r_{b}^2}$ and $0< J_{\varphi}< \sqrt{\frac{1-\Lambda r_{b}^2}{4\pi G}} \frac{1}{r_{b}M} $.  Branch 4A is of the Euclidean soliton, branches 4B and 4C are of the Euclidean BH, and branches 4D and 4E are of the BG instanton.  ($\BS{{\rm Left}}$) The entropy of branch 4D, which is the dominant saddle at high energy, has no bound and increases indefinitely. Unlike that branch, the entropy of 4E has the bound $S= \frac{\pi r_{b}^2}{4G} \left( \sqrt{\frac{12 }{\Lambda r_{b}^2} -3}-1 \right)^2$. ($\BS{{\rm Right}}$) The thermodynamically stable branches are 4A, 4B, and  4E. Although 4A dominates at low temperature, it suffers from quantum tunneling to 4D. Therefore, the system is thermodynamically unstable at low temperature. On the other hand, 4E does not suffer from quantum tunneling instabilities.    }
\label{FIG4no1no8}
\end{center}
\end{figure}
\fi
                                                 %
The thermodynamically stable branches are 4A, 4B, and 4E,  and the unstable ones are 4C and 4D. The entropy and energy of 4D grow indefinitely, so there are no entropy bounds. Although I do not know its implication, the system has an energy gap above $E=\frac{r_{b}}{G}\sqrt{1-\frac{\Lambda}{3}r_{b}^2}$. There are no such energy gaps in the pure gravity case. According to branch 4D, the system is thermodynamically unstable at low temperature; when 4D is dominant, apparently it is unstable, and when 4A is dominant, quantum tunneling may lead to an infinite growth of the cosmological horizon. However, due to 4E, the system is thermodynamically stable at high temperature. The situation is opposite to the case of $-\infty<\Lambda\leq0, \sqrt{\frac{1-\Lambda r_{b}^2 }{4\pi G}} \frac{1}{r_{b}M}<J_{\varphi}$, where the system is thermodynamically stable at low temperature and unstable at high temperature. Another property of the system in a thermal bath is the existence of an effective entropy bound; that is, there exists the maximum entropy in thermal equilibrium, $S=\frac{\pi r_{b}^2}{4G}\left( \sqrt{\frac{12}{\Lambda r_{b}^2}-3 }-1 \right)^2$, which can be reached when we take $T\to \infty$.
~\\
$\bullet$ $\sqrt{\frac{1-\Lambda r_{b}^2}{4\pi G}} \frac{1}{r_{b}M} < J_{\varphi} < \infty $ \\
~\\
We have seen there are two types of BG instanton: \\
$\cdot$ the branch that consists of 4F, 4G, 4H appearing when $-\infty<\Lambda \leq 0, \sqrt{\frac{1-\Lambda r_{b}^2}{4\pi G}} \frac{1}{r_{b} M}<J_{\varphi}< \infty$ \\
$\cdot$  the branch that consists of 4D, 4E appearing when $0<\Lambda \leq \frac{1}{r_{b}^2}, 0<J_{\varphi}<\sqrt{\frac{1-\Lambda r_{b}^2}{4\pi G}} \frac{1}{r_{b}M} $ \\
Roughly, the former causes the instability at high temperature and the latter causes the one at low temperature. In this parameter range, both types exist. As we will see shortly, they make the system thermodynamically unstable for all temperatures. Fig. \ref{FIG4no1no9} shows the behaviors of the free energies and entropies of the branch 4F, 4G, 4H, 4D, and 4E.
                                                 %
\iffigure
\begin{figure}
\begin{center}
	\includegraphics[width=8.cm]{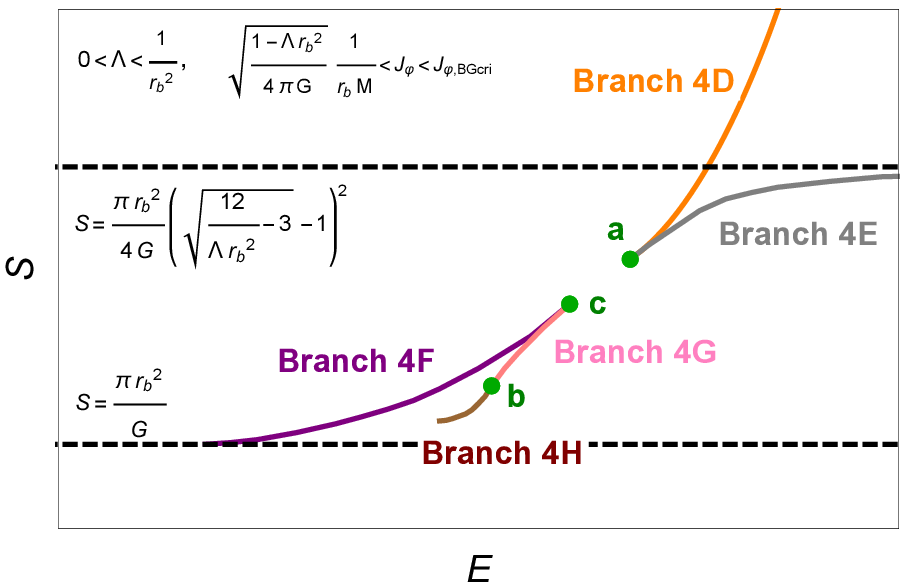} ~~~ \includegraphics[width=8.cm]{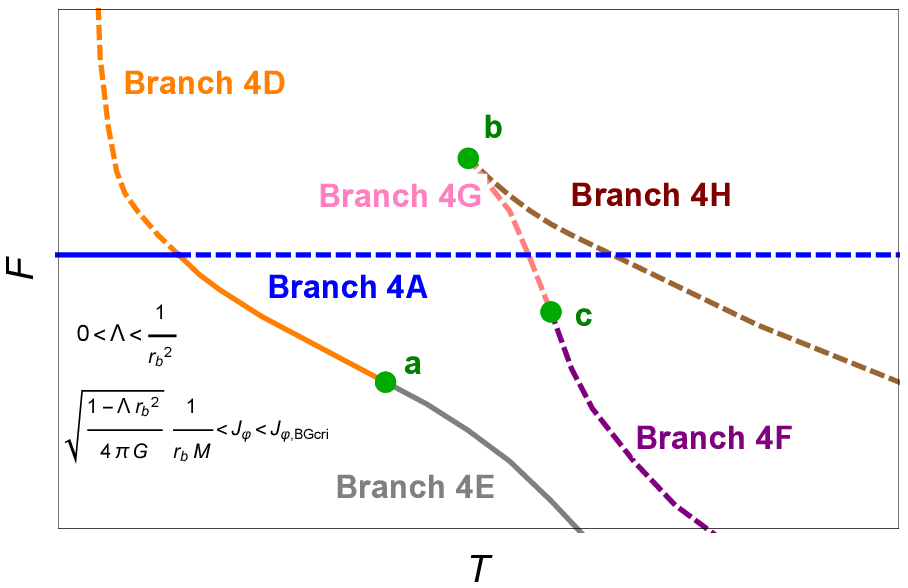} 
	\caption{Qualitative behaviors of entropies (left panel) and free energies (right panel) of BG instanton (and Euclidean soliton) when $0<\Lambda< \frac{1}{r_{b}^2}$ and $\sqrt{\frac{1-\Lambda r_{b}^2}{4\pi G}} \frac{1}{r_{b}M} < J_{\varphi}< J_{\varphi, BGcri} $.  The thermodynamically stable branches are 4A, 4E, and 4G; the others are unstable. As we increase $J_{\varphi}$, the points \textbf{a}, \textbf{b}, and \textbf{c} become close and coincide when $J_{\varphi}$ becomes some critical value $J_{\varphi, BGcri}(r_{b}, \Lambda, M)$. Above that value, the branches are reconnected as in Fig. \ref{FIG4no1no10}. ($\BS{{\rm Left}}$) The dominant saddles are 4F and 4D. There is no entropy bound.  ($\BS{{\rm Right}}$) Although some temperature regions are dominated by thermodynamically stable branches 4A and 4E, they suffer from quantum tunneling instability. Therefore, the system is thermodynamically unstable.   }
\label{FIG4no1no9}
\end{center}
\end{figure}
\fi
                                                 %
Branch 4A, 4B, and 4C cease to exist above some critical value of $J_{\varphi}$. Even when they exists, it does not give the dominant contribution to the free energy, does not affect the thermodynamical stability, and does not relate to the existence of the entropy bound as we saw before. Therefore, I do not draw it in Fig. \ref{FIG4no1no9} (and Fig. \ref{FIG4no1no10}). From the figure, it is clear that there is no entropy bound. In the right panel of the figure, the low and high temperature regions are dominated by branches 4A and 4E, respectively, which are thermodynamically stable branches. Only the middle temperature region is dominated by the thermodynamically unstable branch 4D. However, both the low and high temperature regions suffer from quantum tunneling instability. Eventually, the system is thermodynamically unstable  for all temperature regions. In the figure, the points \textbf{a}, \textbf{b},  and \textbf{c} denote the boundary between 4D and 4E, that between 4G and 4H, and that between 4F and 4G, respectively. As $J_{\varphi}$ is increased, these points become closer. At some critical value, which I call $J_{\varphi, BGcri}(r_{b}, \Lambda, M)$, these points coincide (i.e., branch 4G disappears). Above that value, branches 4D, 4E, 4F, and 4H are reconnected as in Fig. \ref{FIG4no1no10}. 
                                                 %
\iffigure
\begin{figure}
\begin{center}
	\includegraphics[width=8.cm]{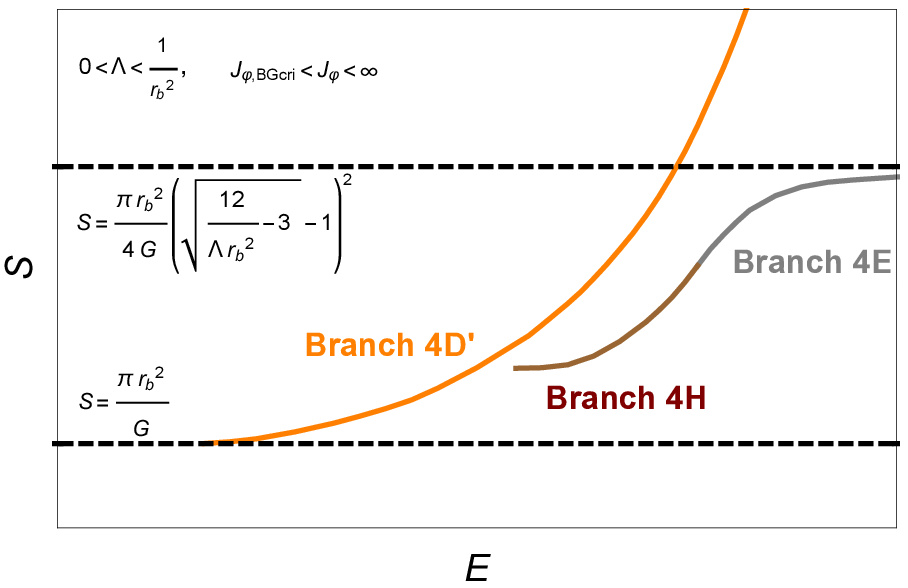} ~~~ \includegraphics[width=8.cm]{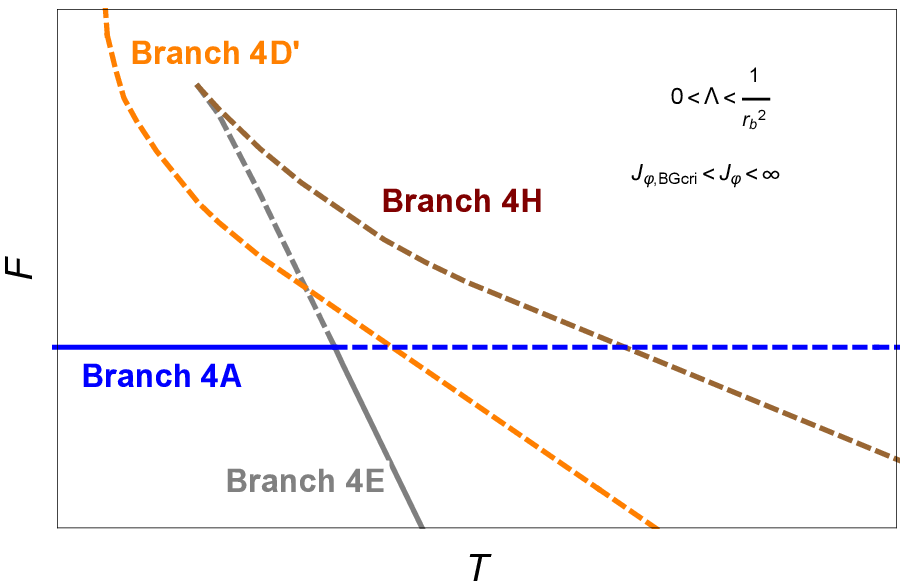} 
	\caption{Qualitative behaviors of entropies (left panel) and free energies (right panel) of BG instanton (and Euclidean soliton) when $0<\Lambda\leq \frac{1}{r_{b}^2}$ and $ J_{\varphi, BGcri}(r_{b}, \Lambda, M) < J_{\varphi}< \infty $. The thermodynamically stable branches are 4A and 4E; branches 4D' and 4H are unstable. ($\BS{{\rm Left}}$) The dominant saddle is 4D'. There is no entropy bound.  ($\BS{{\rm Right}}$) Although the dominant saddles are the thermodynamically stable branches 4A and 4E, they suffer from quantum tunneling instability. Therefore, the system is thermodynamically unstable.}
\label{FIG4no1no10}
\end{center}
\end{figure}
\fi
                                                 %
Branches 4E and 4H are connected, and 4D and 4F are connected. I rename the latter branch 4D'. Owing to the existence of 4D', the system has no entropy bound and is thermodynamically unstable.

\subsubsection{$\frac{1}{r_{b}^2}< \Lambda \leq  \frac{3}{r_{b}^2}$}
When $\Lambda$ is in this range, the behaviors of free energy and entropy of BG instanton branch are qualitatively the same as in the case of pure gravity with $\frac{1}{r_{b}^2}<\Lambda$ and branch 4D',  that is, there is no entropy bound.
Since the free energies of BH branches and PL branches are always subdominant same as before, the system is thermodynamically unstable. 

\subsubsection{$\frac{3}{r_{b}^2}  < \Lambda <\infty$}
This is qualitatively the same as the pure gravity case with $\frac{3}{r_{b}^2} \leq \Lambda$. For any $J_{\varphi}$, the only saddle point is the BG instanton, which is thermodynamically unstable.

\section{Discussion}
\subsection*{Summary}
In this paper, I investigated the thermodynamical properties of the pure gravity system and those of the gravity-scalar system with $S^2 \times \MB{R}$ boundary geometry, assuming that all the Euclidean saddle points within minisuperspace (\ref{EQ2minispGRA}), (\ref{EQ2minispSCA}) contribute to the path integral. Firstly, I investigated the pure gravity case in section 3. As is known in the literature, the pure gravity system with $\Lambda \leq 0$ has the entropy bound $\frac{\pi r_{b}^2}{G}$, is thermodynamically stable, and exhibits a Hawking--Page phase transition \cite{HawkingPage, York2}. Here, $r_{b}$ is the radius of boundary $S^2$. I showed that these properties change drastically when $\Lambda>0$ due to the contribution of a new type of saddle point geometries, which I call the BG instanton; the system ceases to have an entropy bound and becomes thermodynamically unstable. In section 4, I extended the system by  including a scalar field with simple $\varphi^2$ potential. I showed that, when $\Lambda \leq 0$ and the boundary value of scalar field $J_{\varphi}$ is smaller than the critical value $\sqrt{\frac{1-\Lambda r_{b}^2}{4\pi G}} \frac{1}{r_{b}M}$, the system has both an entropy bound and thermodynamical stability. This is because there do not exist BG instantons in this parameter range. However, when $\Lambda>0$ or when $\sqrt{\frac{1-\Lambda r_{b}^2}{4\pi G}} \frac{1}{r_{b}M}<J_{\varphi}$, BG instantons exist and at least one of the properties is (partially) lost. Precisely,  depending on parameters, there can exist two type of BG instanton. The existence of the one makes the system lose the entropy bound and thermodynamical stability at low temperature, and that of the other makes the system lose thermodynamical stability at high temperature. I summarize the thermodynamical properties of the pure gravity system and the scalar-gravity system in Table \ref{TABLE5}.

\begin{table}[htb]
\begin{center}
  \begin{tabular}{|c|c|c||c|c|c|} \hline
    ~ & $\Lambda$ & $(0 \leq) J_{\varphi}$ & Entropy bound &  $\begin{matrix}
{\rm Thermodynamical}  \\
{\rm stability}  \\
\end{matrix} $ & $\begin{matrix}
{\rm Transition}  \\
{\rm temperature} ~ T_{tr} \\
\end{matrix} $ \\ \hline 
    Pure  & \rotatebox[origin=c]{90}{\hspace{1.0cm}} $ \Lambda \leq 0$ & N/A & $\frac{\pi r_{b}^2}{G} $ & stable & $\frac{27}{32\pi r_{b}} \leq T_{tr} < \frac{1}{\pi r_{b}} $ \\ \cline{2-6}
    gravity & \rotatebox[origin=c]{90}{\hspace{0.8cm}}
   $ 0< \Lambda $  & N/A & none & unstable & N/A \\ \hline

      &  & 
$\begin{matrix}
 J_{\varphi} 
\leq \sqrt{\frac{1-\Lambda r_{b}^2}{4\pi G}} \frac{1}{r_{b}M}  \\
\end{matrix} $
\ & $\frac{\pi r_{b}^2}{G} $ & stable & $\sqrt{\frac{2}{3}} \frac{1}{\pi r_{b}} < T_{tr} < \frac{1}{\pi r_{b}} $ \\ \cline{3-6}
\rotatebox[origin=c]{90}{\hspace{1.5cm}} Gravity & $ \Lambda \leq 0$ &   
  $\begin{matrix}
\sqrt{\frac{1-\Lambda r_{b}^2}{4\pi G}} \frac{1}{r_{b}M}  
 < J_{\varphi}  \\
\end{matrix} $
 & $\frac{\pi r_{c,max}^2}{G} \left( > \frac{\pi r_{b}^2}{G} \right) $  & $\begin{matrix}
{\rm low } ~ T: ~ {\rm stable }  \\
{\rm high } ~ T: ~ {\rm unstable }  \\
\end{matrix} $ & N/A   \\ \cline{2-6}

-scalar &  &
$\begin{matrix}
 J_{\varphi} 
\leq \sqrt{\frac{1-\Lambda r_{b}^2}{4\pi G}} \frac{1}{r_{b}M}  \\
\end{matrix} $
 & none  & $\begin{matrix}
{\rm low } ~ T: ~ {\rm unstable }  \\
{\rm high } ~ T: ~ {\rm stable }  \\
\end{matrix} $ & N/A \\ \cline{3-6}
  \rotatebox[origin=c]{90}{\hspace{1.5cm}}   & $ 0< \Lambda \leq \frac{1}{r_{b}^2}$  & 
   $\begin{matrix}
\sqrt{\frac{1-\Lambda r_{b}^2}{4\pi G}} \frac{1}{r_{b}M}  < J_{\varphi}  \\
\end{matrix} $
 & none  & unstable & N/A \\ \cline{2-6}

 &\rotatebox[origin=c]{90}{\hspace{0.8cm}} $ \frac{1}{r_{b}^2} < \Lambda $ &
$0 \leq J_{\varphi}  $
 & none  & unstable & N/A \\ \hline

  \end{tabular}
  \end{center}
  \caption{Summary of thermodynamical properties of pure gravity system and gravity-scalar system. The definitions of $J_{\varphi, BHcri}=J_{\varphi, BHcri}(r_{b}, \Lambda, M)$ and $r_{c, max}=r_{c, max}(r_{b}, J_{\varphi}, \Lambda, M)$ are written somewhere in section 4. Note that, by symmetry, we can restrict $J_\varphi$ to non-negative value. }
  \label{TABLE5}
\end{table}

\subsection*{Other potentials}
Although I consider the simplest potential in this paper, some properties may be common for other potentials. For example, consider a $\MB{Z}_{2}$ symmetric concave potential. If we assume that the scalar field always decreases toward the bolt or the center, as it does for the $\varphi^2$ potential, we can obtain the critical value of $J_{\varphi}$ below which the system is thermodynamically stable when $\Lambda \leq 0$ ($\sqrt{\frac{1-\Lambda r_{b}^2}{4\pi G}} \frac{1}{r_{b}M} $ for the $\varphi^2$ potential case).  For example, when $V(\varphi) = \frac{1}{2n} \lambda_{2n} \varphi^{2n}$, the critical value may be $J_{\varphi}=\left( \frac{n(1-\Lambda r_{b}^2)}{4\pi G \lambda_{2n} r_{b}^2 } \right)^{\frac{1}{2n}}$. We can also say that the minimal transition temperature is $T_{tr}=\sqrt{\frac{2}{3}}\frac{1}{\pi r_{b}}$, irrelevant of the details of the potential as long as it is $\MB{Z}_{2}$ symmetric and concave. This may be obtained by taking the limit of parameters of the potential so that it becomes flat, keeping $J_{\varphi}$ equal to the critical value. Of course, in order to check these matter, more detailed investigations are needed.

Another direction would be to consider double well type potentials, with false and true vacuums. For definiteness, let the true vacuum be a negative value and the false vacuum be a positive value, and let the potential form be suitable for thin-wall approximation. When we choose $J_{\varphi}$ as the true vacuum, there are two types of solution: one is the same as the solution in pure gravity with $\Lambda <0$, and the other is constructed from the $\Lambda < 0$ solution, the $\Lambda > 0$ solution, and a thin wall. The thermodynamical properties of the system would be different from the ones considered in this paper, and it is enough simple to analyze the detail. This will be left for future study.  

\subsection*{Gravity-Maxwell system}
There exist BG instantons also in the gravity-Maxwell system. Consider a Reissner--Nordstrom BH solution and let $r_{in}$ be the radius of the inner horizon and $r_{b}$ be some radius smaller than $r_{in}$. If we Euclideanize the $[r_{b}, r_{in}]$ region with an appropriate circumference, we obtain a BG instanton. Although the thermodynamics of the gravity-Maxwell system was previously investigated in \cite{BradenBrownWhitingYork, CarlipVaidya, Lundgren, BasuKrishnanBalaSubramanian}, the effect of a BG instanton was not considered. It would be interesting to consider its effect and see how its thermodynamical properties are modified.

\subsection*{Integration contour problem}
In this paper, I assumed that all Euclidean saddle points contribute to the partition function. This assumption leads to the inclusion of BG instantons in the path integral and makes the system thermodynamically unstable for some cases. However, as shown in \cite{GibbonsHawkingPerry}, in order to define the Euclidean gravitational path integral, its contour must be purely complex. Therefore, some Euclidean saddles may not contribute and some complex saddles may contribute to the path integral. Following the principle of \cite{HalliwellHartle}, I have to check whether there exist one or more steepest descent contours passing through all Euclidean saddles shown in this paper. I also have to check, if they pass through complex saddles, whether they do not change the thermodynamical properties. These issues will be investigated in \cite{Miyashita3} for the pure gravity case.

\section*{Acknowledgement}
The author is grateful to Shintaro Sato for useful discussion. This work was supported in part by the Waseda University Grant for Special Research Project (No.\ 2020C-775).

\appendix
\section{Numerical solutions in the gravity-scalar system}
In this Appendix, I explicitly show some numerical solutions of the gravity-scalar system. The equation of motion of the system (\ref{EQ2grasca}) is given by
\bea
G_{\mu\nu}+\Lambda  g_{\mu\nu} = 8\pi G T_{\mu\nu} \\
(\square - M^2)\varphi=0 \hspace{1.2cm}
\ena
where the energy momentum tensor is given by
\bea
T_{\mu\nu} = \frac{2}{\sqrt{ g}} \frac{\delta I^{E}_{c, scalar}}{\delta g^{\mu\nu}}
 =  \pd_{\mu}\varphi \pd_{\nu}\varphi - \frac{1}{2}g_{\mu\nu} (g^{\alpha\beta}\pd_{\alpha}\varphi \pd_{\beta} \varphi + M^2 \varphi^2)
\ena
Substituting the ansatz (\ref{EQ4ansatz}) into the above equation, we obtain the following basic equations:
\bea
\delta^{\p} = -4\pi G r (\varphi^\p)^2 \hspace{2.1cm} \notag  \\
f^{\p} + \frac{f-1+\Lambda r^2}{r} = - 4\pi G r \left[ f(\varphi^{\p})^2 + M^2 \varphi^2 \right] \label{EQAbasic} \\
 -f\delta^{\p} \varphi^{\p} + \frac{2}{r} f \varphi^{\p} + f^{\p} \varphi^{\p} + f \varphi^{\p\p} - M^2 \varphi =0 \notag
\ena
There are three types of solutions: Euclidean soliton, Euclidean BH, and BG instanton. I will explain the boundary conditions and show numerical solutions separately.
                                                 %
\iffigure
\begin{figure}
\begin{center}
	\includegraphics[width=7.cm]{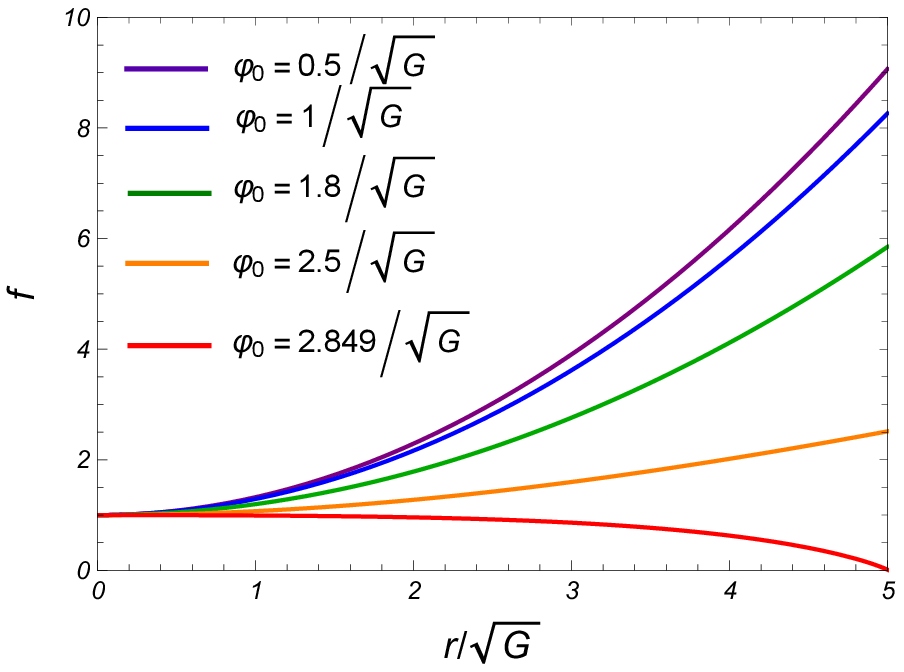} ~~ \includegraphics[width=7.cm]{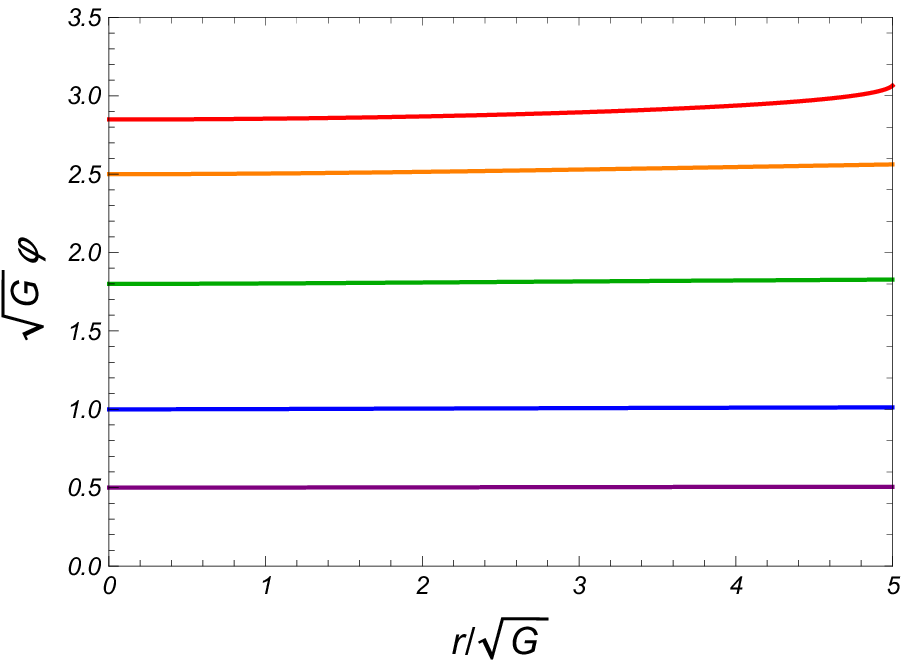} ~~ \includegraphics[width=7.cm]{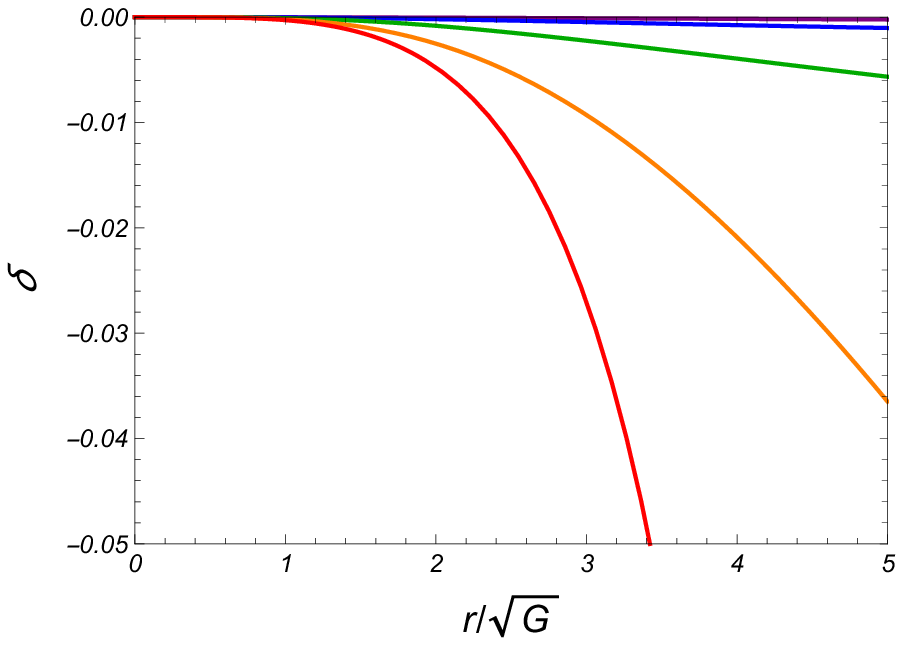} 
	\caption{Solutions $f, \varphi, \delta$ of Euclidean soliton when $\varphi_{0}$ is $0.5/\sqrt{G}, 1/\sqrt{G}, 1.8/\sqrt{G}, 2.5/\sqrt{G},$ and $2.849/\sqrt{G}$ ($r_{b}=5\sqrt{G}, \Lambda= -0.04/G ,M=0.1/\sqrt{G}$).}
\label{FIGAno1}
\end{center}
\end{figure}
\fi
                                                 %
\subsection{Euclidean soliton}
For a solitonic solution, the functions must behave as follows near the center $r=0$ for regularity:
\bea
\delta(r)= - \frac{\pi G M^4 \varphi_{0}^2}{9}r^4 + \cdots \hspace{1.3cm} \\
f(r)=1- \frac{\Lambda + 4\pi G M^2 \varphi_{0}^2 }{3}  r^2 + \cdots \\
\varphi(r)= \varphi_{0} + \frac{M^2}{6} \varphi_{0} r^2 + \cdots \hspace{1.3cm} 
\ena
where $\varphi_{0}$ is a shooting parameter that must be chosen for the functions to satisfy desired boundary conditions at $r=r_{b}$: 
\bea
\varphi(r_{b})=J_{\varphi}
\ena
Note that I set $\delta(0)=0$ using the freedom of rescaling the time coordinate. Examples of the solutions are shown in Fig. \ref{FIGAno1}. 

$\varphi(r)$ always increases toward the boundary from the center, and $f(r_{b})$ decreases as we increase $\varphi_{0}$. The limiting case where $f(r_{b})$ goes to zero will be discussed in A.4.
                                                 %
\iffigure
\begin{figure}
\begin{center}
	\includegraphics[width=7.cm]{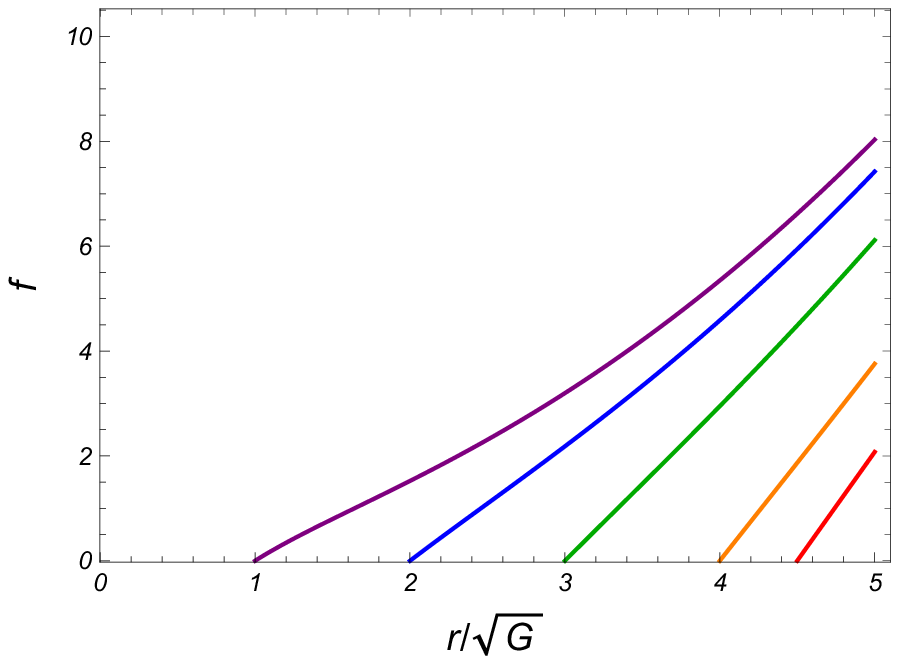} ~~ \includegraphics[width=7.cm]{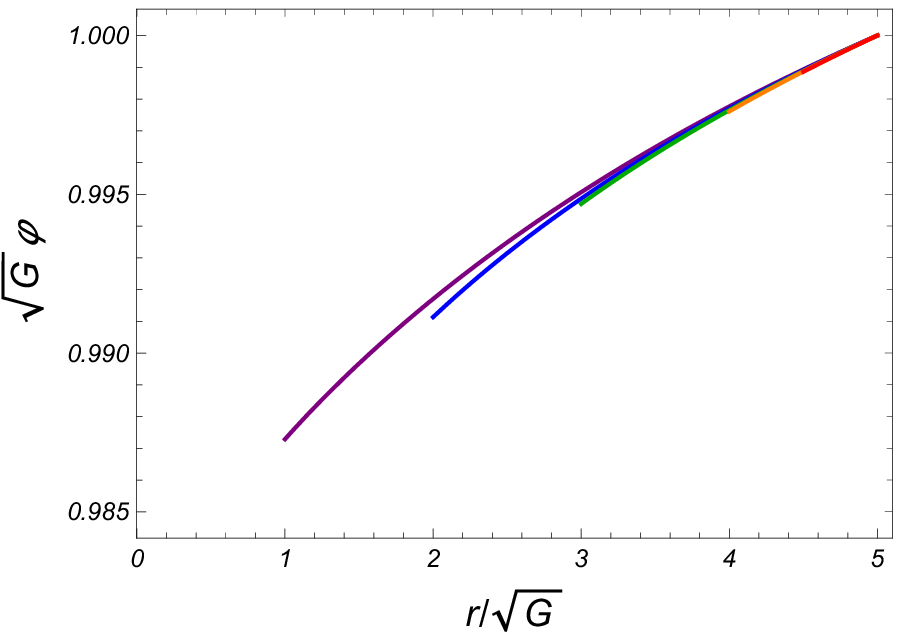} ~~ \includegraphics[width=7.cm]{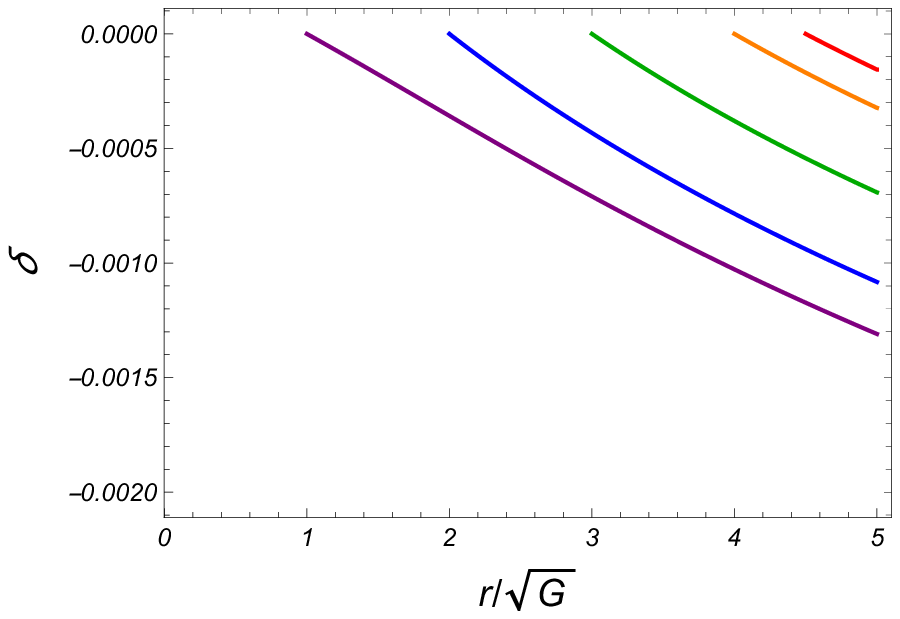} 
	\caption{Solutions $f, \varphi, \delta$ of Eulidean BH when $r_{H}$ is $\sqrt{G}, 2\sqrt{G}, 3\sqrt{G}, 4\sqrt{G},$ and $4.5\sqrt{G}$. $\varphi_{H}$ is chosen so that $J_{\varphi}= 1/\sqrt{G}< \sqrt{\frac{1-\Lambda r_{b}^2}{4\pi G}} \frac{1}{r_{b}M}$  ($r_{b}=5\sqrt{G}, \Lambda= -0.04/G ,M=0.1/\sqrt{G}$).}
\label{FIGAno2}
\end{center}
\end{figure}
\fi
                                                 %
                                                 %
\iffigure
\begin{figure}
\begin{center}
	\includegraphics[width=7.cm]{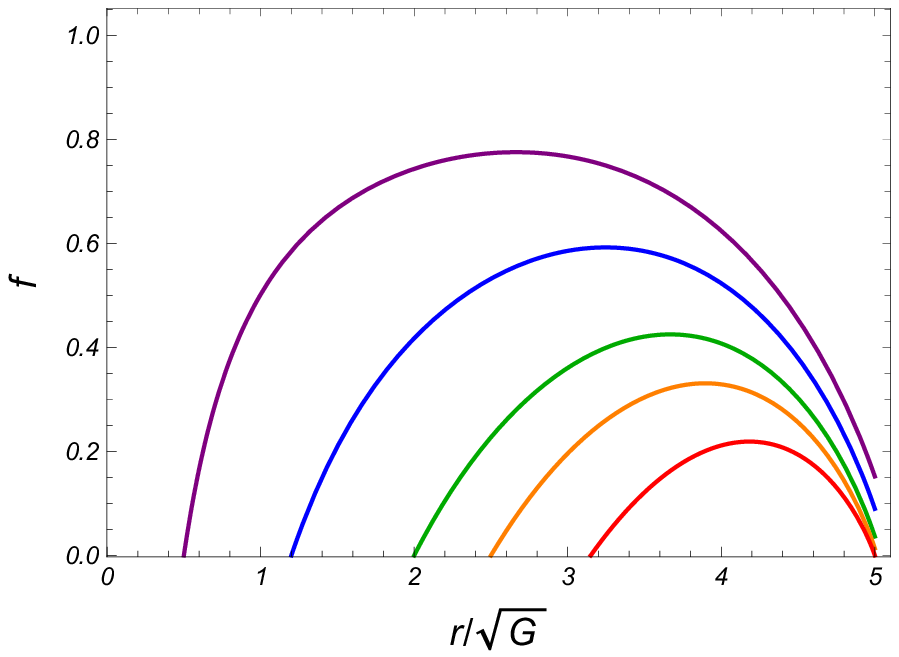} ~~ \includegraphics[width=7.cm]{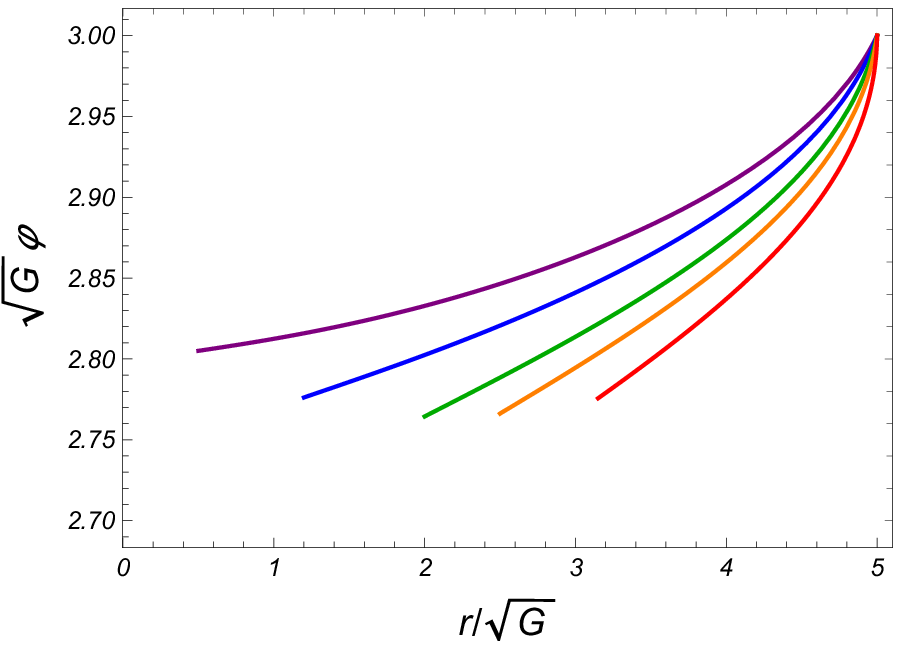} ~~ \includegraphics[width=7.cm]{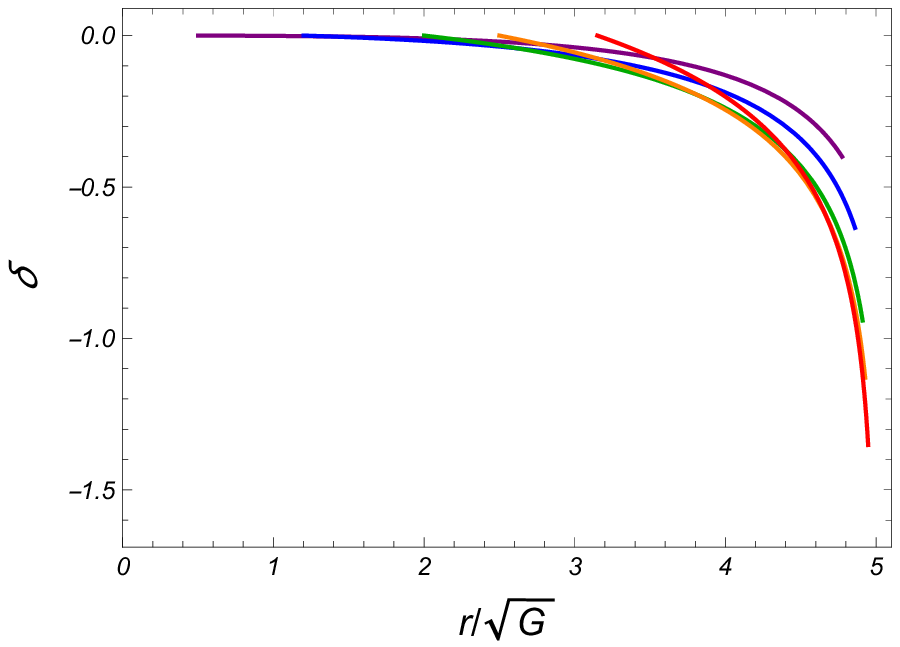} 
	\caption{Solutions $f, \varphi, \delta$ of Eulidean BH when $r_{H}$ is $0.5\sqrt{G}, 1.2\sqrt{G}, 2\sqrt{G}, 2.5\sqrt{G},$ and $3.15\sqrt{G}$. $\varphi_{H}$ is chosen so that $J_{\varphi}= 3/\sqrt{G}> \sqrt{\frac{1-\Lambda r_{b}^2}{4\pi G}} \frac{1}{r_{b}M}$.  ($r_{b}=5\sqrt{G}, \Lambda= -0.04/G ,M=0.1/\sqrt{G}$) The maximum radius is $r_{H, max}(5\sqrt{G}, -0.04/G, 3/\sqrt{G}, 0.1/\sqrt{G}) \simeq 3.15\sqrt{G}$.}
\label{FIGAno3}
\end{center}
\end{figure}
\fi
                                                 %
\subsection{Euclidean BH}
Let the radius of the bolt (horizon) be $r_{H}<r_{b}$ where $f(r)$ is zero. The functions must behave as follows near the bolt for regularity:
\bea
\delta(r)=  -4 \pi G r_{H}\left( \frac{r_{H} M^2 \varphi_{H} }{1-\Lambda r_{H}^2 - 4 \pi G r_{H}^2 M^2 \varphi_{H}^2  } \right)^2 (r-r_{H}) + \cdots  \\
f(r)= \frac{1-\Lambda r_{H}^2 - 4 \pi  G r_{H}^2 M^2 \varphi_{H}^2  }{r_{H}} (r-r_{H}) + \cdots \hspace{2.3cm} \\
\varphi(r)= \varphi_{H} +  \frac{r_{H} M^2 \varphi_{H} }{1-\Lambda r_{H}^2 - 4 \pi G r_{H}^2 M^2 \varphi_{H}^2  } (r-r_{H}) + \cdots \hspace{1.4cm} 
\ena
where $\varphi_{H}$ is a shooting parameter. Therefore, there are two shooting parameters $\{ r_{H}, \varphi_{H} \}$, and they must be chosen for the functions to satisfy desired boundary conditions at $r=r_{b}$:
\bea
\frac{4\pi}{f^{\p}(r_{H})}\sqrt{f(r_{b})}e^{-\delta(r_{b})} = \beta \hspace{0.2cm} \\
\varphi(r_{b})=J_{\varphi}
\ena
I set $\delta(r_{H})=0$ using the freedom of rescaling the time coordinate. When we fix $r_{H}$ and vary $\varphi_{0}$, the behaviors of solutions are similar to those in the Euclidean soliton case (Fig. \ref{FIGAno1}).  In Fig. \ref{FIGAno2}, I show the behavior of solutions with varying $r_{H}$ but fixing $J_{\varphi}$ to a value less than $\sqrt{\frac{1-\Lambda r_{b}^2}{4\pi G}} \frac{1}{r_{b}M}$. In this case, the horizon radius can take any value in $(0, r_{b})$. In Fig. \ref{FIGAno3}, I show the behavior of solutions with varying $r_{H}$ but this time fixing $J_{\varphi}$ to a value greater than $\sqrt{\frac{1-\Lambda r_{b}^2}{4\pi G}} \frac{1}{r_{b}M}$. As we increase the horizon radius, $f(r_{b})$ goes to zero. Therefore, there exists the maximum horizon radius $r_{H, max}(r_{b}, \Lambda, J_{\varphi}, M)$. This limiting case will be discussed in A.4.

                                                 %
\iffigure
\begin{figure}
\begin{center}
	\includegraphics[width=7.cm]{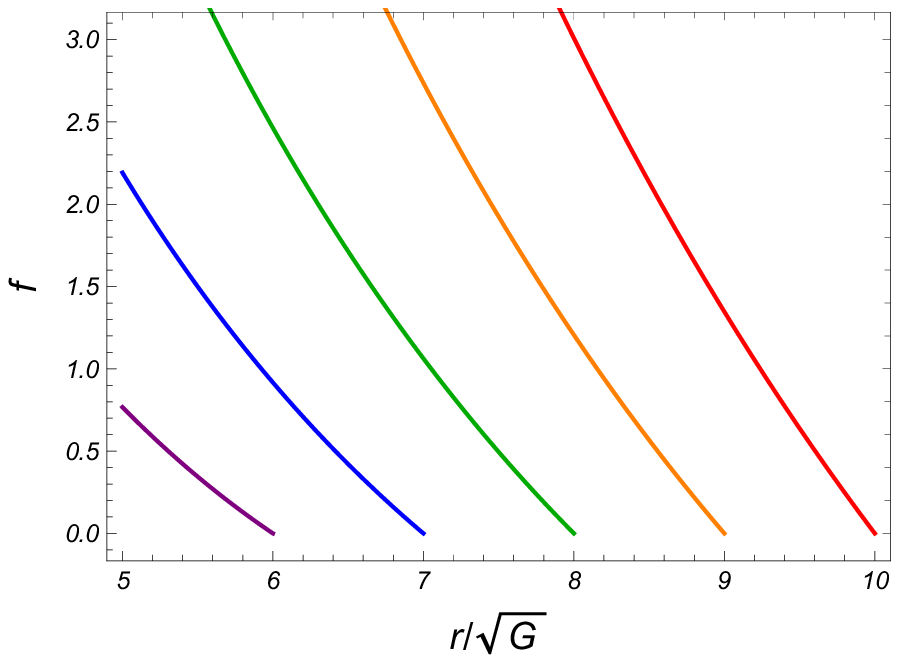} ~~ \includegraphics[width=7.cm]{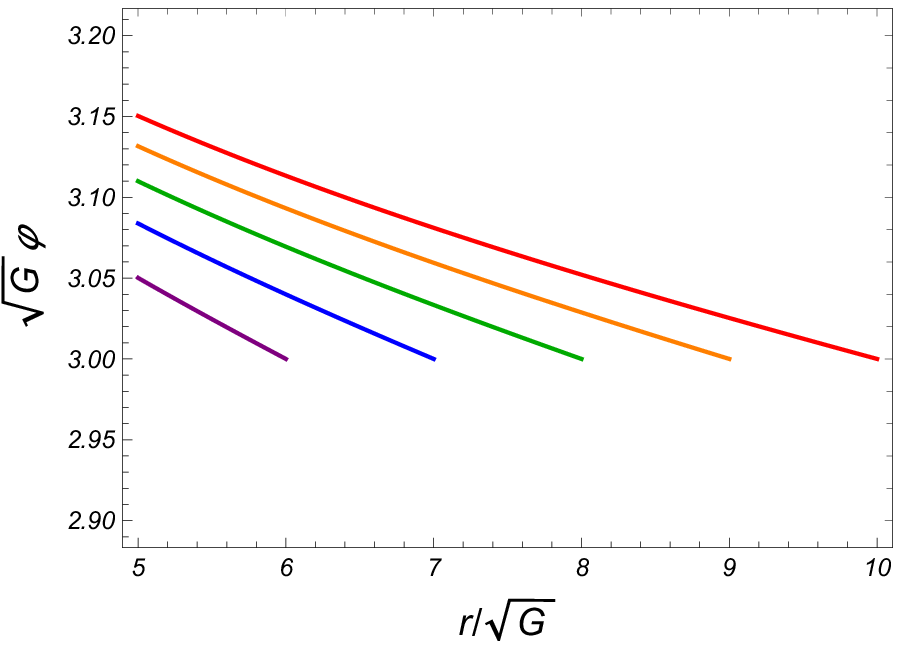} ~~ \includegraphics[width=7.cm]{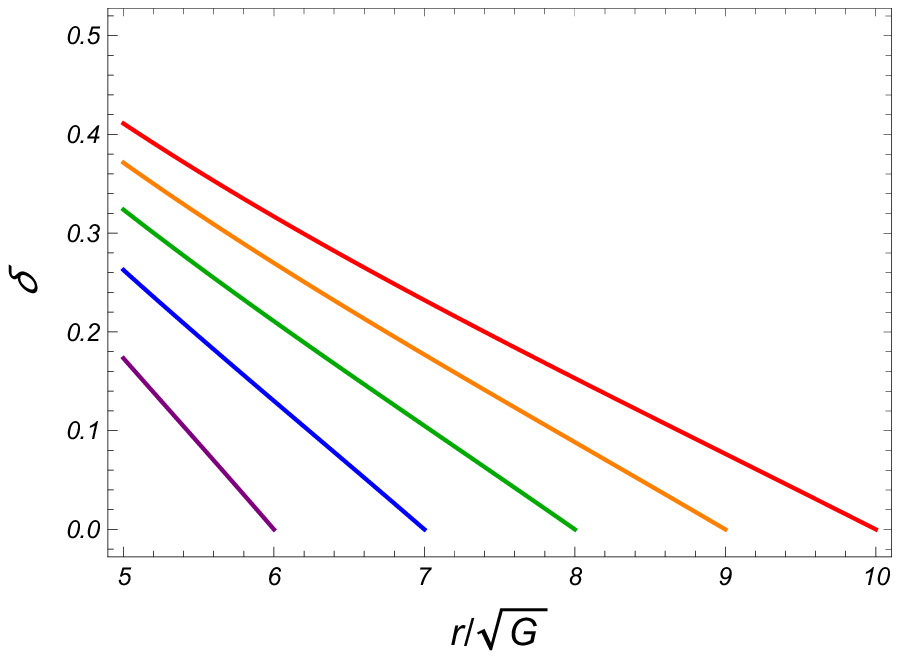} 
	\caption{Solutions $f, \varphi, \delta$ of BG instanton when $r_{c}$ is $6\sqrt{G}, 7\sqrt{G}, 8\sqrt{G}, 9\sqrt{G},$ and $10\sqrt{G}$. $\varphi_{H}$ is set to $3/\sqrt{G}$ ($r_{b}=5\sqrt{G}, \Lambda= -0.04/G ,M=0.1/\sqrt{G}$). }
\label{FIGAno4}
\end{center}
\end{figure}
\fi
                                                 %
                                                 %
\iffigure
\begin{figure}
\begin{center}
	\includegraphics[width=7.cm]{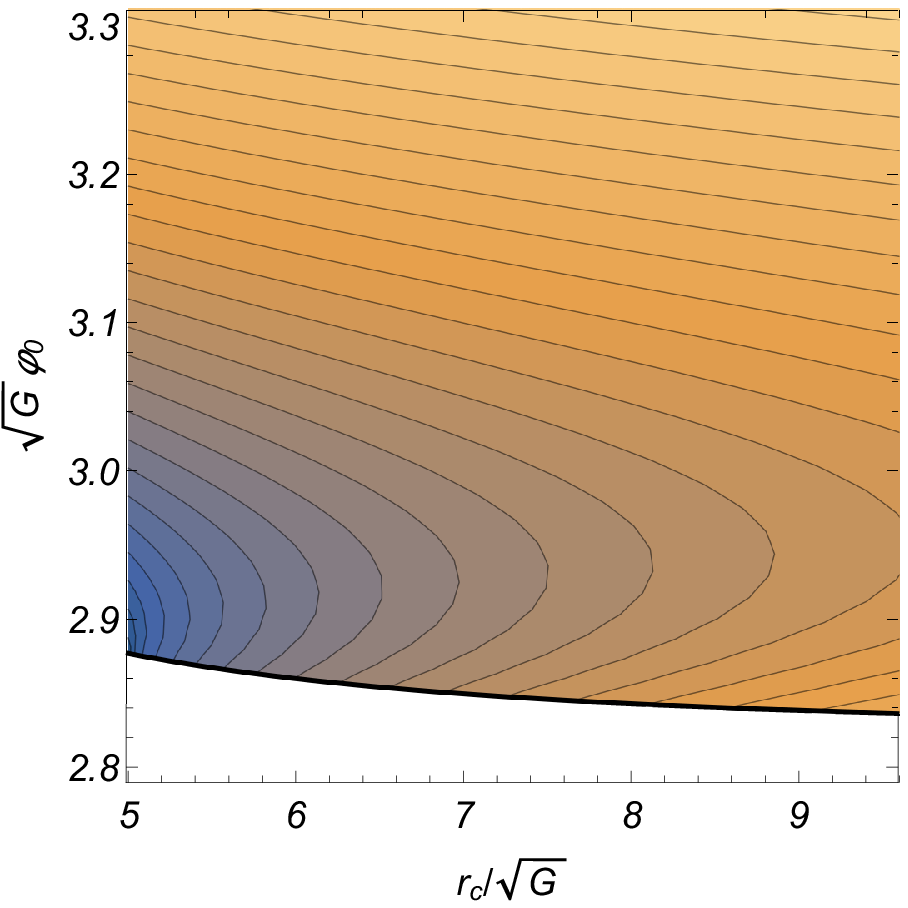} ~~ \includegraphics[width=0.95cm]{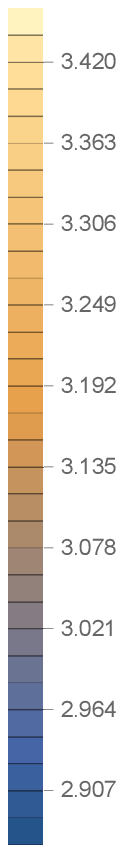} ~~ 
	\caption{Contour plot on $r_{c}-\varphi_{0}$ plane. The height   represents the value of $J_{\varphi}$  ($r_{b}=5\sqrt{G}, \Lambda= -0.04/G ,M=0.1/\sqrt{G}$). }
\label{FIGAno5}
\end{center}
\end{figure}
\fi
                                                 %
\subsection{BG instanton}
Let the radius of the bolt (cosmological horizon) be $r_{c}>r_{b}$ where $f(r)$ is zero. The functions must behave as follows near the bolt for regularity:
\bea
\delta(r)=  -4 \pi G r_{c}\left( \frac{r_{c} M^2 \varphi_{c} }{1-\Lambda r_{c}^2 - 4 \pi G r_{c}^2 M^2 \varphi_{c}^2  } \right)^2 (r-r_{c}) + \cdots  \\
f(r)= \frac{1-\Lambda r_{c}^2 - 4 \pi  G r_{c}^2 M^2 \varphi_{c}^2  }{r_{c}} (r-r_{c}) + \cdots \hspace{2.2cm} \\
\varphi(r)= \varphi_{c} +  \frac{r_{c} M^2 \varphi_{c} }{1-\Lambda r_{c}^2 - 4 \pi G r_{c}^2 M^2 \varphi_{c}^2  } (r-r_{c}) + \cdots \hspace{1.4cm} 
\ena
As in the Euclidean BH case, there are two shooting parameters $\{ r_{c}, \varphi_{c} \}$, and they must be chosen for the functions to satisfy desired boundary conditions at $r=r_{b}$:
\bea
-\frac{4\pi}{f^{\p}(r_{c})}\sqrt{f(r_{b})}e^{-\delta(r_{b})} = \beta \hspace{0.cm} \\
\varphi(r_{b})=J_{\varphi}
\ena
I set $\delta(r_{c})=0$ as before. Examples of the solutions are shown in Fig. \ref{FIGAno4}. 

In subsection 4.1, I characterized BG instantons by $\{ r_{c}, J_{\varphi} \}$ in order to see the connection to boundary quantities. However, these parameters are not good for parameterizing BG instantons. Instead, we can use $\{ r_{c}, \varphi_{0} \}$ for parameterization. Fig. \ref{FIGAno5} shows a contour plot on the $r_{c}-\varphi_{0}$ plane, where the height represents the value of $J_{\varphi}$. The black curve represents $\varphi_{0}= \sqrt{\frac{1-\Lambda r_{c}^2}{4\pi G}} \frac{1}{r_{c} M} $, below which no BG instantons exist. 

Note that Fig. \ref{FIGAno5} is for $\Lambda < 0$, and the situation is more complicated for $\Lambda >0$.

                                                 %
\iffigure
\begin{figure}[h]
\begin{center}
	\includegraphics[width=7.cm]{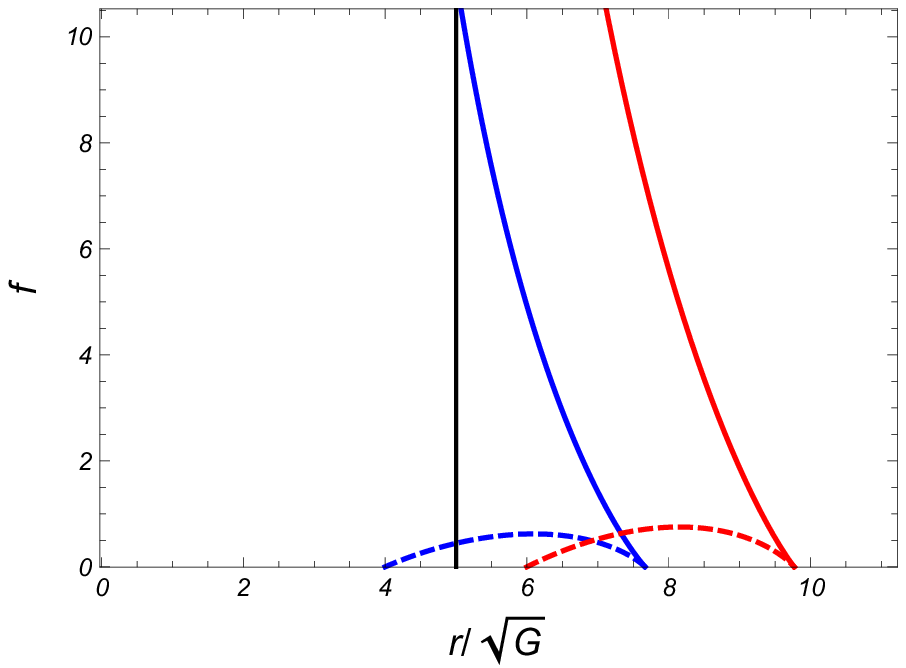} ~~ \includegraphics[width=7.cm]{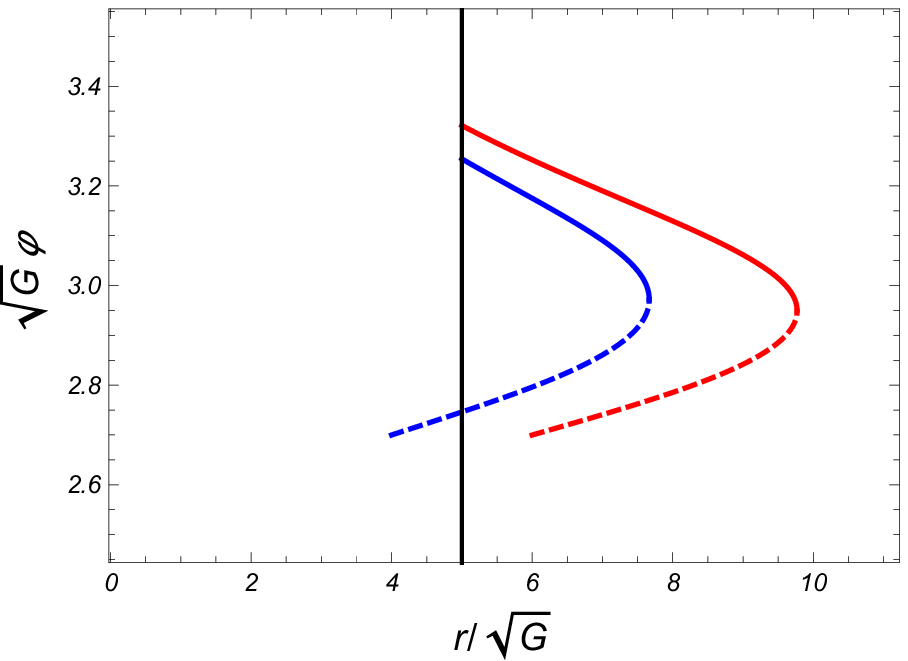} ~~ \includegraphics[width=7.cm]{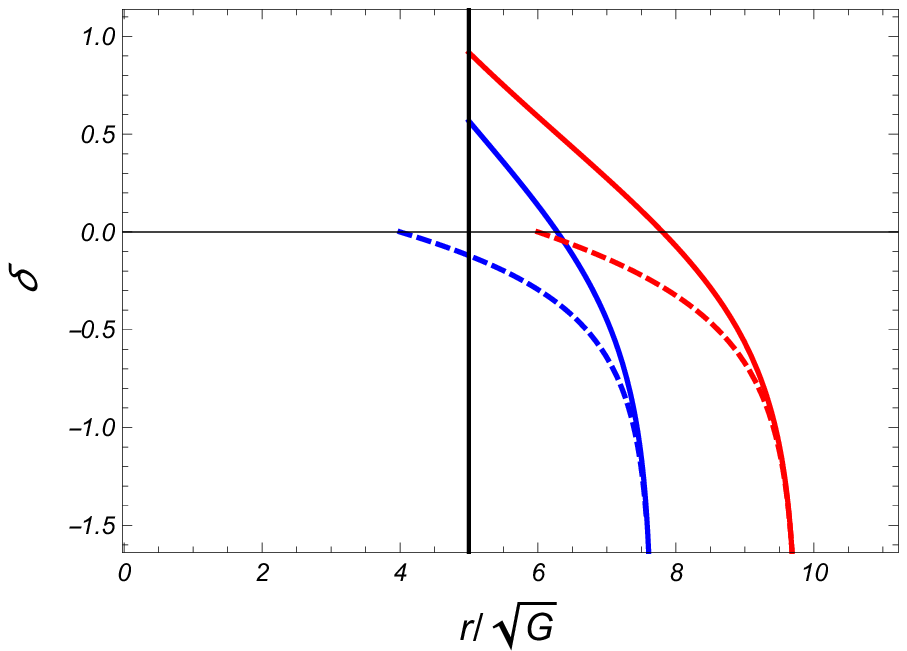} 
	\caption{Solutions $f, \varphi, \delta$ of Euclidean PL-BH and PL-BG instanton when $r_{H}= 4 \sqrt{G}, \varphi_{H}=2.7\sqrt{G} $ for the former and $r_{H}= 6 \sqrt{G}, \varphi_{H}=2.7\sqrt{G} $ for the latter. The vertical line represents the position of the boundary. $r_{max}\simeq 7.665\sqrt{G}$ for the former and $r_{max}\simeq 9.769\sqrt{G}$ for the latter.  ($r_{b}=5\sqrt{G}, \Lambda= -0.04/G ,M=0.1/\sqrt{G}$) }
\label{FIGAno6}
\end{center}
\end{figure}
\fi
                                                 %

\subsection{PL type solution}
When I obtained Euclidean soliton or Euclidean BH solutions, what I  practically did is to solve the equation motion from the center or the bolt to the outward, then stop the integration at some radius $r_{b}$. If we do not stop integration, it eventually hits the point where $f(r)=0$ (See Fig. \ref{FIGAno1}, \ref{FIGAno3}). If we recall the behavior of $f(r)$ of (Euclidean) dS or Schwarzschild dS solution, it may seem an analog of cosmological horizon. However, this is not the case. Let $r_{max}$ be the radius of the point $f(r)=0$, other than the radius of bolt when we consider Euclidean BH. Near $r_{max}$, the solutions may behave as follows;
\bea
\varphi(r)\simeq \varphi_{max} - \frac{1}{\sqrt{2 \pi G r_{max}}} (r_{max}-r)^{\frac{1}{2}} \hspace{3.05cm} \\
f(r) \simeq \left( 8\pi G r_{max} M^2 \varphi_{max}^2 - \frac{2}{r_{b}}\left(1-\Lambda r_{max}^2 \right)  \right) (r_{max}-r) \\
\delta(r) \simeq \delta_{max} + \frac{1}{2} \log (r_{max}-r) \hspace{4.55cm}
\ena
where $\varphi_{max}$ and $\delta_{max}$ are some constants. The solution with $r_{H}=3.15 \sqrt{G}$ in Fig. \ref{FIGAno3} is close to the above solution. Although $f(r)$ goes to zero, the geometry is not cupped off at $r_{max}$ because the $tt$ component of the metric stay finite $e^{-2\delta(r)}f(r) \simeq const. $ there. Therefore, if we use a suitable coordinate, we can extend the solution further and the resulting solution becomes the one shown in Fig. \ref{FIG4no1no1PL}.   The examples of Euclidean PL-BH and PL-BG instanton are shown in Fig. \ref{FIGAno6}.

\appendix

\end{document}